\documentclass[12pt]{article}
\usepackage{graphicx} 
\usepackage{multirow}
\usepackage{subcaption}
\usepackage{listings}
\usepackage{amsmath}
\usepackage{amsfonts}
\usepackage{amscd}
\usepackage{amssymb}
\usepackage{natbib}
\usepackage{url}
\usepackage{bbm}
\usepackage{bm}
\usepackage{array}
\usepackage{geometry}
\usepackage{setspace}
\usepackage{hyperref}
\usepackage{amsmath}
\usepackage{xcolor}
\usepackage{mathtools}
\usepackage{etoc}
\usepackage{booktabs}

\hypersetup{
    colorlinks=true,
    linkcolor={rgb,255:red,0;green,0;blue,139},
    citecolor={rgb,255:red,0;green,0;blue,139},
    urlcolor={rgb,255:red,0;green,0;blue,139}
}

\DeclareMathOperator*{\argmax}{arg\,max}

\def\m{\mathcal}
\def\mb{\mathbb}

\newcommand{\mnorm}[1]{{\vert\kern-0.25ex\vert\kern-0.25ex\vert #1 
    \vert\kern-0.25ex\vert\kern-0.25ex\vert}}
\newcommand{\bmnorm}[1]{{\big\vert\kern-0.25ex\big\vert\kern-0.25ex\big\vert #1 
    \big\vert\kern-0.25ex\big\vert\kern-0.25ex\big\vert}}
\newcommand{\Bmnorm}[1]{{\Big\vert\kern-0.25ex\Big\vert\kern-0.25ex\Big\vert #1 
    \Big\vert\kern-0.25ex\Big\vert\kern-0.25ex\Big\vert}}
\newcommand{\bbmnorm}[1]{{\bigg\vert\kern-0.25ex\bigg\vert\kern-0.25ex\bigg\vert #1 
    \bigg\vert\kern-0.25ex\bigg\vert\kern-0.25ex\bigg\vert}}
\newcommand{\BBmnorm}[1]{{\Bigg\vert\kern-0.25ex\Bigg\vert\kern-0.25ex\Bigg\vert #1 
    \Bigg\vert\kern-0.25ex\Bigg\vert\kern-0.25ex\Bigg\vert}}

\newtheorem{theorem}{Theorem}

\newtheorem{lemma}[theorem]{Lemma}
\newtheorem{corollary}[theorem]{Corollary}
\newtheorem{remark}{Remark}[theorem]

\newtheorem{definition}[theorem]{Definition}

\newtheorem{example}{Example}
\newenvironment{proof}{{\bf Proof:}}{\hfill\rule{2mm}{2mm}}

\usepackage{setspace}

\begin{document}
\makeatletter
\let\oldaddcontentsline\addcontentsline
\def\addcontentsline#1#2#3{} 
\makeatother

\def\spacingset#1{\renewcommand{\baselinestretch}%
{#1}\small\normalsize} \spacingset{1}


\title{Estimating the Number of Components in Finite Mixture Models via Variational Approximation}
\date{}
\author{Chenyang Wang\thanks{cw80@illinois.edu}\, \textsuperscript{1}, Yun Yang\thanks{yy84@umd.edu}\, \textsuperscript{2}\\
  \textsuperscript{1}Department of Statistics, University of Illinois Urbana-Champaign\\
  \textsuperscript{2}Department of Mathematics, University of Maryland, College Park}
\maketitle

\setstretch{1}

\begin{abstract}

This work introduces a new method for selecting the number of components in finite mixture models (FMMs) using variational Bayes, inspired by the large-sample properties of the Evidence Lower Bound (ELBO) derived from mean-field (MF) variational approximation. Specifically, we establish matching upper and lower bounds for the ELBO without assuming conjugate priors, suggesting the consistency of model selection for FMMs based on maximizing the ELBO. As a by-product, we show that the MF approximation inherits the stable behavior of the posterior distribution, which benefits from model singularity and tends to eliminate the extra components under model over-specification. This stable behavior also leads to the $n^{-1/2}$ convergence rate for parameter estimation, up to a logarithmic factor, under model over-specification. Empirical experiments are conducted to validate our theoretical findings and compare with other advanced methods for selecting the number of components in FMMs.
\end{abstract}

\noindent
{\it Keywords:} Finite mixture models; Model selection; 
Variational Inference; Evidence lower bound; Mean-field approximation.


\spacingset{1.25} 
\section{Introduction}
Model selection is a fundamental problem in statistics and machine learning, which attempts to identify the most parsimonious model from a set of candidates to fit observed data. Various criteria have been proposed for this purpose, with the Akaike Information Criterion (AIC) \citep{akaike1973information} and the Bayesian Information Criterion (BIC) \citep{schwarz1978estimating} being two of the most widely used. Under suitable conditions, AIC is known to achieve optimal prediction performance but tends to overestimate the model size. In comparison, BIC, derived as an asymptotic approximation to the log-marginal likelihood (also known as model evidence) in a Bayesian framework, is consistent in selecting the true model (the smallest model containing the true data-generating process) as the sample size grows to infinity under the frequentist perspective. However, BIC’s validity relies on the candidate model being non-singular, meaning the Fisher information matrix is invertible at the true parameter, allowing the maximum likelihood estimator (MLE) to satisfy asymptotic normality. In many practical applications, especially for over-specified models, the Fisher information matrix may become non-invertible, resulting in singular models. In such cases, BIC's penalty term becomes overly stringent. Here, an over-specified model refers to a model that contains the true model but includes more parameters, while an under-specified model excludes the true model.

In this paper, we focus on the model selection problem for a common and representative type of singular models: finite mixture models (FMMs). An FMM consists of finite mixtures of probability distributions from a shared parametric family with a fixed number of components. FMMs are extensively studied in statistical literature \citep{mclachlan2019finite, fruhwirth2006finite, mengersen2011mixtures, schlattmann2009medical}  and widely applied in various fields such as computer vision, astronomy, biology, and economics. The model size of an FMM is determined by the number of components, which is typically unknown in practice. As a result, model selection essentially becomes the task of identifying the minimal number of components required to adequately fit the data. Over-specified FMMs, as the number of components exceeds the true model size, are singular due to parameter non-identifiability, which means that different parameter settings can represent the same distribution. Over-specification also deteriorates the statistical efficiency of parameter estimation; for example, in a two-component Gaussian mixture model where the true model is a single Gaussian, the MLE convergence rate for the location parameter slows from $n^{-1/2}$ to $n^{-1/4}$ \citep{chen1995optimal}, with further deterioration as more components are added \citep{ho2016convergence, heinrich2018strong}.

Several approaches in the literature have been proposed to solve this problem, each offering distinct advantages while also facing some inherent limitations. Likelihood-based methods incorporate penalty terms, typically defined as functions of the mixing weights and component parameters, into the log-likelihood optimized through maximum likelihood estimation (MLE). Among these, the smoothly clipped absolute deviation (SCAD) penalty \citep{chen2009order, huang2017model} is particularly popular. However, because these methods directly penalize the mixing weights, they may overlook clusters associated with small weights. Another line of work involves Bayesian methods using rapidly decaying priors on the number of components \citep{cherief2018consistency, miller2023consistency, ohn2023optimal}. Although these approaches are shown to enjoy model selection consistency, their proofs often rely on strong prior penalties to rule out large models. In practice, this can lead to overly small models, especially when data is limited. In addition, efficiently sampling from the joint posterior over parameters and model configurations can be highly nontrivial. In contrast, our theoretical analysis relies on the intrinsic penalization that results from integration and does not require strong prior penalties to eliminate over-specified models. A third class of methods is based on top-down clustering, where neighboring components are iteratively merged \citep{guha2021posterior, manole2021estimating, do2024dendrogram}. While effective in certain settings, these algorithms can be unstable, require careful tuning of merging or stopping criteria, and may struggle to distinguish clusters with closely located centers.

Extensions of BIC-type model selection criteria to singular models originated from the seminal works of Watanabe \citep{watanabe2001algebraic, yamazaki2003singularities, yamazaki2012stochastic}. These studies employed tools from algebraic geometry, particularly the resolution of singularities, to determine the leading terms in the asymptotic expansion of model evidence, which depends on the real log canonical threshold (RLCT) and its multiplicity (see Section~\ref{sec:sBIC} or \citep{watanabe2018mathematical} for more details). Unlike standard BIC that relies only on the considered model, RLCT requires knowledge of the true parameter value, making it impractical to apply directly.
To address this, \cite{drton2017bayesian} proposed the singular Bayesian information criterion (sBIC), which involves solving a system of equations to approximate model evidences asymptotically. However, sBIC also relies on the known ordering structure of RLCT and its multiplicity over all candidate models, which remains a challenging problem in algebraic geometry. For example, even for Gaussian mixture models (GMMs), their complete characterization is yet unresolved.

In this paper, we propose and analyze a new model selection approach for singular models based on variational Bayes (VB). VB has gained popularity as a scalable alternative to traditional sampling-based algorithms like MCMC to approximate Bayesian computation, offering computational efficiency and strong generalization performance.
Our approach draws inspiration from the recent work of \cite{zhang2023bayesian}, which shows that for non-singular models, the evidence lower bound (ELBO) derived from the mean-field (MF) approximation in VB captures the leading terms in model evidence. Notably, the model selected by the ELBO asymptotically converges to the one chosen by the BIC as the sample size grows. Furthermore, the calculation of ELBO is a natural by-product of the Coordinate Ascent Variational Inference (CAVI) algorithm \citep[see, e.g., Chapter 10 of][]{bishop2006pattern}, eliminating the need for a case-by-case analysis of the effective number of parameters required by singular BIC.
In this work, we extend the use of ELBO based on MF approximation to model selection in finite mixture models (FMMs). Theoretically, we find that although the ELBO does not recover the leading term involving the RLCT in model evidence, it includes a penalty term for over-specification, ensuring consistent model selection.

\subsection{Related literature}
In this subsection, we briefly review more recent works closely related to ours and defer a more extensive discussion to Appendix~\ref{app:review}. For Bayesian finite mixture models (FMMs), \cite{rousseau2011asymptotic} studied the asymptotic behavior of posterior distributions in over-specified Bayesian FMMs, showing that under certain prior conditions, the posterior tends to empty extra components. \cite{Long2013} and \cite{guha2021posterior} derived contraction rates for posterior distributions under model over-specification, with additional reviews on mixture model estimation therein.  
For variational Bayes (VB), recent works \citep{alquier2020concentration, pati2018statistical, yang2020alpha, zhang2020convergence, wang2019frequentist, zhang2023bayesian} established general conditions for estimation consistency and corresponding convergence rates. Notably, \cite{yang2020alpha} and \cite{zhang2023bayesian} developed a theoretical framework covering Bayesian latent variable models, including FMMs. Moreover, \cite{pati2018statistical} demonstrated an $n^{-1/2}$ convergence rate, up to a logarithmic factor, for over-specified FMMs under the Hellinger metric, which plays a key role in our proof.  
For VB applied to Bayesian FMMs, \cite{watanabe2006stochastic} derived upper and lower bounds for the ELBO from the mean-field (MF) approximation for Bayesian GMMs; see Section~\ref{sec:review_VB_FMM}. However, the gap between these bounds does not ensure model selection consistency. \cite{watanabe2007stochastic} extended this to general FMMs, but the gap issue remains unresolved. More recently, \cite{bhattacharya2020evidence} showed that the ELBO from MF approximation recovers the leading RLCT term for singular models in their normal-crossing form. While Hironaka's theorem guarantees the existence of a reparametrization to transform any singular model into its normal-crossing form, explicitly finding such transformations remains highly challenging.

\subsection{Our contributions}
In this study, we examine the large-sample properties of the ELBO from MF variational approximation for FMMs under model {\color{black} over-specification}. We establish matching upper and lower bounds of the ELBO for a general class of FMMs from exponential families. Our refined result implies the consistency of ELBO-based model selection. Unlike prior analyses relying on conjugacy, we assume only smooth and positive priors, and apply Laplace approximation to analyze the conditional posterior of parameters given latent class indicators.  
In our proof we first obtain a lower bound for the ELBO by constructing specific feasible solutions to the VB optimization problem, similar to \cite{watanabe2007stochastic}. The main technical contribution lies in the upper bound analysis, where we leverage the variational risk bound under the Hellinger metric by \cite{pati2018statistical} to identify certain behaviors of mixing weights, yielding a substancially improved upper bound that matches the lower bound up to higher-order terms. Interestingly, unlike singular models in normal-crossing forms \citep{bhattacharya2020evidence}, applying MF to Bayesian FMMs does not recover the leading term in model evidence but still ensures consistent model selection.

As a by-product, we show that MF approximations in Bayesian FMMs exhibit stable behavior, tending to empty extra components under over-specification, similar to the true posterior \citep{rousseau2011asymptotic}. This behavior guarantees an $n^{-1/2}$ convergence rate (up to logarithmic factors) for identifiable components that are not emptied (see Section~3.3). {\color{black} Notably, our results do not rely on strong identifiability assumptions \citep{rousseau2011asymptotic, manole2021estimating} or require a two-step procedure that first selects the true number of components before estimating the model parameters \citep{ho2020robust, manole2022refined}}. Through experiments on GMMs and a real dataset, we empirically validate our theoretical findings and demonstrate that our model selection method is computationally efficient and achieves higher power than existing state-of-the-art methods, correctly selecting the true number of components with fewer samples.

The rest of the paper is structured as follows. Section~\ref{section 2} outlines the notation, provides background knowledge, formulates the problems, and introduce our model selection method for finite mixture models. Section~\ref{section 3} lists our main assumptions and presents the main theoretical results on model selection via ELBO maximization and parameter estimation via the mean-field variational approximation. Finally, Section~\ref{section 4} complements these findings with simulated experiments on GMMs and a real data analysis. Proofs, highlights of technical contributions, extended literature reviews, additional simulation information, concrete examples, and a concluding discussion are deferred to the supplementary material.

\section{Background and New Method} \label{section 2}

In this section, we begin with an overview of the notation, followed by a review of necessary background, and then introduce the problem formulation.

\subsection{Notation}
Throughout the paper, random variables are denoted by capital letters, while their realizations are represented by corresponding lowercase letters. For two probability measures $P$ and $Q$, let $p$ and $q$ be their Radon-Nikodym derivatives with respect to a $\sigma$-finite measure $\mu$, usually chosen as the Lebesgue measure or the counting measure. We denote the total variation and the (squared) Hellinger distance between $p$ and $q$ respectively as $d_{\rm TV}(p,q) = \frac{1}{2} \int |p-q| \,d\mu$ and $h^2(p,q) = \int(\sqrt{p} - \sqrt{q})^2\, d\mu$. Moreover, their Kullback-Leibler (KL) divergence is given by $D_{\rm KL}(p\|q) = \int p\log(p/q)\, d\mu$.
We use $\m{N}(\mu,\Sigma)$ to denote a normal distribution with mean $\mu\in\mb{R}^d$ and variance-covariance matrix $\Sigma\in\mb{R}^{d\times d}$.
Additionally, for a vector $a\in \mb{R}^d$ with entries $\{a_i\}_{i=1}^d$, we use $\|a\|$ to denote its Euclidean $\ell_2$-norm {\color{black} and $\|a\|_\infty\coloneqq \max_{1\le i\le d} |a_i|$ to denote its infinity norm}. We also use $C$ to denote a generic positive constant whose value may vary from line to line. Notationally we write $a\gtrsim b$ to indicate $a\ge Cb$, $a\lesssim b$ to indicate $a\le Cb$ and $a\asymp b$ to mean $a\gtrsim b$ and $a\lesssim b$. 

\subsection{Singular models}
In statistical learning, many classical techniques and conclusions, like Laplace approximation, the convergence rate and asymptotic normality of the MLE, are primarily applicable to regular models. A statistical model  $\m M=\big\{p(\cdot\,|\,\theta)\,\big|\,\theta\in\Theta\big\}$ with parameter space $\Theta$ is considered regular (or non-singular) if $\theta$ is identifiable and $p(\cdot|\theta)$ has a well-defined and positive definite Fisher information matrix $I(\theta) =\mb E_\theta[-\nabla^2 \log p(X\,|\,\theta)]$ for all $\theta \in \Theta$. In contrast, a model is singular when any of these conditions fail.
However, there are many widely used singular models, including finite mixture models, hidden Markov models, factor models, and neural networks. In this article, we focus on finite mixture models, a representative class of singular models that capture complex data distributions through convex combinations of simpler distributions.

Concretely, let us start from a simple parametric family $\m G = \big\{g(\cdot\,;\,\eta)\,\big|\,\eta\in\Omega\big\}$ characterized by $\eta \in \Omega \subset\mb{R}^m$. For an integer $K$, a mixture model with $K$ components in $\m G$ is denoted by $\m M_K =\big\{p(\cdot\,|\,\theta)\,\big|\, \theta=(w_k,\eta_k)_{k=1}^K, \,\eta_k\in\Omega,\,w_k\ge 0, k\in[K] \mbox{ and }\sum_{k=1}^Kw_k=1\big\}$ with
\begin{align}
    p(x\,|\,\theta) = \sum_{k=1}^{K} w_k\,g(x;\,\eta_k),\quad\theta =(w_k,\eta_k)_{k=1}^K \in\mb R^{mK+K},\label{model used}
\end{align}
where $\bm{w} =(w_1,w_2,\ldots,w_K)\in [0,1]^K$ is called the mixing weight (vector) parameter, and, by slightly abusing the notation, we use $\bm{\eta}=(\eta_1,\eta_2,\ldots,\eta_K)\in\Omega^K$ to denote the parameters associated with the $K$ components.
We also denote the corresponding (discrete) mixing distribution as $G(\theta) = \sum_{k=1}^K w_k\,\delta_{\eta_k}$,
where $\delta_{\eta_k}$ denotes the Dirac mass at $\eta_k$.
Generally, mixture models may lose identifiability, since different parameters may lead to the same data distribution $p(\cdot\,|\,\theta)$. Good news is that $G(\theta)$ is always identifiable and there is a one-to-one correspondence between $G(\theta)$ and $p(\cdot\,|\,\theta)$. This serves as one of the key assumptions in our theory (see Section~\ref{sec:assump}). For example, the following two GMMs with different parameter setting both correspond to the standard normal distribution: $p(x\,|\,\theta_1) = 1\cdot \m{N}(0,1) + 0\cdot\m{N}(1,1)$ and $p(x\,|\,\theta_2) = \frac{1}{2}\cdot \m{N}(0,1) + \frac{1}{2}\cdot\m{N}(0,1)$. Fortunately, they still have the same mixing distribution $G(\theta_1) = G(\theta_2)=\delta_0$.

The loss of identifiability often leads to non-invertible Fisher information matrices and decreased efficiency—even when an identifiable submodel is considered, the best estimator may suffer from slow convergence rates. For instance, \cite{chen1995optimal} demonstrated that when the mixing distribution is restricted to the form $G(\theta) = \frac{2}{3}\delta_{-h} + \frac{1}{3}\delta_{2h}$ for $h\in \mb{R}$, and $g$ satisfies certain regular conditions to make the reduced parameter $h$ identifiable, the MLE for $h$ still converges at the slow rate of $n^{-1/4}$ when the true parameter $h^\ast = 0$. Therefore, it is crucial to correctly specify the number of components before inference, motivating the model selection problem in this paper.

\subsection{Model selection criteria for singular models}\label{sec:sBIC}
The Bayesian Information Criterion (BIC) is widely used for model selection for regular parametric models. Let $X^n = (X_1, ..., X_n)$ be $n$ i.i.d. observations from the true model $p^*(x)$, which belongs to a set $\m M$ of candidate models.
Under this setup, the BIC for model $\m M$ is derived as an asymptotic approximation to the so-called model evidence, or the log-marginal likelihood in a Bayesian framework, defined as
\begin{align}
    F(\m{M}) = \log \int_{\Theta_{\m{M}}} \pi(\theta\,|\,\m{M})\,p(X^n\,|\,\theta,\m{M}) \; d\theta =  \underbrace{\log p(X^n\,|\,\widehat\theta_{\m M},\m{M})- \frac{m}{2}\log n}_{\mbox{\scriptsize BIC for model }\m M} \,+\ \m O_P(1),\label{evidence approx}
\end{align}
where $\Theta_\m{M}\in\mb R^m$ denotes its parameter space, $\pi(\theta\,|\, \m M)$ the prior density, $p(X^n\,|\,\theta,\m{M})$ the likelihood function, and 
$\widehat\theta_{\m M}=\argmax_{\theta\in \Theta_{\m M}}p(X^n\,|\,\theta,\m{M})$ is the MLE of $\theta$ under model $\m M$. The BIC consists of two terms: a model goodness-of-fit term and an additional term of $-\frac{m}{2}\log n$ resulting from the integration over $\mb{R}^m$, which penalizes larger models that lead to overfitting. It is classical \citep{schwarz1978estimating} that maximizing BIC enjoys the model selection consistency.
Unfortunately, for singular models with non-invertible Fisher information matrices, the asymptotic expansion~\eqref{evidence approx} no longer holds, as the effective dimension of the parameter space is no longer equal to $m$.

In a seminal work,  \cite{watanabe2001algebraic} employed algebraic geometry tools to show that for a singular model $\m M$, the asymptotic behavior of the log-marginal likelihood can be characterized by
\begin{align}
    F(\m{M}) = \log p^*(X^n) - \lambda_{\m M} \log n + (\nu_{\m M}-1)\log\log n + \m O_P(1),\label{RLCT}
\end{align}
where $\lambda_M$ is called the \emph{real log canonical threshold} (RLCT) of model $\m M$ that determines the effective number of parameters, and $\nu_{\m M}$ is its \emph{multiplicity}. Leveraging this general result, \cite{drton2017bayesian} proposed a model selection method by maximizing a singular Bayesian information criterion (sBIC), computed by solving a system of equations whose unique solution approximates the leading terms in $F(\m{M})$.
However, this approach requires the ordering structure of RLCT and its multiplicity for all candidates, the determination of which is a highly nontrivial problem even in the field of algebraic geometry within pure mathematics.

\subsection{Component Selection in Mixture Models via Variational Approximation}\label{sec:review_VB_FMM}
To handle the analytically intractable integral in $F(\m{M})$, variational inference attempts to approximate $F(\m{M})$ defined in~\eqref{evidence approx} from below by utilizing Jensen's inequality (to $-\log(t)$),
\begin{align*}
    \log \int_{\Theta_{\m{M}}} \frac{\pi(\theta\,|\,\m{M})\,p(X^n\,|\,\theta,\m{M})}{q_\theta(\theta)} \,q_\theta(\theta) \, d\theta \ge \int_{\Theta_{\m{M}}} q_\theta(\theta) \log \frac{\pi(\theta\,|\,\m{M})\,p(X^n\,|\,\theta,\m{M})}{q_\theta(\theta)}\,d\theta\coloneqq \m L(q_\theta),
\end{align*}
where $q_\theta(\theta)$ is from a tractable distribution family $\Gamma$ for approximating the posterior distribution $p(\theta\,|\,X^n,\m M)$. We focus on the commonly used mean-field family $\Gamma_{\rm MF}$ in this work, where $q_\theta$ factorizes over the (blocks of) components of $\theta$. The \emph{evidence lower bound} (ELBO) is then defined as the best approximation to $F(\m{M})$ by maximizing $\m L(q_\theta)$ over $q_\theta\in \Gamma_{\rm MF}$. In the rest of the paper, when a fixed model $\m{M}$ is considered, we suppress $\m{M}$ in our notation for simplicity.

Now let us specialize the above general framework to finite mixture models~\eqref{model used}.
As a common practice, we augment the FMM into a hierarchical model by introducing latent (class indicator) variables to facilitate computation. In fact, a key property utilized in our analysis is that the conditional distribution of observations given latent variables is a regular parametric model. Let $S^n = (S_1, ..., S_n)$ denote the collection of all latent variables, with each $S_i$ taking values from $\{1,..,K\}$, indicating the mixture component that $X_i$ comes from. Then we have the hierarchical form of the model as $[X_i\,|\,S_i=k]\overset{iid}{\sim} g(x\,;\,\eta_k)$ and $S_i$ are i.i.d.~with $P(S_i=k) = w_k$, for $i\in[n]$ and $k\in[K]$. Moreover, we can write
the joint distribution of $X^n$ given $\theta$ as
\begin{align*}
    p(X^n\,|\,\theta) = \sum_{s^n\in[K]^n} p(X^n,\,s^n\,|\theta) = \sum_{s^n\in[K]^n} p(X^n\,|\,s^n,\theta)\,p(s^n\,|\,\theta) = \prod_{i=1}^n\sum_{s_i\in[K]} p(X_i\,|\,s_i,\bm \eta)\,p(s_i\,|\,\bm w).
\end{align*}
Let $Z^n = (\theta,S^n)$ denote the collection of all hidden variables, then the augmented posterior distribution $p(Z^n\,|\,X^n)$ satisfies $p(\theta,s^n\,|\,X^n)\varpropto p(X^n,s^n\,|\,\theta)\,\pi(\theta)$. Now the variational inference uses $q_{Z^n}\in\Gamma$ to approximate the joint posterior distribution $p(Z^n\,|\,X^n)$ (by replacing $\theta$ with $Z^n$ therein).
{\color{black}
To preserve within-block parameter dependence while maintaining a balance between flexibility and tractability, we adopt the widely used block mean-field (MF) approximation, under which the variational distribution is given by: 
\begin{align}
    \Gamma_{\rm MF} = \big\{ q_{Z^n} = q_{\theta} \otimes q_{S^n} \big\}. \label{mean-field}
\end{align}  
The optimal variational distribution $\widehat{q}_{Z^n} = \widehat{q}_{\theta} \otimes \widehat{q}_{S^n}$, used to approximate the true posterior $p(Z^n\,|\,X^n)$, is obtained by minimizing the KL divergence between $q_{Z^n}$ and the true posterior.
The existence of an optimal solution is always guaranteed, as the set of product measures is weakly closed and the KL divergence objective has weakly compact sub-level sets. However, uniqueness is generally not guaranteed—particularly in non-convex optimization problems such as the mean-field approximation for mixture models, where label switching can lead to multiple equivalent solutions. To account for this, we define the set of all minimizers of the KL divergence as
    \begin{align*}
        \widehat{\m{Q}}_{Z^n} \coloneqq  \mathop{\arg\min}\limits_{q_{Z^n} \in \Gamma_{\rm MF}} D_{\rm KL}\big( q_{Z^n}(\cdot) \,\big\|\, p(\cdot \,|\, X^n) \big).
    \end{align*}  
In what follows, we use $\widehat{q}_{Z^n}$ to denote an arbitrary element of $\widehat{\m{Q}}_{Z^n}$.
Now the ELBO is defined as
\begin{align}
    \m{L}(q_{Z^n}) = \int q_{Z^n}(z^n) \log \frac{p(X^n, s^n | \theta)\, \pi(\theta)}{q_{Z^n}(z^n)} \, dz^n,\label{elbo}
\end{align}
where the integral symbol includes both integration over the continuous variable $\theta$ and summation over the discrete variables $s^n$. Then minimizing the KL divergence is equivalent to maximizing the ELBO over $q_{Z^n}$ since
$$
D_{\rm KL}\big(q_{Z^n}(\cdot)\,\big\|\,p(\cdot\,|\,X^n)\big) = \log \int p(X^n|\theta)\,\pi(\theta)\,d\theta - \m{L}(q_{Z^n}),
$$
where the first term on the right hand side represents the true evidence, which is a constant with respect to $q_{Z^n}$.
In the context of finite mixture models, $\m{L}(\widehat{q}_{Z^n})$, the ELBO evaluated at its maximizer, can serve as a surrogate for the model evidence, provided it retains some key characteristics. By definition, $\mathcal{L}(\widehat{q}_{Z^n})$ attains the same value for all $\widehat{q}_{Z^n} \in \widehat{\mathcal{Q}}_{Z^n}$. Therefore, the ELBO is well-defined and is not affected by the choice of a particular element from this set.
 Suppose we have a known upper bound $K_{\max}$ of the true component number $K^*$, and use $\widehat {\m L}_K = \m L_K(\widehat{q}_{Z^n})$ to denote the optimal ELBO value under the FMM with $K$ components, then our selected number of components $\widehat K$ via ELBO maximization is defined as
\begin{align}\label{eqn:K_est}
    \widehat K = \argmax_{K\le K_{\max}} \widehat {\m L}_K.
\end{align}
In cases where multiple solutions minimize $\widehat{\m{L}}_K$, we define $\widehat{K}$ as the smallest such solution. However, our theorem in the next section ensures that, with high probability, the solution is unique.}

Regarding the block MF approximation, the following theorem gives a necessary condition for $\widehat{q}_{Z^n}$ to be the maximizer of the ELBO, as stated in Theorem 2.1 by \cite{beal2003variational}. This theorem provides a system of distributional equations determining $\widehat{q}_\theta(\theta)$ and $\widehat{q}_{S^n}(s^n)$, which plays an important role and will be repeatedly used in our proofs. {\color{black} We provide its proof in Appendix~\ref{proof of Thm1}.}
\begin{theorem}\label{seperation}
    Under the block MF approximation, the variational posterior $\widehat{q}_{Z^n} = \widehat{q}_{\theta}\otimes\widehat{q}_{S^n}$ satisfies
    \begin{align}
        & \widehat{q}_\theta(\theta) = \frac{1}{\widehat C_r}\pi(\theta)\exp\bigg\{\sum_{s^n\in [K]^n}{\widehat{q}_{S^n}(s^n)}\log p(X^n, s^n\,|\,\theta)\bigg\} \label{VBtheta},\\
      &  \mbox{and}\quad \widehat{q}_{S^n}(s^n) = \frac{1}{\widehat C_Q}\exp\bigg\{\int{\widehat{q}_\theta(\theta)}\log p(X^n, s^n\,|\,\theta)\, d \theta\bigg\},\label{VBsn}
    \end{align}
    where $\widehat C_r$ and $\widehat C_Q$ are the normalization constants.
\end{theorem}

{\color{black} It is important to note that the variational family $\Gamma_{\text{MF}}$ is defined solely through the block factorization $q(Z^n) = q(\theta) \otimes q(S^n)$. While this definition does not explicitly impose a finer factorization structure within $q_{S^n}$ or $q_\theta$, the specific forms and factorization properties of the \emph{optimal densities} arise as consequences of the model structure and the prior assumptions.
Specifically, since the observations $X^n$ are i.i.d., the complete-data log-likelihood $\log p(X^n, s^n \mid \theta)$ decomposes into a sum over observations $i=1,\dots,n$. As a result, the optimal variational posterior for the latent variables given in (8) naturally factorizes as $\widehat{q}_{S^n}(s^n) = \prod_{i=1}^n \widehat{q}_{S_i}(s_i)$. 
Furthermore, under independent priors for $\bm{w}$ and each component parameter $\eta_k$, the optimal parameter posterior preserves this independence structure, yielding $\widehat q_\theta(\theta) = \widehat q_{\bm w}(\bm{w}) \prod_{k=1}^K \widehat q_\eta(\eta_k)$.}

Under the above settings, \cite{watanabe2007stochastic} derived the upper and lower bounds for $\m{L}(\widehat{q}_{Z^n})$ under a symmetric Dirichlet prior Diri$(\phi_0,\ldots,\phi_0)$, $\phi_0 > 0$, on the mixing weights $\bm w$ and a conjugate prior for the parameters $\bm \eta$ of mixture components from an exponential family. Specifically, if we denote the variational posterior mean (estimator) of $\theta$ as $\overline\theta \coloneqq  \int_\Theta \theta \,\widehat{q}_{\theta}(\theta)\, d\theta$, then \cite{watanabe2007stochastic} proved that (after some notation adaptions)
\begin{align*}
     \log p^*(X^n) - \underline\lambda\log n + \m O_P(1) \le \m{L}(\widehat{q}_{Z^n})\le \log p^*(X^n) -\overline{\lambda}\log n + \log \frac{p(X^n\,|\,\overline{\theta})}{p^*(X^n)} +\m O_P(1)
\end{align*}
as sample size $n$ tends to infinity, and the leading coefficients $\underline\lambda$ and $\overline{\lambda}$ are given by
\begin{align*}
        \underline\lambda = \begin{cases}
        (K-K^*)\phi_0 + \frac{mK^*+K^*-1}{2}, & \phi_0\le\frac{m+1}{2},\\
        \frac{mK+K-1}{2}, & \phi_0>\frac{m+1}{2};
        \end{cases}\quad\mbox{and}\quad \overline\lambda = \begin{cases}
        (K-1)\phi_0 + \frac{m}{2}, & \phi_0\le\frac{m+1}{2},\\
        \frac{mK+K-1}{2}, & \phi_0>\frac{m+1}{2},
        \end{cases}
\end{align*} 
where $K^*$ denotes the true number of components in the underlying data-generating process.
When $\phi_0 < (m+1)/2$, a gap exists between $\underline{\lambda}$ and $\overline{\lambda}$. Notably, $(m+1)/2$ also serves as a threshold for $\phi_0$, below which the desirable stable behavior of the posterior distribution and its variational approximation (see Corollary~\ref{empty out rate}) to empty the extra components can be maintained. This stability further improves parameter estimation efficiency (see Theorem~\ref{para root n}).  
In contrast, when $\phi_0 > (m+1)/2$ and $K$ is over-specified, two or more components can be very close with non-negligible weights each, leading to unstable and less accurate parameter estimation (e.g., see numerical results in Section~\ref{sec: section 4.1}). In practice, $\phi_0 \le 1$ is typically preferred in the literature. For example, a defaulted choice of $\phi_0$ in topic models \citep{blei2003latent} is $K^{-1}$, encouraging sparse mixing weights $\bm w$ for better interpretation. Conversely, a larger $\phi_0$ promotes denser mixing weights; in particular, $\text{Diri}(\phi_0,\ldots,\phi_0)$ tends to concentrate around uniform weights $(K^{-1},\ldots,K^{-1})$ as $\phi_0 \to \infty$.  
Therefore it is an important theoretical question whether this gap can be eliminated for small $\phi_0$, so that model selection based on maximizing the ELBO is provably consistent. In this work we bridge this gap by proving matching lower and upper bounds for ELBO under any positive $\phi_0$. Our numerical results in Section~\ref{sec: section 4.1} also suggest that {\color{black} a suitably small $\phi_0$ ($\phi_0< (m+1)/2$) is preferable for accurate model selection and parameter estimation.}

\section{Theoretical Results}\label{section 3}
In this section, we present our main theoretical results. Appendix~\ref{sec:tech} highlights several of our technical contributions. Here, we assume that the dataset $X^n = (X_1, ..., X_n)$ consists of $n$ observations from $\mb R^d$ that are i.i.d.~samples from a fixed true data-generating process $p^*(x) = \sum_{k=1}^{K^*} w_k^*g(x;\eta_k^*)$,
where $K^*$ denotes the true component number, and $\theta^* =  (w_k^*,\,\eta_k^*)_{k=1}^{K^*} \in \mb{R}^{mK^* + K^*}$ denotes the (unique) true parameters under the mixture model~\eqref{model used} with $K^*$ components, where all $\eta_k$'s are distinct and $\min_{k\in[K^*]} w_k^\ast>0$. Let $G^*$ denote the corresponding mixing measure for the truth. Throughout this section, we consider a finite mixture model with {\color{black} $K$ components, where $K \le K_{\max}$, for some known upper bound $K_{\max}$}. In addition, we assume the parametric family $\m G = \big\{g(\cdot\,;\,\eta)\,\big|\,\eta\in\Omega\big\}$ for the mixture components to be an exponential family in the canonical form, that is, $g(x;\eta) = \exp\big\{\eta^T T(x)-T_0(x)-A(\eta)\big\}$, where $\eta\in\mb R^m$ is the natural parameter, $T(x)$ is the sufficient statistic and $A(\eta)$ is the log-partition function. The Fisher information matrix for this exponential family $I(\eta) = \nabla^2 A(\eta)$ is positive definite, which implies the convexity of $A(\eta)$ over its domain. Consequently, the log-likelihood function $\log g(x;\eta)$ is concave with respect to $\eta$ for all $x$. One important assumption of our theory (see Section~\ref{sec:assump} below) is the relationship between the total variation and Wasserstein-type distances in FMMs, which transforms the closeness of mixture models in density functions into closeness in their mixing measure or model parameters. Such a relationship can be verified for most commonly used FMMs; see Appendix~\ref{examples} for concrete examples. An $r$-th order Wasserstein distance (or simply $r$-Wasserstein distance) between two mixing measures $G(\theta) = \sum_{k=1}^K w_k\,\delta_{\eta_k}$ and $G(\theta') = \sum_{k=1}^{K'} w'_k\,\delta_{\eta'_k}$ is defined as:
\begin{align}
    W_r\big(G(\theta),G(\theta')\big) = \Big(\inf_{q_{ij}}\sum_{i,j}q_{ij}|\eta_i - \eta_j'|^r\Big)^{1/r},\label{define wasserstein}
\end{align}
where the infimum is taken over all nonnegative probability masses $\big\{q_{ij}:\, i\in[K],\, j\in[K']\big\}$ satisfying two marginal constraints $\sum_{i=1}^K q_{ij} = w'_j$ for $j\in[K']$ and $\sum_{j=1}^{K'} q_{ij} = w_i$ for $i\in[K]$.

\subsection{Assumptions} \label{sec:assump}
We consider the symmetric Dirichlet distribution Diri$(\phi_0,\ldots,\phi_0)$, $\phi_0 > 0$ as our prior for the mixing weights $\bm w$, whose density takes the form of
$\pi_{\bm w}(\bm w) = \frac{\Gamma(K\phi_0)}{\Gamma(\phi_0)^K}\prod^K_{k=1} w_k^{\phi_0 - 1}$, for all $\bm w=(w_1,\ldots,w_K)\in[0,1]^K$ such that $\sum_{k=1}^Kw_k=1$, where $\Gamma(\cdot)$ is the Gamma function. 
{\color{black} Note that our results can be extended to other priors $\pi'_{\bm w}$ on $\bm w$, provided that their density ratio with respect to the Dirichlet prior is bounded. Specifically, if there exist constants $R > L > 0$ such that $\pi'_{\bm w}(\bm w)/\pi_{\bm w}(\bm w) \in [L, R]$, then the main theorem continues to hold. As detailed in Remark~\ref{rmk: extended prior}, under this condition the theorem remains valid with modified constant terms.}
{\color{black} For the prior on $\bm \eta$, we make the following assumption:

\medskip
\noindent {\bf Assumption P:} The prior on $\bm \eta$ is independent and identical for each $\eta_k$, i.e., $\pi_{\bm \eta}(\bm \eta) = \prod_{k=1}^K\pi_\eta(\eta_k)$, $\eta_k\in\Omega$. And there exist two positive constant $B_2\ge B_1>0$ such that $\pi_\eta(\eta)\in[B_1,B_2]$, for any $\eta\in\Omega$. Moreover, $\pi_\eta(\eta)$ is twice continuously differentiable on $\Omega$.
\smallskip

\noindent
This assumption does not require the use of conjugate priors on $\eta$, and can be applied to any general twice-differentiable prior distributions. Our second assumption concerns finite mixture model.}

\noindent {\bf Assumption A:} Recall that $p^*$ is the true data-generating probability with mixing measure $G^*$ and $K^*$ components.
\begin{enumerate}
    \item (\emph{Parameter space compactness}) The parameter space $\Omega$ of mixture components is a compact subset of $\mb{R}^m$. {\color{black} In particular, this implies the existence of a constant $L < \infty$ such that $\sup_{\eta, \eta' \in \Omega} \|\eta - \eta'\| \le L$.}
    \item (\emph{Regularity of mixture component family}) {\color{black} On $\Omega$, the log-partition function $A(\eta)$ is four times continuously differentiable and $\|\nabla_\eta A(\eta)\|_\infty \leq B_3$ for some positive $B_3$.} Moreover, there exist two positive constants $a$ and $b$ such that $y^T I(\eta)\,y\in[a,b]$ for all $\|y\| = 1$ and all $\eta\in\Omega$, where $I(\eta) = \nabla^2A(\eta)$ is the Fisher information matrix for the mixture components.
    \item {\color{black} (\emph{A known upper bound on the number of components}) There exists a known and finite upper bound $K_{\max}$ of the number of components, i.e., $K^* \le K_{\max}<\infty$.}
    \item (\emph{Identifiability of mixing measure}) The total variation between any FMM $p(x|\theta)$ from the same family with $K(\ge K^*)$ components, and the true model $p^\ast$ achieves\\ $d_{\rm TV}\big(p(\cdot\,|\theta),p^*(\cdot)\big) = 0$ if and only if its mixing measure satisfies $G(\theta) = G^*$. Also, there exists a positive integer $r$ and positive constants $\varepsilon$ and $c_0$ depending on $G^*$ and $K$ such that 
    $d_{\rm TV}\big(p(\cdot\,|\theta),p^*(\cdot)\big)\ge c_0 W_r^r\big(G(\theta),G^*\big)$ for any $p(\cdot\,|\theta)$ whose mixing measure $G(\theta)$ satisfies $W_r\big(G(\theta),G^*\big)<\varepsilon$.
\end{enumerate} 
\smallskip

\noindent
Assumption A1 is commonly adopted in the literature for technical simplicity when analyzing singular models.
{\color{black} Assumption A2 is a mild condition with \(\Omega\) being compact, satisfied by all non-singular (i.e., non-curved) exponential families. Here, we assume four-times continuous differentiability in order to obtain a global error bound on the compact set and to track the constant term in the ELBO expansion uniformly, without dependence on the location of the variational posterior mode. This assumption, together with Assumption P, allows us to apply the Laplace approximation to analyze each distribution over the mixing component parameter \(\eta_k\) within the mean-field approximation.}
Assumption A3 is standard in model selection algorithms for mixture models~\citep[e.g.,][]{manole2021estimating, do2024dendrogram}. Additionally, assuming a finite $K_{\max}$ enables the use of a union bound to analyze the log-likelihood ratio in Corollary~\ref{coro:ELBO} with a controlled error bound.

Assumption A4 requires partial identifiability for the true model $p^*$. It states that for any mixing measure $G(\theta)$ sufficiently close to $G^*$, the total variation of the density functions is lower bounded by some Wasserstein-type distance between the mixing measures. For example, if the family ${g(x;\eta), \eta\in\Omega}$ is second-order identifiable (see Definition 3.2 in \citep{ho2016strong}), in other words, $g(x;\eta)$ and its derivatives with $\eta$ up to the second order are linearly independent, then A4 holds with $r = 2$. For families that are not identifiable in the second-order, such as the location-scale Gaussian distributions, verification of A4 can be done case-by-case, where $r$ may also depend on $K^*$ and $K$ but can be uniformly established with $K\le K_{\max}$. See Appendix~\ref{examples} for more examples that satisfy these assumptions.
{\color{black} With this assumption, a total variation error bound on $p(\cdot\,|\,\theta)$ can be used to control the Wasserstein distance between the mixing measures $G(\theta)$ and $G^\ast$, which is relevant for parameter estimation. In particular, this implies that even in an overfitted model, the mean-field approximation will capture at least $K^*$ non-empty components.}

It is worth noting that, with Assumptions A1 and A4, the (localized) inequality in the second part of A4 can be made global. Specifically, there exists a constant $c_0'$ such that $d_{\rm TV}(p(\cdot\,|\theta), p^*(\cdot)) \ge c_0' W_r^r(G(\theta), G^*)$ for all possible $\theta$. The proof is straightforward. Let $\m{A}\coloneqq \{\theta: W_r(G(\theta), G^*) \ge \varepsilon, \bm\eta \in \Omega^K\}$. Since $W_r(G(\theta), G^*)$ is continuous in $\theta$, $\m{A}$ is a compact set and $d_{\rm TV}(p(\cdot\,|\theta), p^*(\cdot)) > 0$ on $\m{A}$ by the first part of A4. For the considered exponential family, the total variation is also continuous with $\theta$. Therefore, it attains a strictly positive infimum on $\m{A}$, which we denote as $\tau \coloneqq  \inf_{\theta \in \m{A}} d_{\rm TV}(p(\cdot\,|\theta), p^*(\cdot))$. Meanwhile, from A1, we have $W_r^r(G(\theta), G^*) \le L^r$. Therefore, on $\m{A}$, we also have $d_{\rm TV}(p(\cdot\,|\theta), p^*(\cdot)) \ge (\tau/L^r) W_r^r(G(\theta), G^*)$. Letting $c_0' = \min\{c_0, \tau/L^r\}$, it then follows from above and the second part of Assumption A4 that for any $\theta$, we have:
\begin{align}
    d_{\rm TV}\big(p(\cdot\,|\theta), p^*(\cdot)\big)\ge c_0' W_r^r\big(G(\theta),G^*\big),\label{tv greater wr}
\end{align}
which provides a global inequality relating the two distances over a given compact set. The above argument allows us to avoid assuming a reverse inequality
$d_{\rm TV}(p(\cdot\,|\theta),p^*(\cdot)) \le c_0 W_r^\alpha(G(\theta),G^*)$ for some $\alpha > 0$ and for any $p(\cdot\,|\theta)$ with mixing measure $G(\theta)$ in order to extend the local inequality into a global one, as was done in Corollary 3.1 of \cite{ho2016strong}.

\subsection{Large sample properties of ELBO} 
Now we are ready to present our first main result as follows, which provides two-sided bounds to the ELBO $\m{L}(\widehat q_{Z^n})$ with matching leading terms. Recall that we use $\overline\theta \coloneqq  \int \theta \,\widehat{q}_{\theta}(\theta)\, d\theta$ to denote the variational point estimator for $\theta$.

\begin{theorem}[Finite-sample two-sided bound of ELBO]\label{main theorem}
   Suppose Assumptions {\color{black}P and A1--A4} hold and $K\ge K^*$. {\color{black} For sufficiently large $n$}, there exist constants $C_1$ and $C_2$ (expressions provided in Appendix~\ref{proof Thm2}) independent of $n$ such that, with probability at least $1 - n^{-1/2}$, we have
    \begin{align*}
       -\lambda\log n + C_1 + \m O(n^{-1})\le \m{L}(\widehat{q}_{Z^n}) - \log p^*(X^n) \le -\lambda\log n +  \log\frac{p(X^n\,|\,\overline{\theta})}{p^*(X^n)} +  C_2 + \m O(\sqrt{\log n/n}),
    \end{align*}
where $\lambda$ is given by
    \begin{align}
        \lambda = \begin{cases}
        (K-K^*)\phi_0 + \frac{mK^*+K^*-1}{2}, & \phi_0<\frac{m+1}{2},\\
        \frac{mK+K-1}{2}, & \phi_0\ge\frac{m+1}{2}.\label{lambda}
        \end{cases}
    \end{align}
\end{theorem}

{\color{black}
Concrete expressions for the constants $C_1$ and $C_2$, along with a detailed discussion, are provided in Appendix~F.4. Notably, both constants share a common term, $C_{\phi_0} := \frac{K-1}{2} \log(2\pi) + \log\frac{\Gamma(K\phi_0)}{\Gamma(\phi_0)^K}$, which arises from the KL divergence between the variational approximation to the mixing weight posterior and its Dirichlet prior.
{\color{black}  To understand the effect of $\phi_0$, we analyze the asymptotic behavior of $C_{\phi_0}$ in two limiting regimes:

\noindent
\textbf{1. The Singular Limit ($\phi_0 \to 0$).} 
Using the Laurent expansion of the Gamma function near zero, $\Gamma(x) = \frac{1}{x} - \gamma + \mathcal{O}(x)$, where $\gamma$ is the Euler–Mascheroni constant. Then we obtain the approximation
\(
\log\frac{\Gamma(K\phi_0)}{\Gamma(\phi_0)^K} \approx -\log(K\phi_0) - K(-\log \phi_0) = (K-1)\log \phi_0 - \log K.
\)
Since $\log \phi_0 \to -\infty$, this term acts as a dominant negative penalty that scales linearly with $K$. This strong penalty on complexity dominates other constant factors arising from the Laplace approximation and the prior density on $\eta$, pushing the ELBO maximization toward models with smaller $K$. This provides a theoretical explanation for the sparsity-inducing behavior observed in the singular regime.

\noindent
\textbf{2. The Regular Limit ($\phi_0 \to \infty$).} 
Using Stirling's approximation, $\log \Gamma(x) \approx (x - 1/2)\log x - x + \frac{1}{2}\log(2\pi)$. This yields the approximation $\log\frac{\Gamma(K\phi_0)}{\Gamma(\phi_0)^K} \approx K \phi_0 \log K$. However, Theorem~\ref{main theorem} indicates that the penalty coefficient $\lambda$ is also maximized in this regime. This creates a competition between the penalty $-\lambda \log n$ and the reward $C_{\phi_0}$. Specifically, when $\phi_0$ is close to the threshold $(m+1)/2$, the maximized $\lambda$ imposes a heavy penalty that may lead to underestimation. Conversely, when $\phi_0$ becomes excessively large, the rapidly growing constant term dominates other constant factors and counteracts the penalty, leading to instability where model selection becomes highly sensitive to the sample size $n$.}
    
These theoretical predictions align with our numerical results in Section~\ref{section 4}, which suggest that setting $\phi_0$ around $1$ (a uniform prior) provides the most stable performance. Furthermore, in Section~\ref{sec:para_est}, we demonstrate the additional advantages of choosing $\phi_0$ below the threshold $(m+1)/2$ for improved parameter estimation.}

There remains a gap $\Delta_n=\log p(X^n\,|\,\overline{\theta}) -\log p^*(X^n)$ in Theorem~\ref{main theorem}, which is related to the likelihood ratio test statistic ${\rm LRT} = \log p(X^n\,|\,\widehat \theta\,) -\log p^*(X^n)$ through $\Delta_n\leq {\rm LRT}$, with $\widehat \theta =\argmax_{\theta\in\Theta}\log p(X^n\,|\,\theta)$ denoting the MLE of $\theta$. 
According to the singular learning theory \citep[Chapter 6.1,][]{watanabe2009algebraic}, this ${\rm LRT}$, under suitable conditions, weakly converges to a limiting random variable defined through the supreme of a Gaussian process as $n\to\infty$, by using the invariance property of MLE and a resolution of singularity technique developed in \cite{watanabe2009algebraic}, which transform the log-likelihood function of the singular model into its canonical normal crossing form; see Appendix~\ref{app:coro:ELBO} for further details.

\begin{corollary}[Asymptotic expansion of ELBO]\label{coro:ELBO}
    Under Assumptions {\color{black} P and A1--A4}, the asymptotic expansion of ELBO for $K\ge K^*$ has the following form:
    \begin{align*}
        \m{L}(\widehat{q}_{Z^n}) = \log p^*(X^n)-\lambda\log n + \m O_P(1),\quad \mbox{as }n\to\infty.
    \end{align*}
\end{corollary}
Recall that $\widehat K$ denotes the estimated number of components via ELBO maximization defined in~\eqref{eqn:K_est}, then we have the following corollary.
\begin{corollary}[Model selection consistency via ELBO maximization]\label{consistency in model selection}
    Suppose Assumptions {\color{black} P and A1--A4} hold, then the estimated number of components $\widehat K$  in~\eqref{eqn:K_est} via ELBO maximization satisfies
    \begin{align*}
        \mb{P}\big(\widehat K= K^*\big|\,X^n\big)\; {\rightarrow} \;1,\quad  \mbox{as }n\to\infty.
    \end{align*}
    This means that model selection via ELBO maximization in FMMs is (asymptotically) consistent.
\end{corollary}

\subsection{Parameter estimation in finite mixtures via variational approximation}\label{sec:para_est}
Theorem~\ref{main theorem} reveals that the large-sample behavior of the ELBO exhibits two distinct regimes, determined by the prior hyperparameter $\phi_0$ on the mixing weight $\bm w$. When $\phi_0 \ge (m+1)/2$, the effective number of parameters in the variational approximation $\widehat{q}_{Z^n}$ is $(mK + K - 1)$, matching the degrees of freedom of $\theta = (\bm w, \bm\eta)$ (with one lost due to the constraint $\sum_{k=1}^K w_k = 1$). This corresponds to the regular regime, where the ELBO asymptotically aligns with the standard BIC.

A more interesting case arises when $\phi_0 < (m+1)/2$, corresponding to the singular regime, as the effective number of parameters $2\lambda$ is strictly smaller than $(mK + K - 1)$. As a by-product of our proof, the following corollary shows that in this regime, the variational approximation $\widehat{q}_{\bm w}$ to the posterior distribution on the mixing weight has an interesting stable behavior of emptying out the $(K-K^*)$ redundant components when $K$ is over-specified.

\begin{corollary}[Stability of mixing weights]\label{empty out rate}
    Suppose Assumptions {\color{black} P and A1--A4} hold and there exists a constant $C_3$ and a sequence $\kappa_n$ such that $\Delta_n\le C_3\log\log n$ with probability {\color{black} $1-\kappa_n$}. Then {\color{black} for sufficiently large $n$} it holds with probability at least {\color{black} $1 - n^{-1/2} - \kappa_n$} that: 1.~when $\phi_0<(m+1)/2$ and for $\rho_1 = 2(C_3 + 1)/(m + 1-2\phi_0)$,
    \begin{align*}
        \inf_{\sigma\in\m{S}_K} \sum_{k = K^*+1}^K \overline w_{\sigma(k)}< \frac{(K-K^*)\left[(\log n)^{\rho_1} + \phi_0\right]}{n},
    \end{align*}
    where $\m{S}_K$ is the set of all permutations over $[K]$; 2.~when $\phi_0>(m+1)/2$ and for $\rho_2 = 2(C_3 + 1)/(2\phi_0 - m -1)$,
    \vspace{-0.75cm}
    \begin{align*}
         \inf_{k\in[K]} \overline w_k \ge \frac{1}{\left(\log n\right)^{\rho_2} }+\frac{\phi_0}{n}.
    \end{align*}
\end{corollary}
{\color{black}}
Our condition $\Delta_n\le C_3\log\log n$ requires a finite sample analysis of the likelihood ratio test statistic under singularity, and can be verified in many concrete examples~\citep{watanabe2006stochastic,watanabe2007stochastic,rotnitzky2000likelihood,Drton2009,liu2003asymptotics,mitchell2019hypothesis}, {\color{black} where $\kappa_n$ can be taken as $(\log n)^{-1}$}; also see Appendix~\ref{app:review} for a brief review.
This corollary indicates that for small $\phi_0$, redundant mixture components are emptied out at a rate of $n^{-1}$ (up to $\log n$ factors), which is faster than the usual parametric rate of $n^{-1/2}$. This super-efficiency is a unique property in the singular regime, where certain components of the posterior can converge faster, depending on the behavior of the prior near the boundary of the parameter space \citep{Green2014}. By contrast, in the regular regime with $\phi_0 > (m+1)/2$, the mixing weight estimates tend to be spread across all $K$ components exhibiting less stability and losing the ``super-efficiency'' property. {\color{black} We observe that when $\phi_0 < (m+1)/2$, decreasing $\phi_0$ results in a smaller $\rho_1$ and a larger exponent on the $\log n$ term in the convergence rate. This observation is consistent with the analysis following Theorem~\ref{main theorem}, suggesting that one should avoid choosing an overly small $\phi_0$. When $\phi_0 > (m+1)/2$, increasing $\phi_0$ reduces $\rho_2$, resulting in more uniform weight allocation across all components and deteriorate the estimation accuracy of the mixing weights. Our empirical results in Figure~\ref{fig: fig5} of Section~\ref{sec: section 4.1} also support these findings.} A related result on the robustness of the true posterior $p(w\,|\,X^n)$ was established by \cite{rousseau2011asymptotic} who proved an $n^{-1/2}$ convergence rate (up to logarithmic terms) for small $\phi_0$, under stronger second-order identifiability assumptions. Corollary~\ref{empty out rate} shows that variational approximations similarly inherit the stability and super-efficiency properties of the posterior in finite mixture models.

When $\phi_0 < (m+1)/2$,  the rapid emptying of redundant components ensures that only the $K^*$ statistically identifiable components retain non-negligible weights. A natural question is whether the parameters associated with these true components---those retaining positive weights---converge at the parametric rate $n^{-1/2}$ (up to logarithmic factors). The answer is affirmative. To formalize this result, we first introduce a precise notion of identifiability.

\begin{definition}[Weak identifiability]
    The family $\{g(x;\eta),\eta\in\Omega\}$ is said to be weakly identifiable, or identifiable in the first order, if $g(x;\eta)$ is differentiable in $\eta$ and for any finite $K$ different $\{\eta_k\}_{k=1}^K$, for some $\alpha_k\in\mb{R}$ and $\beta_k\in\mb{R}^d$, the following equation holding for almost every $x$,
    \begin{align*}
        \sum_{k=1}^{K} \Big[\alpha_k g(x;\eta_k) + \beta_k^T\nabla_\eta g(x;\eta_k) \Big]= 0,
    \end{align*}
    implies $\alpha_k = 0$ and $\beta_k = 0$ for all $1\le k\le K$.
\end{definition}
This weak identifiability, or identifiability in the first order, is the least stringent form of identifiability condition for parameter estimation in FMMs in the literature and is satisfied by most of the commonly used distributions (e.g., see \cite{ho2016convergence}). The term ``weak'' is used in contrast to the so-called strong identifiability condition, which is also called identifiability in higher orders (with orders greater than one). For example, a one-dimensional family $\{g(x;\eta), \eta\in\Omega\}$ is said to be identifiable in $J$-th order (or $J$-strongly identifiable) if for any finite $K$ different $\{\eta_k\}_{k=1}^K$, and some $\alpha_k\in\mb{R}$ and $\beta_k^{(j)}\in\mb{R}^d$, $j\in[J]$, the following equation holds for almost all $x$,
\begin{align*}
        \sum_{k=1}^{K}\Big[\alpha_k g(x;\eta_k) + \sum_{j=1}^J\beta_k^{(j)} g^{(j)}(x;\eta_k) \Big]= 0,
\end{align*}
implies $\alpha_k = 0$ and $\beta_k^{(j)} = 0$ for all $1\le k\le K$ and $j=1,\ldots,J$. Here, $g^{(j)}(x;\eta_k)$ denotes the $j$-th order derivative of $g(x;\eta)$ with respect to $\eta$.

Note that strong identifiability always implies weak identifiability. A strong identifiability condition of order two is commonly imposed in the literature to characterize convergence rates in location or scale mixture models \citep{ho2016strong,jordan1999introduction,heinrich2018strong,manole2021estimating}. For weakly identifiable models, however, convergence rates must typically be established on a case-by-case basis (e.g., \cite{ho2016convergence}) by proving an inequality akin to Assumption A4, which relates the total variation distance to a Wasserstein-type distance. The resulting rates generally depend on the order $r$ in this inequality (see Assumption A4) and are often slower than the parametric rate $n^{-1/2}$.
In contrast, our result below demonstrates that, owing to the stable behavior of mixing weights in the singular regime (Corollary~\ref{empty out rate} with $\phi_0 < (m+1)/2$),weak identifiability combined with Assumption A4---regardless of the order $r$---still guarantees a parametric convergence rate for the parameters associated with the identifiable mixture components. Specifically, the variational posterior estimator of $\theta$ achieves an $n^{-1/2}$ rate (up to logarithmic factors) even for weakly identifiable finite mixture models, such as the location-scale Gaussian mixture.
{\color{black} While previous works \citep{heinrich2018strong, ho2020robust, manole2022refined, do2024dendrogram} have also established this convergence rate, their results typically rely on strong identifiability (second order or higher). In contrast, our framework relaxes this to weak identifiability (first-order) and requires Assumption A4 to bridge the distance between distributions and the distance between parameters, which can be verified to hold for weakly identifiable models of interest, such as location-scale Gaussian mixtures.}

\begin{theorem}[Convergence rate of component parameters]\label{para root n}
    Suppose the same assumptions as in Corollary~\ref{empty out rate} and the family $\{g(x;\eta),\eta\in\Omega\}$ is weakly identifiable. For any fixed $\phi_0<(m+1)/2$,
    {\color{black} denote $\rho_1': = 1\vee\rho_1 + 1$}, then there exists a large constant $M$ {\color{black} independent of $n, K$ and $\phi_0$}, such that it holds for sufficiently large $n$ with probability at least {\color{black} $1-n^{-1/2} - \kappa_n$},
    \begin{align*}
        \inf_{\sigma\in \m{S}_K}\sup_{1\le k\le K^*}|\overline{\eta}_{\sigma(k)} - \eta_k^*| \le \frac{M(\log n)^{\rho_1'/2}}{\sqrt{n}},
    \end{align*}
     where $\m{S}_K$ is the set of all permutations over $[K]$.
\end{theorem}
Theorem~\ref{para root n} can be interpreted as a consequence of Corollary~\ref{empty out rate}. As the mixing weights of redundant components decrease to zero at sufficiently fast rate despite the over-specification, we can effectively treat the working models as well-specified with the correct number of components. In such well-specified cases, \cite{ho2016strong} derived in Theorem 3.1 that $d_{\rm TV}(p(\cdot\,|\theta), p^*(\cdot)) \ge c_0 W_1(G(\theta), G^*)$ when $W_1$ is small, provided that the family is weakly identifiable. This leads to a nearly parametric convergence rate for estimating the parameters of each component, thanks to the nearly parametric convergence rate under the total variation metric. This argument illustrates how singularity helps with the stability behavior of mixing weights and enhances parameter estimation efficiency. In Figure~\ref{fig: fig5} we validate the theoretical correctness by examining the behavior of $W_1$.

\section{Numerical Studies}\label{section 4}
In this section, we provide simulation studies to further validate our theoretical results. We illustrate and compare the two predicted regimes, namely, the singular regime where $\phi_0<(m+1)/2$ and the regular regime where $\phi_0>(m+1)/2$, in terms of model selection and parameter estimation performance. In addition, we evaluate the proposed ELBO-based model selection approach against state-of-the-art methods for determining the number of components in finite mixture models, and demonstrate its practical utility on a real dataset.

{\color{black} Throughout this work, the ELBO is optimized via coordinate-ascent variational inference (CAVI) \citep{bishop2006pattern}. 
Since our analysis is built on exponential-family models in canonical form, we place conjugate priors on the component parameters so that the variational updates admit closed forms. 
For Gaussian mixtures, we use a normal prior on the component means in the location-only setting (fixed covariance), while in the location--scale setting (unknown covariance) we use a Normal--inverse-Wishart prior, which has been adopted in related applications \citep{heins2024gradient}.}

\subsection{Impact of \texorpdfstring{$\phi_0$}{phi0} on ELBO in finite mixture models}\label{sec: section 4.1}
To study the effectiveness of using the ELBO $\m{L}(\widehat q_{Z^n})$ analyzed in Theorem~\ref{main theorem} for model selection, and to examine how different choices of $\phi_0$ may impact the results, we conduct numerical experiments using Gaussian mixtures under different settings.

\begin{figure}[t]
  \centering
  \begin{minipage}{0.37\textwidth}
    \centering
    \includegraphics[width=\linewidth]{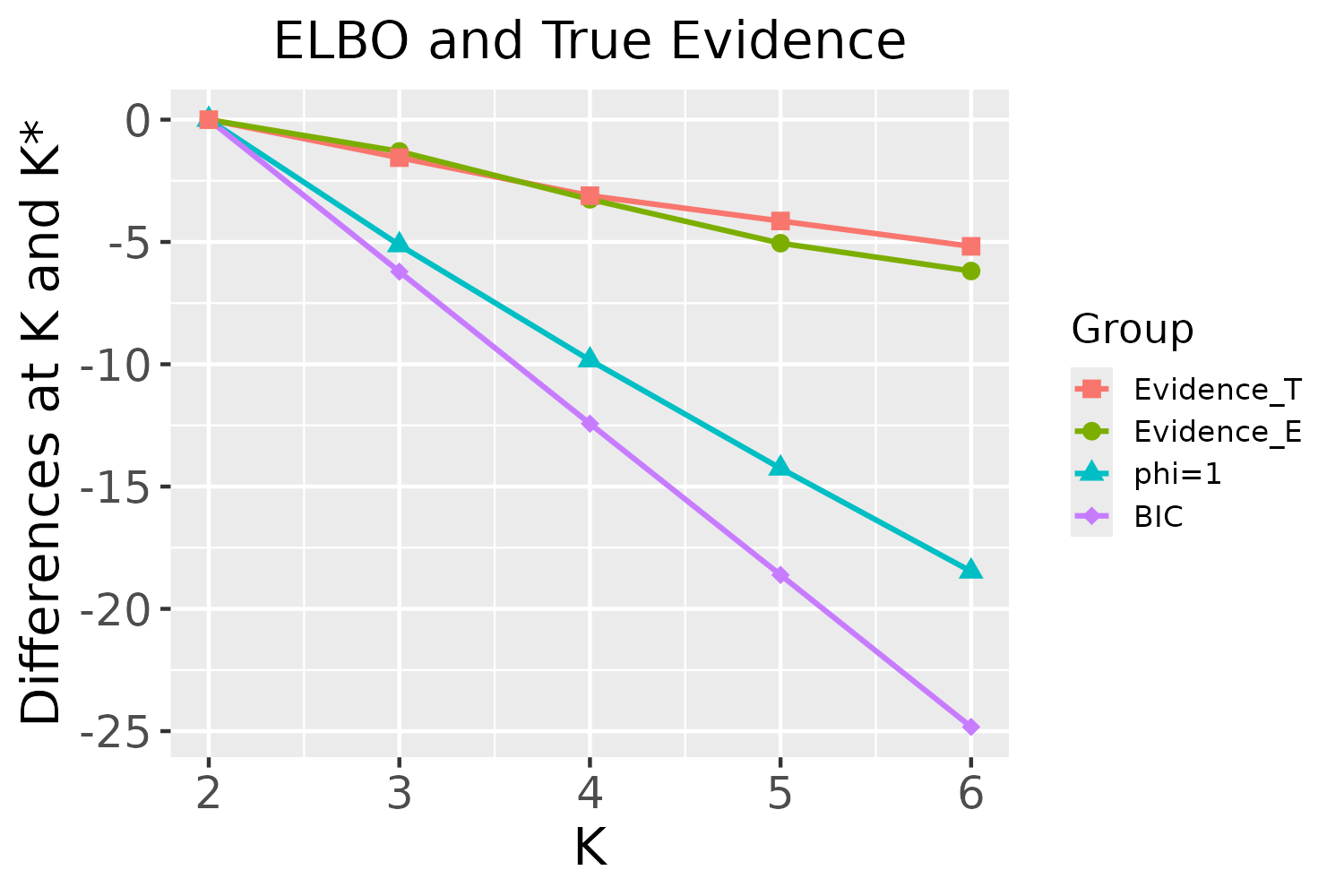}
    \caption{Comparison of penalties in ELBO, Evidence, and BIC.}\label{fig: evidence}
  \end{minipage}
  \hfill
  \begin{minipage}{0.60\textwidth}
    \centering
    \includegraphics[width=\linewidth]{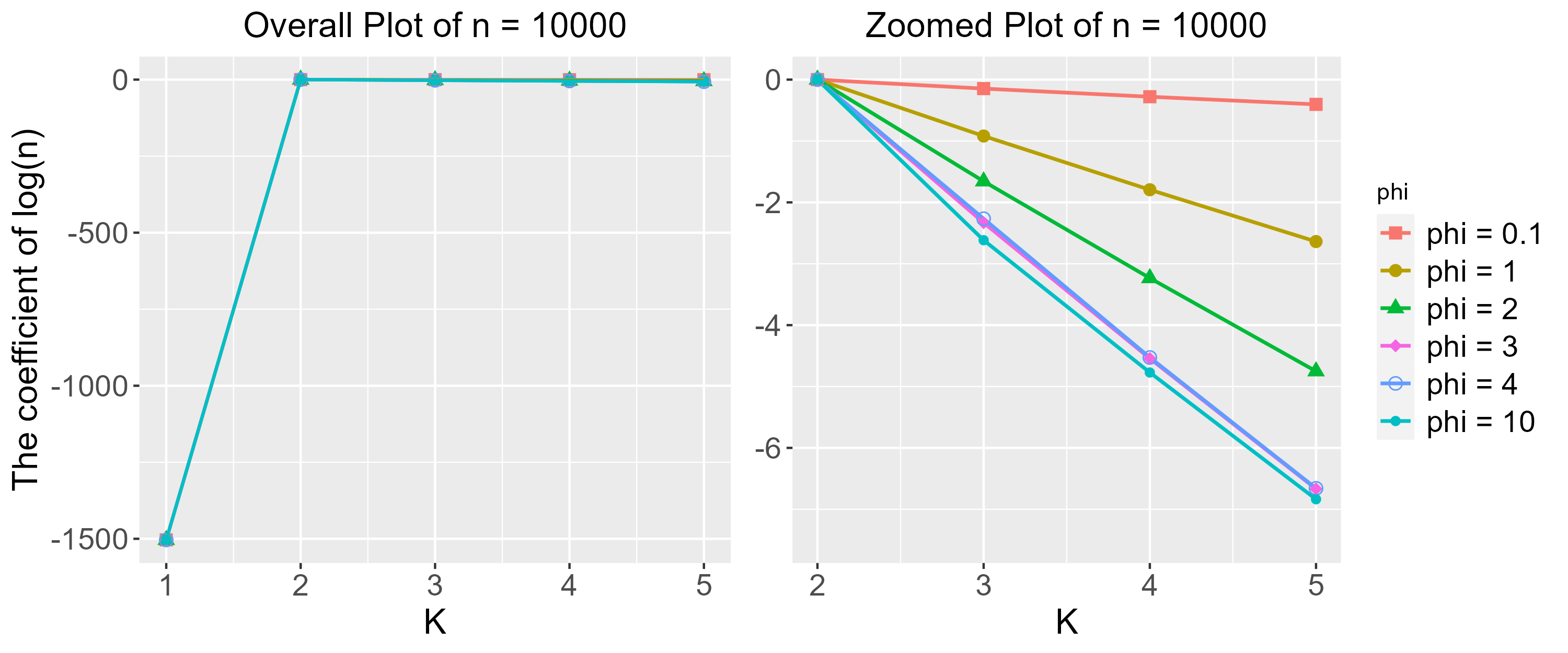}
    \caption{Plots of $(\widehat {\m L}_K - \widehat {\m L}_{K^*})/\log n$ versus $K$ with fixed $n=10000$.}\label{fig: fig1}
  \end{minipage}
  
\end{figure}

\vspace{-0.5cm}
\paragraph{Difference between ELBO and true evidence.}
First we compare the differences between ELBO, the true evidence, and BIC. \cite{aoyagi2010bayesian} derived the RLCT, which determines the (asymptotic) leading coefficient for the true evidence, only for univariate location Gaussian distribution with uniform prior ($\phi_0 = 1$) on mixing weights, so we set $m = d = 1$, then $(m+1)/2 = 1$. We set $K^* = 2$, $w^* = (0.5, 0.5)$, and $\mu^* = (-2, 2)$. The RLCT values in \eqref{RLCT} are given as
\begin{align*}
    \lambda_{\m M} = K^* - 1 + \frac{j + j^2 + 2(K-K^*+1)}{4(j+1)} \ \ \mbox{with} \ \ j = \max\{i:\,i+i^2\le2(K-K^*+1)\}. 
\end{align*}
{\color{black} We note that the ELBO is not the model evidence itself, but a computationally tractable surrogate (a lower bound) rather than an exact substitute for the evidence. We visualize these quantities in Figure~\ref{fig: evidence} by plotting the ELBO, BIC, as well as the empirical evidence (denoted as ${\rm Evidence_{\rm E}}$) derived by nested sampling \citep{skilling2006nested} and the theoretical evidence (denoted as ${\rm Evidence_{\rm T}}$) by plugging-in the theoretical RLCT value. At a sample size of $n = 500$, we subtracted the values corresponding to $K = 2$ for all methods to facilitate comparison. It can be observed that the penalty for evidence is evidently smaller than for ELBO. However, due to the impact of constant terms in ELBO, at this sample size, the downward trend of ELBO at $\phi_0 = (m+1)/2$ is still weaker than that of BIC.

While nested sampling provides a benchmark estimate of the evidence, it is substantially more computationally expensive and does not scale well to higher-dimensional or repeated experiments; in contrast, the ELBO produced by CAVI can be computed orders of magnitude faster. We now also report runtime comparisons in~\ref{runtime}.}

\begin{figure}[t]
    \centering
    \includegraphics[width=0.85\textwidth]{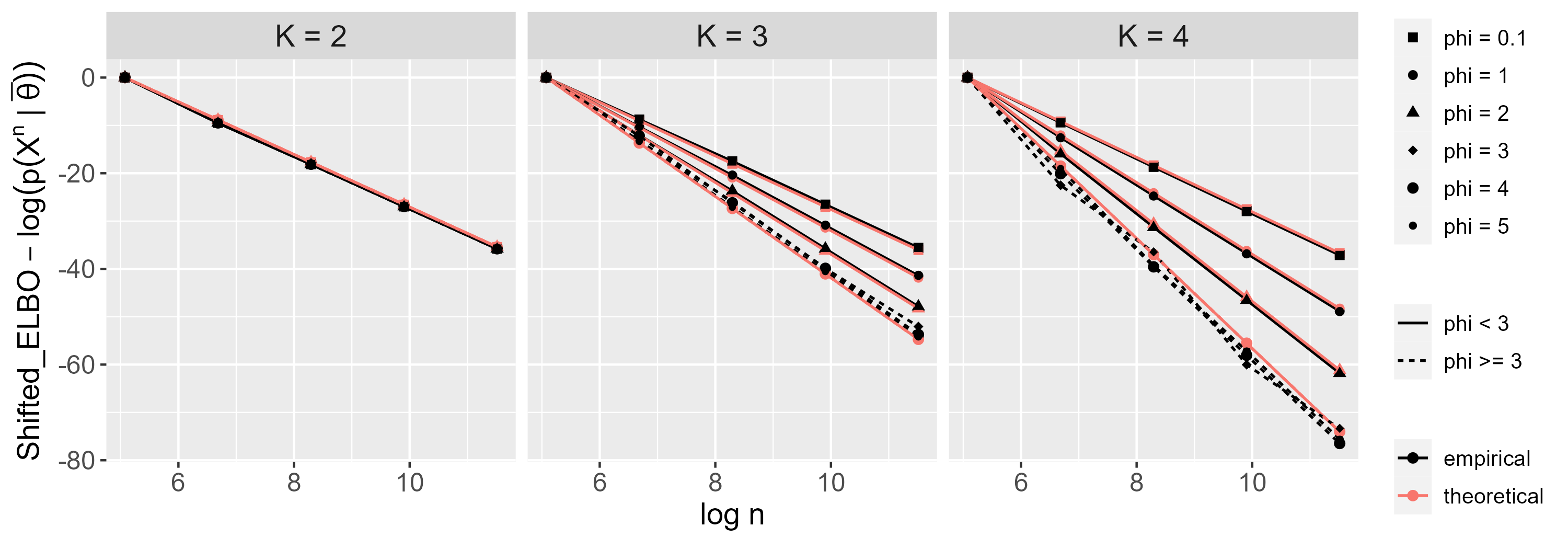}
    \caption{Plots of $\widehat {\m L}_K - \log p(X^n|\overline{\theta})$ for different values of $\phi_0$ at fixed $K$, with all starting points shifted to zero for better comparison.}
    \label{fig: fig3}
\end{figure}

\vspace{-0.5cm}
\paragraph{Validation of the relationship between $\lambda$, $K$ and $\phi_0$.}
Here we verify the expression of $\lambda$ given in \eqref{lambda}. The simulated data $X^n$ is generated from a $m$-dimensional location-Gaussian mixture model with $K^* = 2$ components.
We use the identity matrix $\bm{I}$ as the variance-covariance matrix for the mixture components and treat it as known. The true mixing weights are $w_1^* = w_2^* = 1/2$ and the two location parameters are $\eta_1^* = -2/\sqrt{m}\cdot\bf{1}$ and $\eta_2^* = 2/\sqrt{m}\cdot\bf{1}$, where $\bf{1}$ denotes the $m$-dimensional all one vector. We employ a symmetric Dirichlet prior Diri$(\phi_0,\ldots,\phi_0)$ for the mixing weight $w$ and an independent $\m N(\bm{0},\,\bm{I})$ prior for the location parameters $\{\eta_k\}_{k=1}^K$. We consider $K_{\max}=5$ candidate GMMs with component number $K\in\{1,2,\ldots,K_{\max}\}$. Since the CAVI algorithm can be sensitive to the initialization, {\color{black} we fit the data with $K_{\rm int}$ components using the EM algorithm, then use the responsibility matrix (representing the posterior probabilities that each data point belongs to each component) as the initialization for $S^n$, for each $K_{\rm int} \in \{1, 2, \ldots, K\}$, and then iteratively update $q_\theta(\theta)$ and $q_{S^n}(s^n)$ until the ELBO value converges.} For each $K$, we select the largest value obtained among different initializations as our final output of $\m{\widehat L}_K$, the ELBO value for the model with $K$ components. Now we set $m=d=5$.

In the first setting, fix the sample size at $n=10^4$ and let $K$ change under fixed $\phi_0$. We choose {\color{black} $\phi_0 \in \{0.1,1, 2,3,4,10\}$}, where $(m+1)/2=3$ corresponds to the critical threshold distinguishing the singular and regular regimes, as predicted in Theorem~\ref{main theorem}. In particular, our theory predicted that for $K>K^*$, the ELBO value $\widehat {\m L}_K$ satisfies $\widehat {\m L}_K - \widehat {\m L}_{K^*} =\m O_P(1) - \min\{\phi_0,\,3\}\, (K-K^*)\,\log n.$
Consequently, when plotting $(\widehat {\m L}_K - \widehat {\m L}_{K^*})/\log n$ against $K$, the resulting curve will approximate a straight line with a slope of $-\min\{\phi_0,\,3\}$. The numerical results shown in Figure~\ref{fig: fig1} confirm this theoretical prediction. As we can see, the lines become steeper as $\phi_0$ increases until $\phi_0 = 3$, after which they remain unchanged, as the lines for $\phi_0 = 3,4$ and $10$ almost overlap.

{\color{black} In the second scenario, we use the same true model as above (with $d = m = 5$) and consider settings with fixed $K$ and varied $n$ to numerically verify the theoretical dependence of $\lambda$ on $\phi_0$. According to Theorem~\ref{main theorem}, if we plot $\m{L}(\widehat q_{Z^n}) - \log p(X^n|\overline\theta)$ against $\log n$, we expect to see a linear relationship with a slope of $-\lambda$, which equals $(K-K^*)\phi_0 + (mK^* + K^* -1)/2$ for $\phi_0 < 3$ and $(mK + K -1)/2$ for $\phi_0 \ge 3$. We set $n\in\{160,800,4000,20000,100000\}$, and shift each line determined by $\phi_0$ so that the value at $n = 160$ equals zero, which allows for a clearer comparison of the slopes. The corresponding results are summarized in Figure~\ref{fig: fig3}. We also mark the theoretical values with red lines. As we can see, the experimental results align closely with the theoretical values, as $\phi_0$ exceeding the threshold 3, slopes of the lines remain essentially unchanged (under the same $K$). The close alignment with the theoretical values is because taking difference between results for different $n$ under the same $K$ eliminates the constant terms related to $K$. These empirical observations are again consistent with the predictions from our theory.}

{\color{black} We further confirm that the weakly identified location--scale Gaussian family exhibits the same qualitative patterns predicted by our theory. See Appendix~\ref{app:scale} for details.
}

\begin{figure}[t]
  \centering
  \begin{minipage}{0.64\textwidth}
    \centering
    \includegraphics[width=\linewidth]{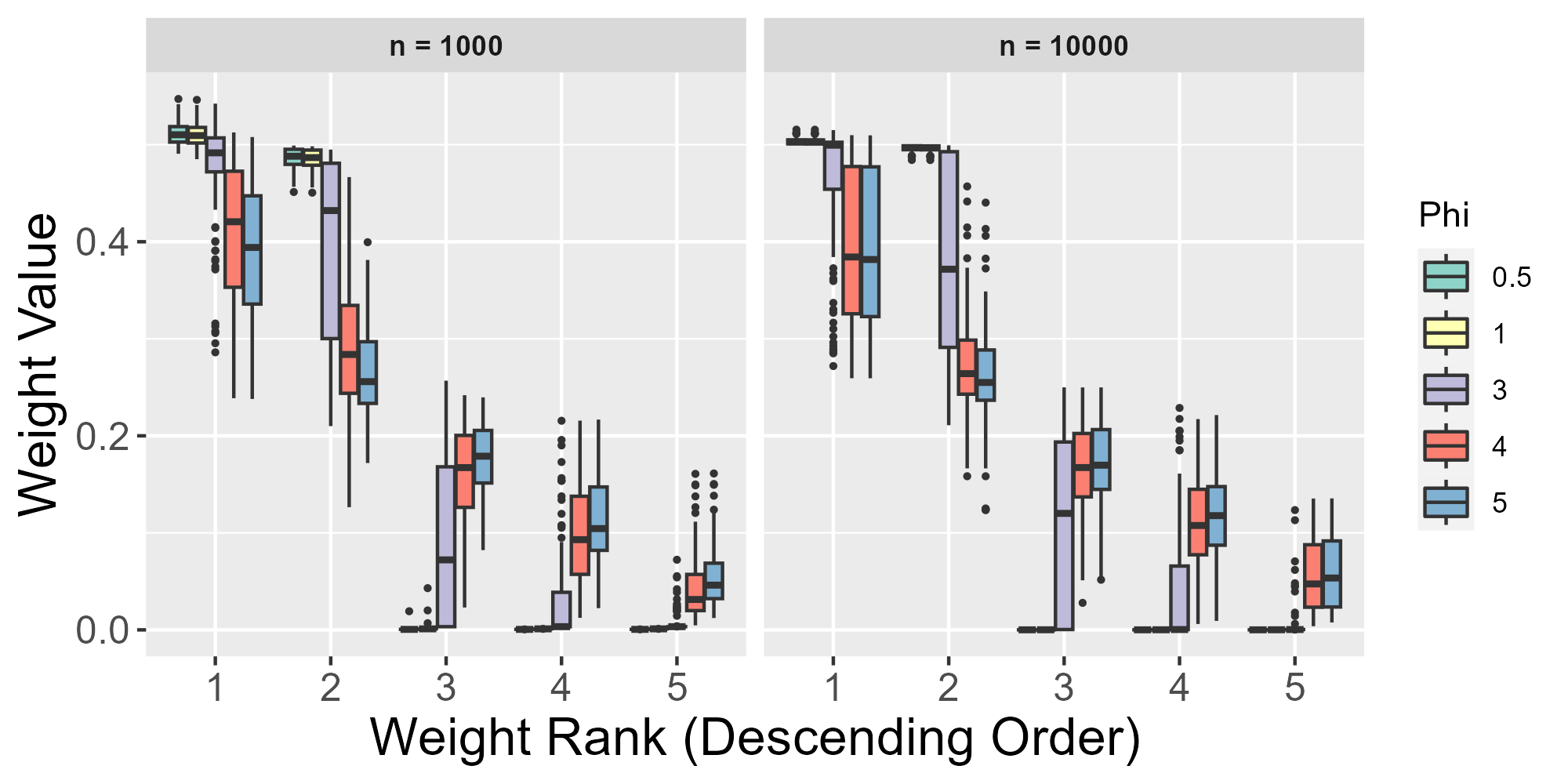}
    \end{minipage}
  \hfill
  \begin{minipage}{0.32\textwidth}
    \centering
    \includegraphics[width=\linewidth]{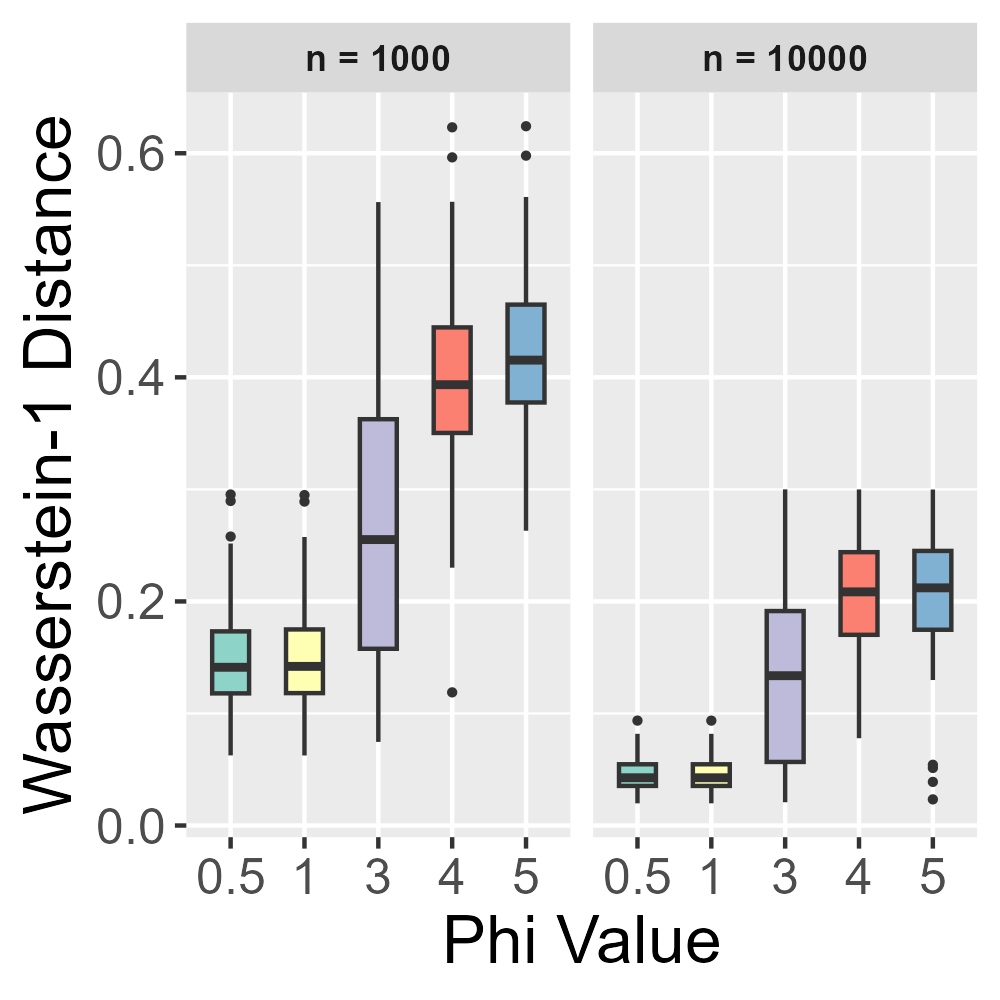}
  \end{minipage}
  \caption{Box plots of $\overline{w}_k$ (descending order) and $W_1(G^*, G(\overline{\theta}))$ for different $\phi_0$ with $K = 5$.}
  \label{fig: fig5}
\end{figure}

\vspace{-0.5cm}
\paragraph{Verification of the asymptotic behavior of parameters.}
Finally, to evaluate the stability of the variational approximation $\widehat q_{\bm w}$, which tends to eliminate redundant components under over-specified models ($K > K^*$) with small $\phi_0$, we revisit the true model from Figure~\ref{fig: fig1}. Specifically, we report the variational posterior means of the mixing weights $\overline{\bm w}$ under $K = 5$ and $K^* = 2$. 
Figure~\ref{fig: fig5} presents box plots of the sorted $\overline{w}_k$ across 100 repetitions for each setting. The behavior of $\overline{w}_k$ is consistent with our predictions from Corollary~\ref{empty out rate}: for $\phi_0 < 3$, the weights concentrate on two components, whereas for $\phi_0 > 3$, they are distributed across all five components.
{\color{black} When $\phi_0\in\{0.5,1\}$, increasing $n$ from $10^3$ to $10^4$ reduces the variational posterior mean weights of the three redundant components by roughly one order of magnitude (from about $5\times10^{-4}$ and $10^{-3}$ to $5\times10^{-5}$ and $10^{-4}$), consistent with an $n^{-1}$ emptying-out rate.}
Notably, at $\phi_0 = 3$, the deviations are the largest, indicating maximal instability at $\phi_0 = (m+1)/2$. These findings further support using $\phi_0<(m+1)/2$ to enhance robustness and accuracy in parameter estimation.

Additionally, Figure~\ref{fig: fig5} reports the estimated Wasserstein distances $W_1$ between the estimated mixing measure $G(\overline \theta)$ and the true $G^*$ assessing the convergence of $\overline \eta_k$. It is clear that for $\phi_0 < 3$, the values of $W_1$ are significantly smaller and more stable compared to $\phi_0 \ge 3$. For example, when $n = 1000$, the average $W_1$ values for $\phi_0 = 0.5$ and $1$ are 0.148 and 0.149, respectively, while the numbers for $\phi_0 = 3,4$ and $5$ are $0.272, 0.395$ and $0.422$. When $n$ increases to $10000$, the mean $W_1$ of $\phi_0 = 0.5$ and $1$ both decrease to 0.045, approximately $30\%$ of the original, which verifies the $n^{-1/2}$ convergence rate. Again, $W_1$ oscillates the most at $\phi_0 = 3$, consistent with the instability observed $\overline{w}_k$.
We also observe that both the weights assigned to the redundant components and the magnitude of $W_1$ tend to increase on average as $\phi_0$ becomes larger, corroborating the roles of $\rho_1$ and $\rho_2$ in Corollary~\ref{empty out rate}. 
{\color{black} The numerical values underlying Figure~\ref{fig: fig5} are provided in Appendix~\ref{app:raw fig5}, which more directly illustrates the $n^{-1}$ emptying-out of redundant mixing weights and the $n^{-1/2}$ convergence of point estimates under small $\phi_0$.}

\begin{figure}[t]
    \centering
    \includegraphics[width=\textwidth]{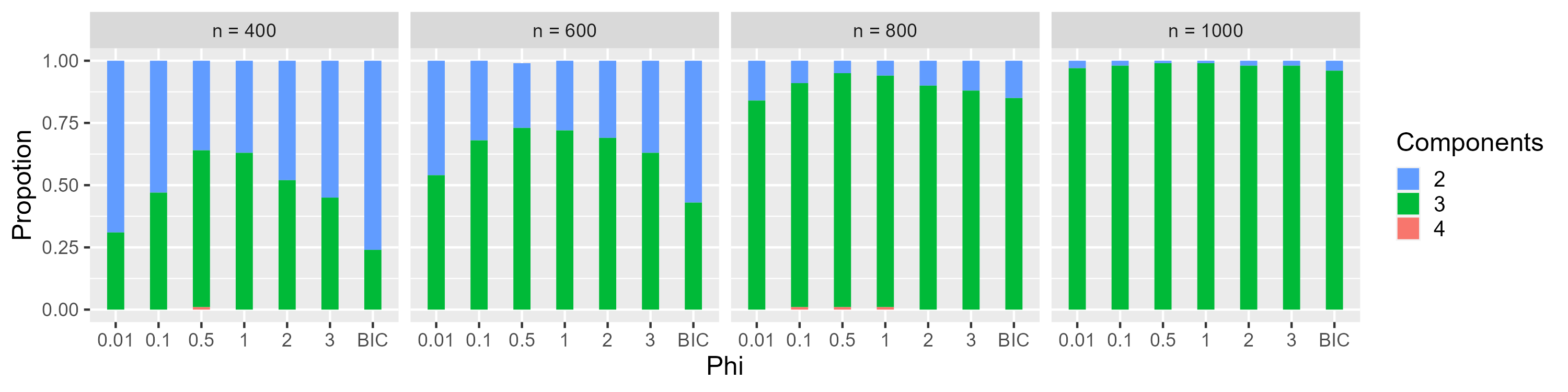}
    \caption{The stacked bar plots of selected components under different $\phi_0$ values.}
    \label{fig: fig4}
\end{figure}

\vspace{-0.5cm}
\paragraph{Efficiency of different choices of $\phi_0$.}
Furthermore, to compare the efficiency of model selection caused by varying $\phi_0$, we designed a set of experiments with bivariate location-scale Gaussian mixtures. The true distribution is with $K^*=3$, where the parameter dimension in each component is $m = 5$ ($2$ for the means and $3$ for the covariance matrix). The true parameters are as follows:
\begin{align*}
    \bm w^* = (0.075, 0.075, 0.85), \quad \bm \mu^* = \{(-1, -0.8)^T, (1, -0.8)^T, (0.8, 0.8)^T\}\\
    \bm \Sigma^* = \left\{\Bigg(\begin{array}{cc} 0.25 & 0.05 \\ 0.05 & 0.25 \end{array}\Bigg), 
    \Bigg(\begin{array}{cc} 0.25 & -0.05 \\ -0.05 & 0.25 \end{array}\Bigg), 
    \Bigg(\begin{array}{cc} 0.25 & 0 \\ 0 & 0.2 \end{array}\Bigg)\right\}
\end{align*}
Under sample sizes $n \in \{400, 600, 800, 1000\}$, we conducted 100 repeated experiments for $\phi_0 \in \{0.01, 0.1, 0.5, 1, 2, 3\}$ and recorded the frequency of selection, shown in Figure~\ref{fig: fig4}. As analyzed below Theorem~\ref{main theorem}, both excessively small and large $\phi_0$ can lead to underestimation, so it is evident that accuracy is highest when $\phi_0 = 1/2$ and $\phi_0 = 1$. Thus, $\phi_0$ values between $1/2$ and $1$ are recommended, where the penalty is moderate, and constant terms remain stable with $K$. When $K^*$ is known, $1/K^*$ is a common Dirichlet prior choice, balancing sparsity and prior strength. Here, $1/2$ is close to $1/3$ ($K^* = 3$), preserving this property well. $\phi_0 = 1$ (a uniform prior on the simplex) also performs well, suggesting that when $K^*$ is uncertain, a non-informative prior is reasonable. Section~\ref{section 4.2} further demonstrates the robustness of $\phi_0 = 1$, reinforcing its practical applicability.
We also recorded BIC's selection results. When the sample size is small, ELBO with $\phi_0 = 3$ achieves notably higher accuracy than BIC due to the constant terms mitigating the penalty. As $n$ increases, their accuracy gap narrows, suggesting that their behavior becomes more similar.

\subsection{Comparison with other model selection methods}\label{section 4.2}

\begin{figure}[t]
    \centering
    \includegraphics[width=\textwidth]{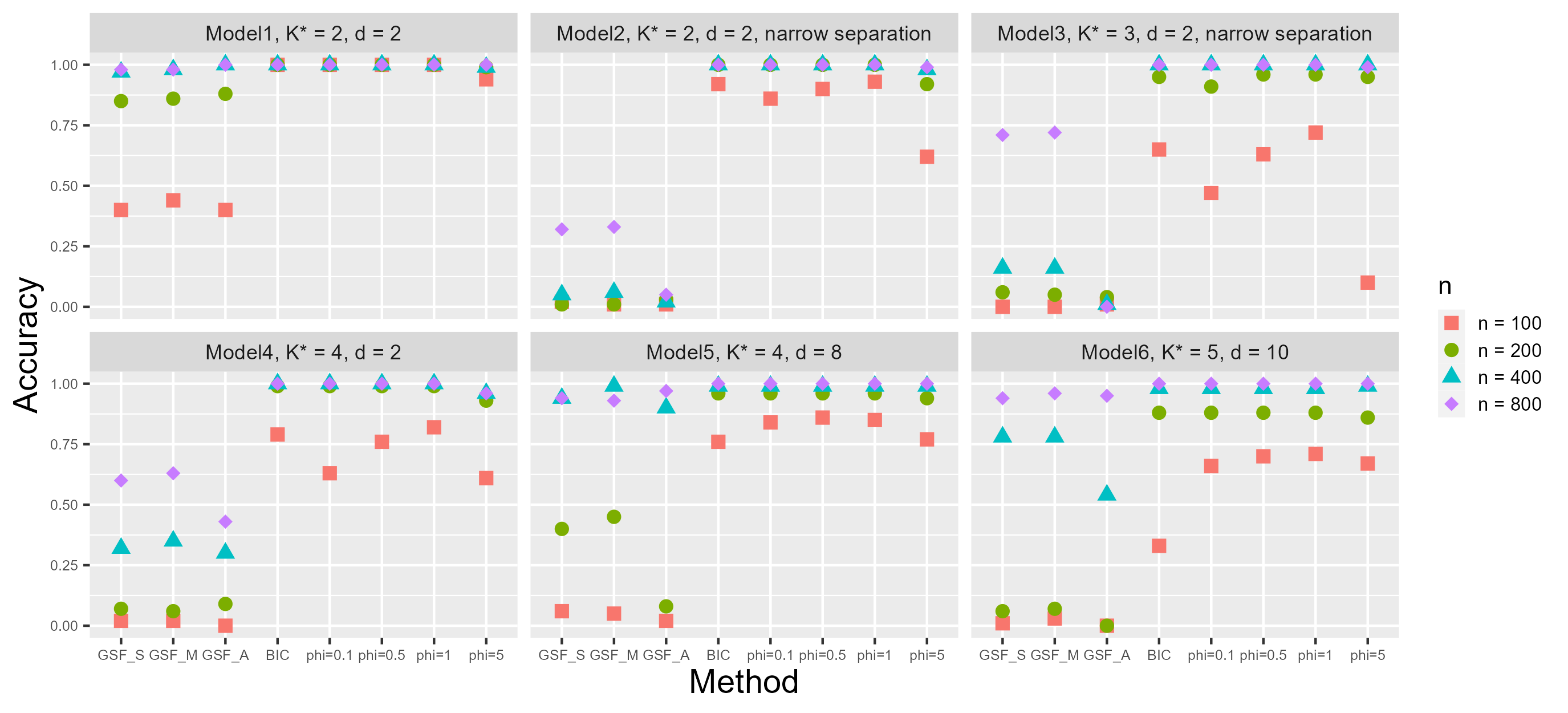}
    \caption{Proportion of correctly selected numbers under different settings and methods.}
    \label{fig: comparasion}
\end{figure}

Recently, \cite{manole2021estimating} proposed Group-Sort-Fuse (GSF), a method for model selection in FMMs, which demonstrates higher consistency and efficiency than existing approaches such as the Merge-Truncate-Merge (MTM) algorithm of \cite{guha2021posterior}. GSF is a penalized likelihood approach that simultaneously estimates the number of components and the mixing measure $G(\theta)$. Three penalties are considered in their work: Smoothly Clipped Absolute Deviation (SCAD), Minimax Concave Penalty (MCP), and Adaptive Lasso (ALasso). Their study shows that GSF outperforms MTM in terms of accuracy and computational efficiency, as MTM relies on posterior sampling and incurs higher computational costs. Accordingly, in this study, we focus on comparing three methods: ELBO maximization, BIC, and GSF with the three penalties (denoted as ${\rm GSF_S}$, ${\rm GSF_M}$ and ${\rm GSF_A}$) in a location Gaussian mixture model. {\color{black} The results are presented in Figure~\ref{fig: comparasion}, showing the percentage of correct identifications out of 100 repetitions.}

Since GSF is only effective for strongly identifiable families, we designed experiments with various location-Gaussian mixtures.
We conducted experiments with $K^* = 2$ to $5$ and $m = d$ from $2$ to $10$. It can be observed that ELBO consistently outperforms GSF across all settings. Models 1 and 2 both have $K^* = 2$ and $d = 2$, but Model 2 features closer component separation, illustrating GSF’s sensitivity to poorly separated mixtures.
In Models 3 through 6, with $K^* = 3, 4, 4$ and $5$ and $d = 2, 2, 8$ and $10$ respectively, GSF performs poorly at sample sizes of $100$ and $200$, frequently underestimating the number of components. BIC tends to favor smaller models when $n$ is small due to its overly conservative penalty, a limitation that becomes more pronounced as $m$ increases. For instance, in Model $1-4$, BIC outperforms ELBO with $\phi_0 = 0.1$ and $5$ when $n = 100$. However, in Model 5, with $\phi_0 = 5 > (m+1)/2 = 4.5$, BIC underperforms to all ELBO-based methods, yielding similar results only to ELBO with $\phi_0 = 5$. In Model 6, where $\phi_0 = 5 < (m+1)/2 = 5.5$, BIC’s disadvantage becomes more evident: at $n = 100$, BIC achieves only 33\% accuracy, whereas all ELBO-based methods exceed 65\%.

\subsection{Single-Cell RNA Sequencing Data \citep{zheng2017massively}}

\begin{figure}[t]
  \centering
  \begin{minipage}{0.62\textwidth}
    \centering
    \includegraphics[width=\linewidth]{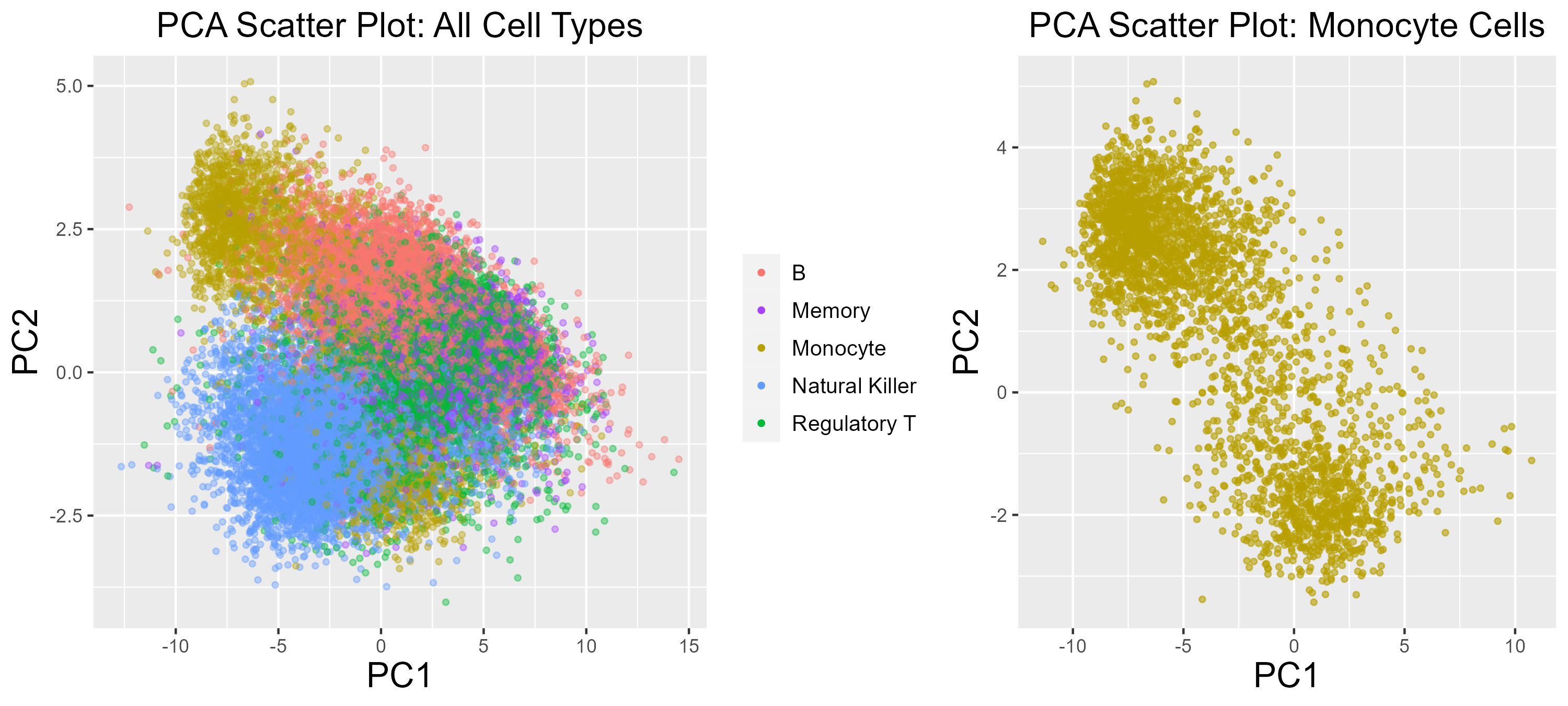}
    \end{minipage}
    \hfill
    \begin{minipage}{0.31\textwidth}
    \centering
    \includegraphics[width=\linewidth]{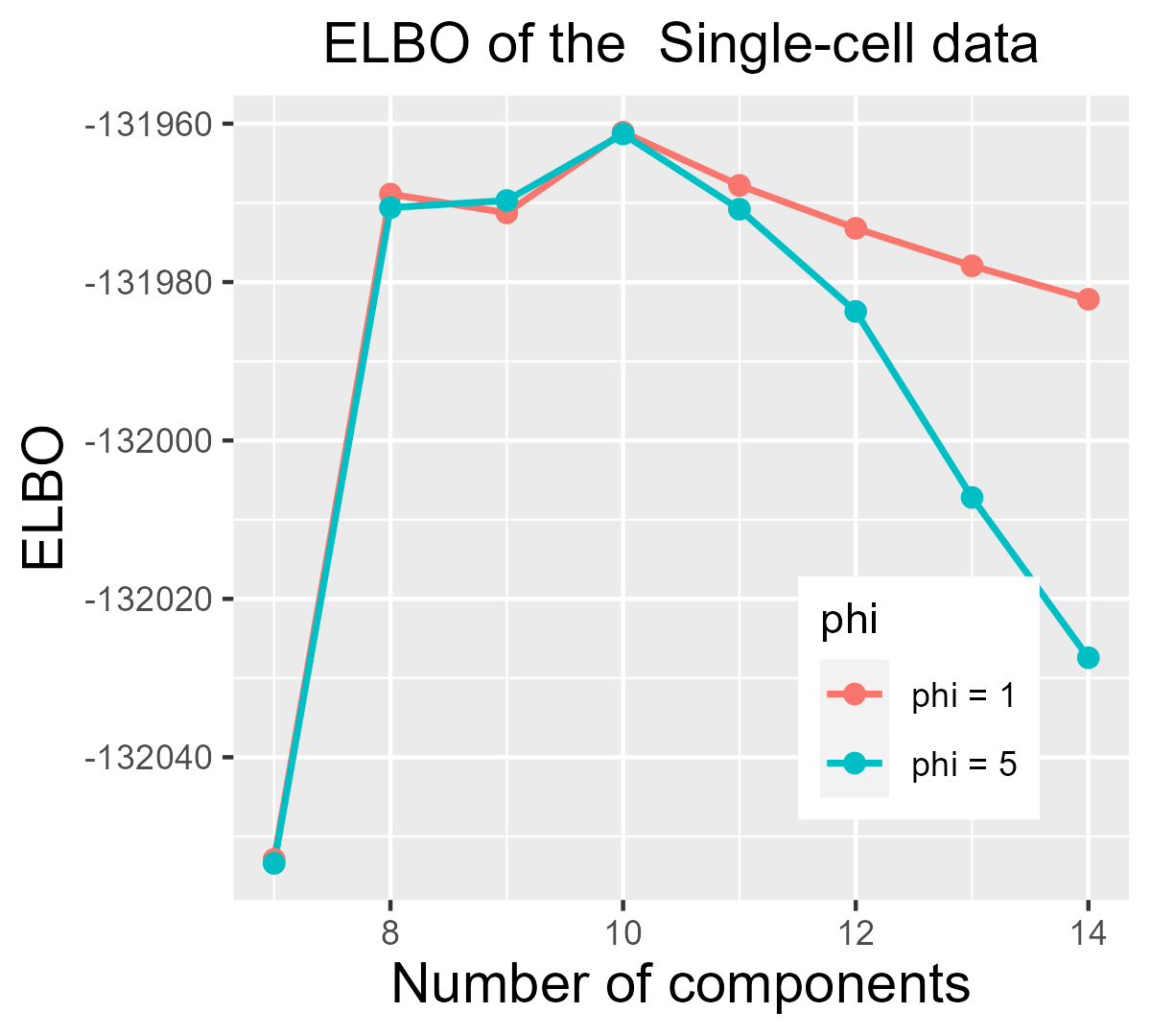}
    \end{minipage}
  \caption{Scatter plot of the first two principal components and results based on ELBO. Clearly, gene expression sub-clusters exist within the same cell type.}
  \label{fig: fig7}
\end{figure}

{\color{black} In this subsection, we apply our method to the Single-cell RNA-sequencing dataset from \cite{zheng2017massively}. We retained five cell types: Memory T cells, B cells, Regulatory T Cells, Natural Killer cells, and Monocytes and selected the 50 genes with the highest expression frequency and applied a pseudo-transformation, yielding a dataset of 26,794 cells and 50 genes. 
Principal Component Analysis (PCA) reveals an elbow point at three dimensions; thus, we retain the first three principal components and model the data as a three-dimensional location-scale Gaussian mixture, corresponding to $m=9$ with a threshold $(m+1)/2 = 5$ for regime classification.

The ELBO-based methods (with $\phi_0 = 1$ and $5$) both selected 10 components, while BIC selected 8. 
Although GSF is theoretically applicable only to location Gaussian mixtures with identical covariance matrices, we evaluate it using the two penalties that exhibited the best performance in previous experiments. Both GSF methods demonstrate high sensitivity to initialization. Using the \texttt{normalLocOrder} function and $30$ repeated runs, MCP most frequently selects $11$ components ($9$ out of $30$ runs), with a mean of $10.9$ and standard deviation $1.4$; SCAD most frequently selects $13$ components ($7$ runs), with a mean of $11.2$ and standard deviation $1.7$.

Despite retaining five main cell types, heterogeneity in biological states, such as differentiation stages, cell cycle states, or spatial locations, may give rise to sub-clusters in gene expression. For example, Monocytes are clearly separated into at least two sub-clusters along the first two principal components (Figure~\ref{fig: fig7}). 
Notably, \cite{zheng2017massively} (Figure~3) applied K-means clustering on t-SNE embeddings and identified 10 clusters, suggesting that $10$ is a reasonable and biologically meaningful partition, consistent with the selection made by the ELBO-based method. Detailed clustering results for the $10$ components selected by ELBO are provided in Appendix~\ref{clustering}, alongside the $8$ clusters identified by BIC. BIC merges Memory T cells with Regulatory T cells and therefore fails to assign them to distinct clusters. Furthermore, BIC interprets one of the three peaks observed in NK cells as an artifact caused by cluster overlap. However, contour analysis shows that this peak has a higher density than the sum of the other two, supporting its interpretation as a distinct subpopulation, which ELBO correctly captures while BIC does not.

We also compared the computational efficiency of different methods on this real dataset, with results summarized in Appendix~\ref{runtime}. ELBO achieves a runtime comparable to BIC (implemented with the EM algorithm), averaging $496$ seconds for a sample size of $5000$, compared to $336$ seconds for BIC. The additional computational cost arises from matrix inversion steps required in location-scale Gaussian mixtures; this additional cost can be avoided in simpler models such as location-only Gaussians or multinomials. Furthermore, as an approximation to the true marginal likelihood, ELBO can be computed substantially faster than nested sampling, making it practical for large-scale datasets and complex models.
}

\bibliographystyle{apalike}

\setlength{\bibsep}{1pt}
\small
\bibliography{reference}

\newpage

\makeatletter
\let\addcontentsline\oldaddcontentsline
\makeatother

\clearpage
\appendix

\begin{center}
    {\LARGE \bf Supplementary Materials for ``Model Selection for Finite Mixture Models via Variational Approximation''}
\end{center}
\spacingset{1.25}

\begingroup
\setcounter{tocdepth}{2}  
\tableofcontents
\endgroup
\clearpage

\section{More Literature Review}\label{app:review}
In this appendix, we offer additional details on some of the literature reviewed in the main paper, along with information on several more related works.

\medskip
\noindent {\bf Variational Bayes.}
Variational inference was introduced by \cite{jordan1999introduction} for probability density approximation with intractable integrals and point estimation for parameter determination. In variational inference, the posterior distribution is approximated by the closest member relative to the Kullback-Leibler (KL) divergence in a specified family. Among the various approximating schemes, mean-field approximation, where the variational family adopts a factorized form, emerges as the most prevalent type of variational inference. It is characterized by its conceptual simplicity, ease of implementation, and particular suitability for addressing problems that involve a high number of latent variables. \cite{pati2018statistical} relates the Bayes risk to the variational solution for a general distance metric. In addition, for non-singular models, which include finite mixture models with a known component number, \cite{han2019statistical} proved that the convergence rate of point estimators based on the variational posterior is $n^{-1/2}$, and they also provide the asymptotic normality of the optimal mean-field approximation centered at the maximum likelihood estimation.

\medskip
\noindent {\bf Asymptotic properties of likelihood ratio tests in mixture models.}
The use of maximum likelihood estimator (MLE) to fit mixture models has received a lot of attention since the 1960s. \cite{dempster1977maximum} introduced a general iterative approach, the expectation–maximization (EM) algorithm, for computing the MLE in latent variable models. The convergence properties of the MLE for the mixture problem were then theoretically established. Meanwhile, the testing of homogeneity, in other words, determining whether the mixture model has only one component or more, became an important research question. For non-singular models, when under the null hypothesis the parameter is constrained in a subspace of the whole parameter space, the limit distribution of the likelihood ratio testing statistic (LRTS) $\Lambda_n$ is a chi-square distribution. 
Unlike in non-singular model testing, \cite{hartigan1985failure} observed that under homogeneity, the statistic diverges to infinity if the whole parameter space is unrestricted. \cite{ghosh1984asymptotic} developed the asymptotic theory for the distribution of the LRTS under this setting. They showed that in the limit, $2\log\Lambda_n$ is distributed as the square of the supremum of a centered Gaussian process ${W_S:S\in\m{S}}$ admitting continuous sample paths. Also, \cite{bickel1993asymptotic} investigated the null behavior of the LRTS for this model. At the beginning of the 21st century, more mixture models were considered, and the divergence rate of the LRTS with an unconstrained parameter space was verified. For example, normal mixture models in \cite{liu2004asymptotics} and gamma mixture models in \cite{liu2003testing} were shown to diverge at the rate $\m{O}(\log\log n)$ with probability at least $1-\m O(1/\log n)$. 

On the other hand, mixture models with restricted parameter spaces were also explored. While the limit distribution is still related to a Gaussian process, this time the index set $\m{S}$ is compact, so the LRTS is bounded in probability. The general results for models with restricted parameter spaces under loss of identifiability are provided by \cite{liu2003asymptotics}, and the application to finite mixture models can also be found in Section 4 of the same paper. {\color{black} In deriving the index set $\m{S}$, they relied on strong identifiability, which highlights the challenges of dealing with weakly identifiable families, such as location-scale Gaussian mixtures. Later, \cite{azais2009likelihood} provided results under more general conditions, which at least encompass the homoscedastic location-scale Gaussian family. Specifically, under the assumption of a compact parameter space and other general conditions, the LRT converges to the supremum of a Gaussian process with a compact index set.} However, although their method may test the true model against the alternative, it requires prior knowledge to set the null hypothesis, and the exact form of the Gaussian process still requires a case-by-case analysis.

\section{Discussion}\label{section 5}
In this paper, we advocate the use of ELBO from mean-field variational Bayes for model selection in finite mixture models. We analyze the large-sample properties of ELBO by proving matching lower and upper bounds that improve upon existing results by utilizing the consistency property of variational Bayes. As a direct consequence, we have established an asymptotic expansion for the ELBO and proved the consistency of model selection through ELBO maximization. As a by-product of our proof, we show that the stable behavior of the posterior distribution, which benefits from model singularity, can be inherited by the mean-field variational approximation to the posterior. We also derived a parametric convergence rate of $(\log n)^{\rho_1'/2}/\sqrt{n}$ for component parameters. 

{\color{black} \subsection{On the choice of $K_{\max}$}\label{app:disc_kmax}
Our theory assumes that candidate models are indexed by $K\in\{1,\ldots,K_{\max}\}$ for a known finite upper bound $K_{\max}$. This assumption is mainly used to simplify the analysis by allowing uniform control of the remainder terms in the ELBO expansion across candidate models. If $K_{\max}$ were unknown (or allowed to grow arbitrarily), then for each fixed $K$ the ELBO expansion still applies, but the effective sample size required for reliable selection would increase with the complexity of the candidate model class.

A key mechanism can be seen from the constant terms in the ELBO expansion. For a symmetric Dirichlet prior with parameter $\phi_0$, the constant $C_{\phi_0}(K)
=\log\frac{\Gamma(K\phi_0)}{\Gamma(\phi_0)^K}$ is convex in $K$ for any fixed $\phi_0$ and increases rapidly with $K$. As discussed below Theorem~\ref{main theorem}, a Stirling-type approximation suggests that, for fixed $\phi_0$, its contribution grows on the order of $\phi_0\,K\log K$ as $K\to\infty$. In parallel, Laplace-type approximation errors typically increase with the number of mixture components. Consequently, when $K_{\max}$ is very large, the leading penalty term $\lambda\log n$ must dominate increasingly large constant and remainder contributions in order for the ELBO to favor the true model over highly overfitting alternatives. One necessary scaling that needed for uniform dominance of the leading $\lambda\log n$ term is
\[
K\log K = o(\log n),
\]
which formalizes the intuition that allowing much larger candidate models slows down finite-sample separation between the true model and complex overfitting models.

From a practical perspective, a conservative upper bound is often available from domain knowledge, even when $K^*$ is unknown. When such a bound is genuinely unavailable, one may adopt a simple search heuristic by increasing $K$ sequentially and monitoring the ELBO: if the ELBO does not exceed its current maximum for a fixed number of consecutive increments (e.g., 5--10), the search can be terminated, and the selected model is taken as the $K$ attaining the largest ELBO among the explored candidates. While such stopping rules are heuristic, they can be useful in practice when the candidate model class cannot be specified a priori.
}
\subsection{Future directions}\label{app:disc_future}
Several directions merit further investigation. First, it would be valuable to extend the current development to other singular models such as hidden Markov models, neural networks, and factor models, where variational methods are widely used but large-sample model selection theory remains incomplete. Second, a finite-sample analysis of ELBO-based selection would require a more refined characterization of likelihood ratio type statistics under singularity. In particular, obtaining explicit non-asymptotic control beyond leading-order terms would clarify how quickly the asymptotic regime emerges and how the constants and remainder terms affect practical sample sizes; see our brief comments following Corollary~\ref{empty out rate}.

A further promising direction is to establish model selection consistency for finite mixture models based on the evidence derived from the full posterior distribution. Our current proof technique is tailored to the mean-field variational approximation, which yields closed-form first-order optimality conditions and thereby simplifies the analysis. While we expect that the true posterior evidence should also yield model selection consistency in the same spirit, proving this rigorously would likely require fundamentally different tools, potentially involving algebraic geometry and singular learning theory, as in \citet{watanabe2001algebraic}; see also \citet{drton2017bayesian} for discussions on large-sample expansions of the evidence in general singular models.

\section{Applications of theory to concrete examples}\label{examples}
In this subsection, we verify Assumption A and the weak identifiability condition for several representative finite mixture models, and determine their effective dimensions, $\lambda$, arising in the ELBO from the MF approximation.

\begin{example}[Location Gaussian mixture model]\label{example 1}
    In this example, we consider the location Gaussian mixture model where the density function of each component is from the univariate location family of Gaussian distributions $\{\m{N}(\mu,\sigma^2);\mu\in \Omega\}$, where $\sigma^2$ is assumed to be known and the canonical parameter in this exponential family is simply $\eta \coloneqq  \mu$. Moreover, the Fisher information matrix is $1/\sigma^2$ times the identity matrix. For simplicity, if we assume $\sigma^2 = 1$ without loss of generality, then
    $$g(x;\mu) = \frac{1}{\sqrt{2\pi}}\exp\left\{-\frac{x^2}{2} + \mu x -\frac{\mu^2}{2}\right\}.$$
    According to Theorem 3 in \cite{chen1995optimal}, $g(x;\mu)$ is identifiable in the second order, which also implies the weak identifiability; in fact, it is also identifiable in any finite order according to \cite{heinrich2018strong}. Therefore, Assumption A4 is satisfied with $r = 2$ by Theorem 3.2 (a) from \cite{ho2016strong}, and our Theorem~\ref{main theorem} applies with $\lambda$ being 
    \begin{align*}
        \lambda = \begin{cases}
        (K-K^*)\phi_0 + \frac{2K^*-1}{2}, & \phi_0\le\frac{1}{2},\\
        \frac{2K-1}{2}, & \phi_0>\frac{1}{2}.
        \end{cases}
    \end{align*}
\end{example}

\begin{example}[Scale exponential mixture model]
    In this example, we take $g(x;\eta) = \eta \,e^{-\eta x} = \exp\{-\eta x + \log\eta \}$ as the (scale) exponential distribution family with rate parameter $\eta^{-1}$), whose fisher information is $1/\eta^2$. According to Theorem 2.4 in \cite{heinrich2018strong}, $g(x;\mu)$ is also identifiable in any finite order. Therefore, Assumption A4 is satisfied with $r = 2$ and the weak identifiability also holds. As a result, Theorem~\ref{main theorem} also applies to the scale exponential mixture model with the same $\lambda$ as Example \ref{example 1}.
\end{example}

\begin{example}[Location-scale Gaussian mixture model]
    We consider the classical form of the univariate location-scale Gaussian distribution family where
    \begin{align*}
        g(x;\mu,\sigma^2) = \exp\left\{-\frac{x^2}{2\sigma^2} + \frac{x\mu}{\sigma^2} - \frac{\mu^2}{2\sigma^2}-\frac{1}{2}\log (2\pi\sigma^2)\right\}.
    \end{align*}
    This family is no longer strongly identifiable due to the algebraic relationship between the two partial derivatives (up to the second order) 
    \begin{align*}
        \frac{\partial^2 g}{\partial\mu^2}(x;\mu,\sigma^2) = 2\frac{\partial g}{\partial\sigma^2}(x;\mu,\sigma^2).
    \end{align*}
    For this parametric family, Proposition 2.2 from \cite{ho2016convergence} shows that if $G(\theta)$ has at most $K$ components, then for any $G(\theta)$ such that $W_{\overline r}(G(\theta),G^*)$ is sufficiently small, we have
    \begin{align*}
        d_{\rm TV}(p(\cdot\,|\,\theta),p^*(\cdot))\gtrsim W_{\overline r}^{\overline r}(G(\theta),G^*),
    \end{align*}
    for some constant $\overline{r}\ge 2$ depending only on $K-K^*$. This verifies Assumption A4 with $r=\overline r$. Hence, we can apply Theorem~\ref{main theorem} to conclude that the effective dimension $\lambda$ in the ELBO is ($m=2$)
    \begin{align*}
        \lambda = \begin{cases}
        (K-K^*)\phi_0 + \frac{3K^*-1}{2}, & \phi_0\le\frac{3}{2},\\
        \frac{3K-1}{2}, & \phi_0>\frac{3}{2}.
        \end{cases}
    \end{align*}
  Moreover, although the location-scale Gaussian distribution family is not strongly identifiable, it is weakly identifiable, as demonstrated in \cite{ho2016strong}. Consequently, all results from Section~\ref{sec:para_est} regarding parameter estimation via mean-field approximation remain applicable.
\end{example}

\begin{example}[Multinomial mixture model] In the final example, we consider a mixture of discrete distributions, specifically, the multinomial mixture model that is widely used in topic modeling. In particular, we take $g(x;\eta) = \binom{M}{x_1,\,\ldots,\,x_d}\prod_{j = 1}^d \eta_j^{x_j}$, $x_j\in[M]$, $\sum_{j=1}^d x_i=M$. This is the probability mass function of the multinomial distribution Mult$\big(M,(\eta_1,\ldots,\eta_d)\big)$ with a known $M$ and parameters $(\eta_1,\ldots,\eta_d)$, where $\sum_{j=1}^d\eta_j =1$ and has an effective dimension of parameters $m = d-1$. A sufficient condition for this multinomial mixture model with $K$ components to be strongly identifiable is $3K -1\le M$ according to Corollary 1 in \cite{manole2021estimating}. Therefore, Assumption A4 holds with $r=2$ and our Theorem~\ref{main theorem} applies with 
    \begin{align*}
        \lambda = \begin{cases}
        (K-K^*)\phi_0 + \frac{dK^*-1}{2}, & \phi_0\le\frac{d}{2},\\
        \frac{dK-1}{2}, & \phi_0>\frac{d}{2},
        \end{cases}
    \end{align*}
   where one degree of freedom is lost due to the linear constraint $\sum_{i  = 1}^d\eta_i = 1$. The weak identifiability is again implied by the strong identifiability.
\end{example}

\section{More Numerical Results}

{\color{black} \subsection{Verification of the expression of $\lambda$ for location--scale Gaussian mixtures}\label{app:scale}}

In this subsection, we numerically verify the predicted dependence of $\lambda$ on $K$ and $\phi_0$ for a weakly identifiable location--scale Gaussian mixture model. 
We consider a two-dimensional ($d=2$) Gaussian mixture with true number of components $K_0=2$ and equal mixing weights $w^*=(0.5,0.5)$. 
The two component means are set to $\mu_1^*=(-2,-2)$ and $\mu_2^*=(2,2)$, and both covariance matrices are $\Sigma_1^*=\Sigma_2^*=I_2$. 
The fitted model allows up to $K_{\max}=5$ components. 
We sweep over $\phi_0 \in \{0.1,1,2,3,4,5\}$ and sample sizes $n \in \{2000,5000,10000,20000,40000,80000\}$.
For the bivariate location--scale setting, the parameter dimension of each component is $m=5$ as stated in Section~\ref{sec: section 4.1}. 
The corresponding threshold is $(m+1)/2=3$.

\begin{figure}[t]
  \centering
  \begin{minipage}{0.44\textwidth}
    \centering
    \includegraphics[width=\linewidth]{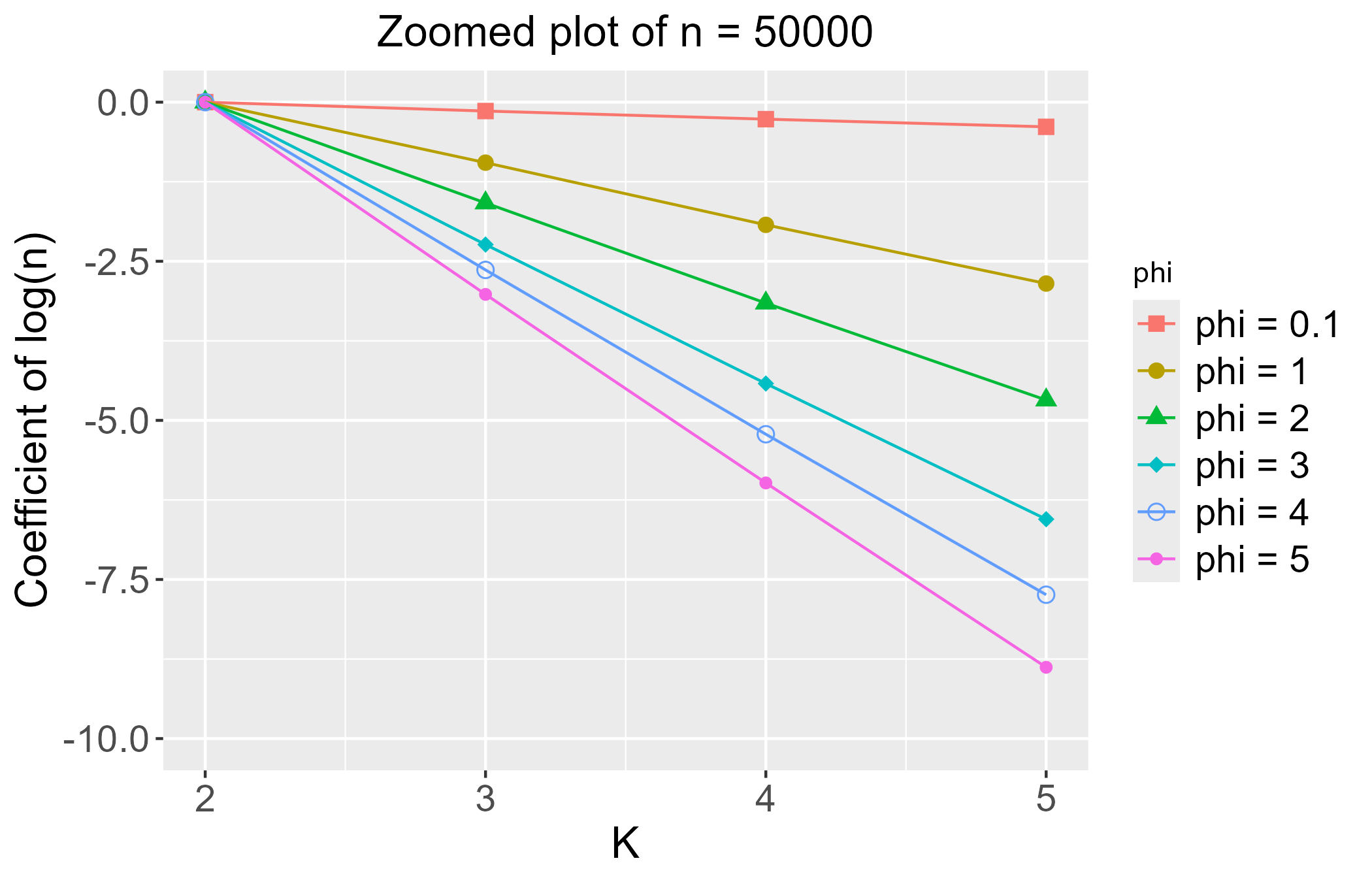}
    \caption{Plots of $(\widehat{\m L}_K-\widehat{\m L}_{K^\star})/\log n$ versus $K$ with fixed $n=50000$ in the location--scale setting.}\label{fig:fig1 scale}
  \end{minipage}
  \hfill
  \begin{minipage}{0.54\textwidth}
    \centering
    \includegraphics[width=\linewidth]{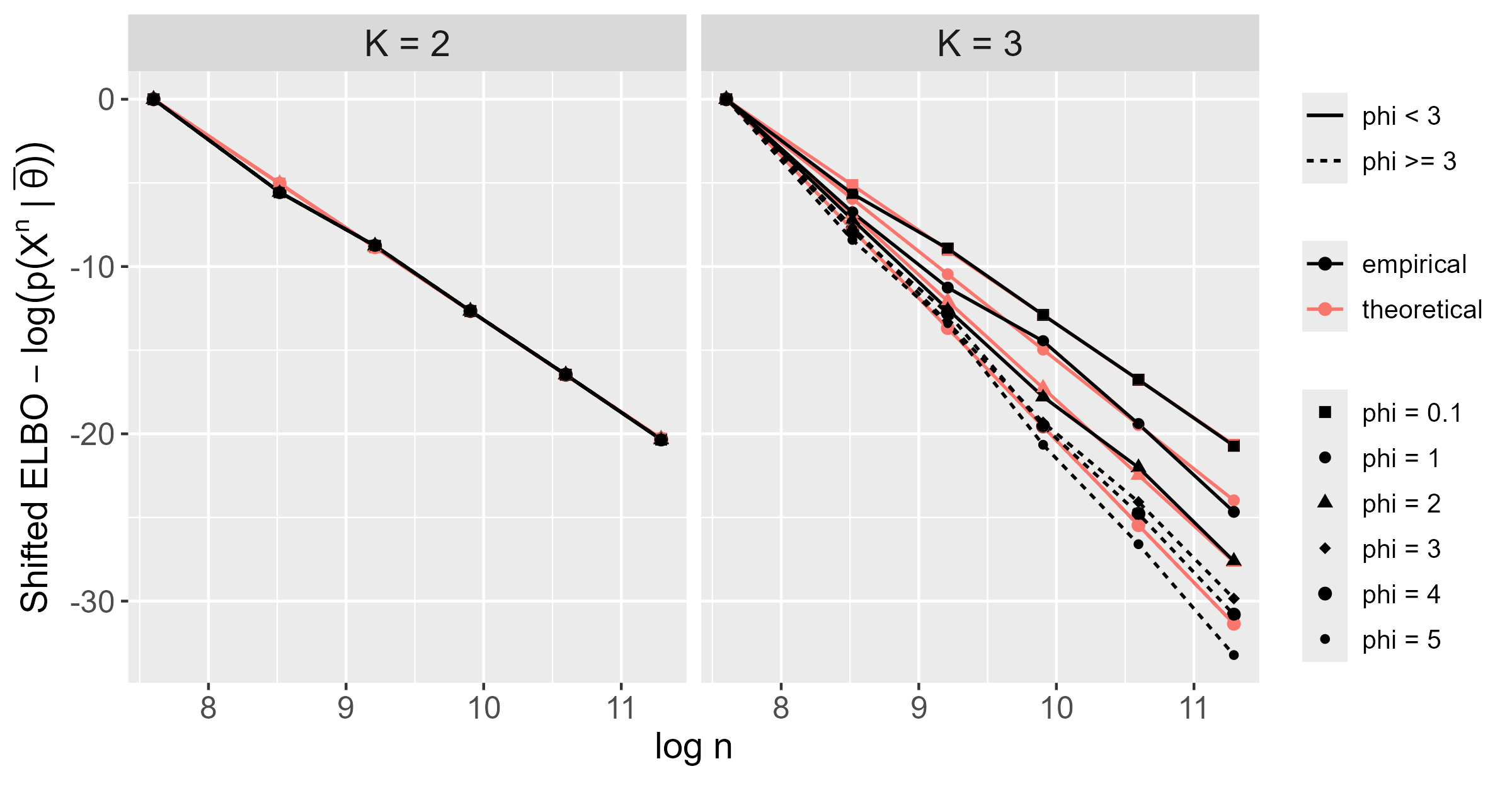}
    \caption{Plots of $\widehat{\m L}_K - \log p(X^n \mid \overline{\theta})$ versus $n$ for different $\phi_0$ at fixed $K$ in the location--scale setting, with the starting values aligned to zero.}\label{fig:fig3 scale}
  \end{minipage}
  
\end{figure}

Figure~\ref{fig:fig1 scale} shows the plots of $(\widehat{\m L}_K-\widehat{\m L}_{K^*})/\log n$ versus $K$ with $n=50000$ fixed. 
Figure~\ref{fig:fig3 scale} shows the plots of $\widehat{\m L}_K-\log p(X^n\mid \overline{\theta})$ versus $n$ for different values of $\phi_0$ at fixed $K$ in the location--scale setting; for visualization, each curve is shifted so that it starts at zero.

Empirically, at Figure~\ref{fig:fig1 scale} and at $K=3$ in Figure~\ref{fig:fig3 scale}, when $\phi_0 \ge 3$ (i.e., above the threshold), the slopes deviate more noticeably from the theoretical line, compared with the location-only case. 
A plausible explanation is that the location--scale model is more complex, and for large $\phi_0$ the variational weights spread over all candidate components; this can increase instability of the CAVI updates and make the algorithm more likely to get trapped in local minima. Alternatively, to better match the theoretical behavior in this more challenging setting, a larger sample size may be required.
This observation again provides practical motivation for choosing $\phi_0$ below the threshold.

{\color{black}\subsection{Raw data for Figure~\ref{fig: fig5}}\label{app:raw fig5}}

\noindent
In this subsection, we report the numerical summaries underlying Figure~\ref{fig: fig5}, including the sample mean and standard deviation of $\overline w_k$ and $W_1$, as shown in Tables~\ref{tab:mixing_weights} and~\ref{tab:w1_distance}. When $\phi_0\in\{1/2,1\}$, increasing the sample size from $n=10^3$ to $n=10^4$ reduces the mean weights assigned to the three redundant components by roughly one order of magnitude (about a factor of $10$), which is consistent with the emptying-out rate predicted by Corollary~\ref{empty out rate}. Meanwhile, the mean $W_1$ decreases to about $30\%$ of its value at $n=10^3$, in agreement with the $n^{-1/2}$ parametric convergence rate stated in Theorem~\ref{para root n}.

\begin{table}[t]
\centering
\small
\setlength{\tabcolsep}{6pt}
\renewcommand{\arraystretch}{1.15}
\caption{Estimated mixing weights $\{\overline w_k\}_{k=1}^5$ under $K=5$, $K^\ast=2$. Entries are mean(std) over 100 repetitions.}
\label{tab:mixing_weights}
{\footnotesize \begin{tabular}{lccccc}
\toprule
\multicolumn{6}{c}{$n=1000$} \\
\midrule
$\phi_0$ & $\overline w_1$ & $\overline w_2$ & $\overline w_3$ & $\overline w_4$ & $\overline w_5$ \\
\midrule
$\frac12$ &
$0.512(0.0110)$ &
$0.486(0.0110)$ &
$6.98e\!-\!4(2.00e\!-\!3)$ &
$5.11e\!-\!4(4.46e\!-\!7)$ &
$5.11e\!-\!4(4.46e\!-\!7)$ \\
$1$ &
$0.511(0.0113)$ &
$0.485(0.0110)$ &
$1.72e\!-\!3(4.60e\!-\!3)$ &
$1.05e\!-\!3(2.12e\!-\!6)$ &
$1.05e\!-\!3(2.12e\!-\!6)$ \\
$3$ &
$0.473(0.0566)$ &
$0.396(0.0916)$ &
$0.0901(0.0868)$ &
$0.0332(0.0527)$ &
$7.67e\!-\!3(0.0122)$ \\
$4$ &
$0.408(0.0713)$ &
$0.289(0.0688)$ &
$0.161(0.0510)$ &
$0.0983(0.0537)$ &
$0.0431(0.0342)$ \\
$5$ &
$0.388(0.0696)$ &
$0.264(0.0494)$ &
$0.177(0.0389)$ &
$0.115(0.0450)$ &
$0.0551(0.0314)$ \\
\midrule
\multicolumn{6}{c}{$n=10000$} \\
\midrule
$\phi_0$ & $\overline w_1$ & $\overline w_2$ & $\overline w_3$ & $\overline w_4$ & $\overline w_5$ \\
\midrule
$\frac12$ &
$0.504(3.25e\!-\!3)$ &
$0.496(3.25e\!-\!3)$ &
$5.12e\!-\!5(1.36e\!-\!8)$ &
$5.12e\!-\!5(1.36e\!-\!8)$ &
$5.12e\!-\!5(1.36e\!-\!8)$ \\
$1$ &
$0.503(3.25e\!-\!3)$ &
$0.496(3.25e\!-\!3)$ &
$1.05e\!-\!4(6.55e\!-\!8)$ &
$1.05e\!-\!4(6.55e\!-\!8)$ &
$1.05e\!-\!4(6.55e\!-\!8)$ \\
$3$ &
$0.461(0.0712)$ &
$0.384(0.0959)$ &
$0.109(0.0928)$ &
$0.0391(0.0658)$ &
$6.15e\!-\!3(0.0205)$ \\
$4$ &
$0.395(0.0795)$ &
$0.273(0.0553)$ &
$0.167(0.0478)$ &
$0.108(0.0490)$ &
$0.0566(0.0385)$ \\
$5$ &
$0.392(0.0801)$ &
$0.261(0.0573)$ &
$0.171(0.0455)$ &
$0.115(0.0460)$ &
$0.0605(0.0401)$ \\
\bottomrule
\end{tabular}}
\end{table}

\begin{table}[t]
\centering
\small
\setlength{\tabcolsep}{6pt}
\renewcommand{\arraystretch}{1.15}
\caption{Estimated Wasserstein distance $W_1(G(\overline\theta),G^\ast)$ under $K=5$, $K^\ast=2$. Entries are mean(std) over 100 repetitions.}
\label{tab:w1_distance}
{\footnotesize \begin{tabular}{lccccc}
\toprule
$n$ & $\phi_0=\frac12$ & $\phi_0=1$ & $\phi_0=3$ & $\phi_0=4$ & $\phi_0=5$ \\
\midrule
$1000$  & $0.148(0.0463)$ & $0.150(0.0468)$ & $0.272(0.124)$ & $0.395(0.0816)$ & $0.422(0.0680)$ \\
$10000$ & $0.0452(0.0145)$ & $0.0452(0.0145)$ & $0.133(0.0771)$ & $0.207(0.0455)$ & $0.207(0.0525)$ \\
\bottomrule
\end{tabular}}
\end{table}

\subsection{Table~\ref{comparasion}, settings for Figure~\ref{fig: comparasion}}\label{app:table}

\begin{table}[t]
\begin{spacing}{1}
\begin{tabular}{cccccc}
\hline
Model & $w^*_1,\eta^*_1$  & 
$w^*_2,\eta^*_2$   & 
$w^*_3,\eta^*_3$   &
$w^*_4,\eta^*_4$   &
$w^*_5,\eta^*_5$ \\
\hline
1     & 0.3, $\left ( \begin{matrix}
0 \\
0 \\
\end{matrix} \right)$  
& 0.7, $\left ( \begin{matrix}
2 \\
2 \\
\end{matrix} \right)$ & & &  \\
\vspace{5pt}
2     & 0.5, $\left ( \begin{matrix}
\sqrt{2} \\
0 \\
\end{matrix} \right)$ &
0.5, $\left ( \begin{matrix}
0 \\
\sqrt{2} \\
\end{matrix} \right)$  &   &  &  \\
\vspace{3pt}
3    & 1/3, $\left ( \begin{matrix}
-1 \\
-1 \\
\end{matrix} \right)$  &
1/3, $\left ( \begin{matrix}
1 \\
-1 \\
\end{matrix} \right)$  &
1/3, $\left ( \begin{matrix}
0 \\
1 \\
\end{matrix} \right)$  &  &  \\
\vspace{3pt}
4     & 0.25, $\left ( \begin{matrix}
0 \\
2 \\
\end{matrix} \right)$   &
0.25, $\left ( \begin{matrix}
0 \\
-2\\
\end{matrix} \right)$ &
0.25, $\left ( \begin{matrix}
2 \\
0 \\
\end{matrix} \right)$   &
0.25, $\left ( \begin{matrix}
-2 \\
0 \\
\end{matrix} \right)$  &  \\
\vspace{3pt}
5     & 0.2, $\left ( \begin{matrix}
2.5 \\
0 \\
0 \\
0 \\
0 \\
0 \\
0 \\
0 \\
\end{matrix} \right )$ &
0.2, $\left ( \begin{matrix}
0 \\
2.5 \\
0 \\
0 \\
0 \\
0 \\
0 \\
0\\
\end{matrix} \right )$ &
0.2, $\left ( \begin{matrix}
0 \\
0 \\
2.5 \\
0 \\
0 \\
0 \\
0 \\
0 \\
\end{matrix} \right )$ &
0.4, $\left ( \begin{matrix}
0 \\
0 \\
0 \\
2.5 \\
0 \\
0 \\
0 \\
0 \\
\end{matrix} \right )$  & \\      
\vspace{3pt}
6     & 0.2, $\left ( \begin{matrix}
2.5 \\
0 \\
0 \\
0 \\
0 \\
0 \\
0 \\
0 \\
0 \\
0 \\
\end{matrix} \right )$ &
0.2, $\left ( \begin{matrix}
0 \\
2.5 \\
0 \\
0 \\
0 \\
0 \\
0 \\
0 \\
0 \\
0\\
\end{matrix} \right )$   &
0.2, $\left ( \begin{matrix}
0 \\
0 \\
2.5 \\
0 \\
0 \\
0 \\
0 \\
0 \\
0 \\
0 \\
\end{matrix} \right )$   &
0.2, $\left ( \begin{matrix}
0 \\
0 \\
0 \\
2.5 \\
0 \\
0 \\
0 \\
0 \\
0 \\
0 \\
\end{matrix} \right )$ &
0.2, $\left ( \begin{matrix}
0 \\
0 \\
0 \\
0 \\
2.5 \\
0 \\
0 \\
0 \\
0 \\
0 \\
\end{matrix} \right )$  \\      
\hline
\end{tabular}
\end{spacing}
\caption{Parameter settings for the multivariate Gaussian mixture models}\label{comparasion}
\end{table}

\subsection{Clustering of cell types in real data analysis}\label{clustering}

\begin{figure}[]
  \centering
  \begin{minipage}{0.49\textwidth}
    \centering
    \includegraphics[width=\linewidth]{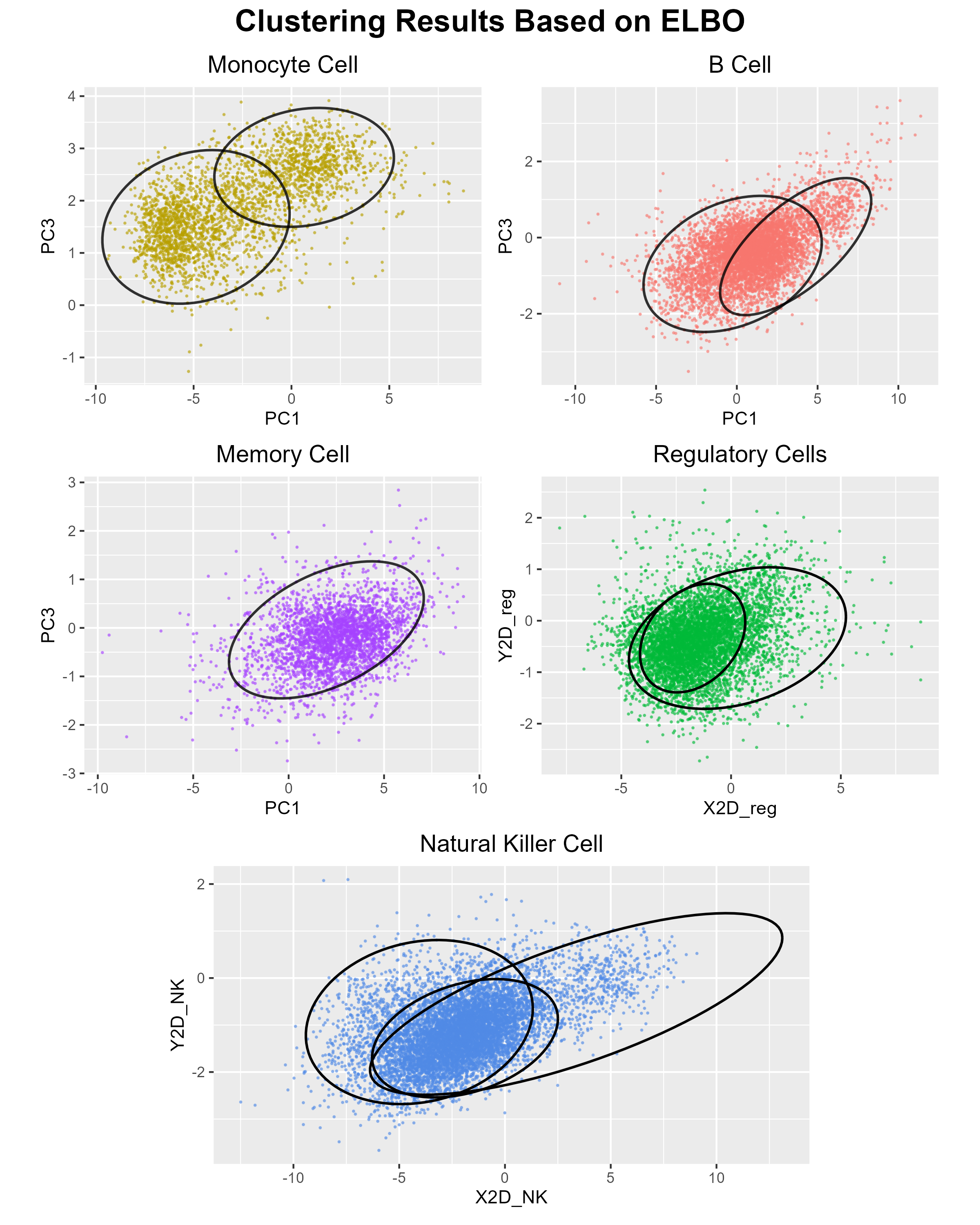}
    \end{minipage}
    \hfill
    \begin{minipage}{0.49\textwidth}
    \centering
    \includegraphics[width=\linewidth]{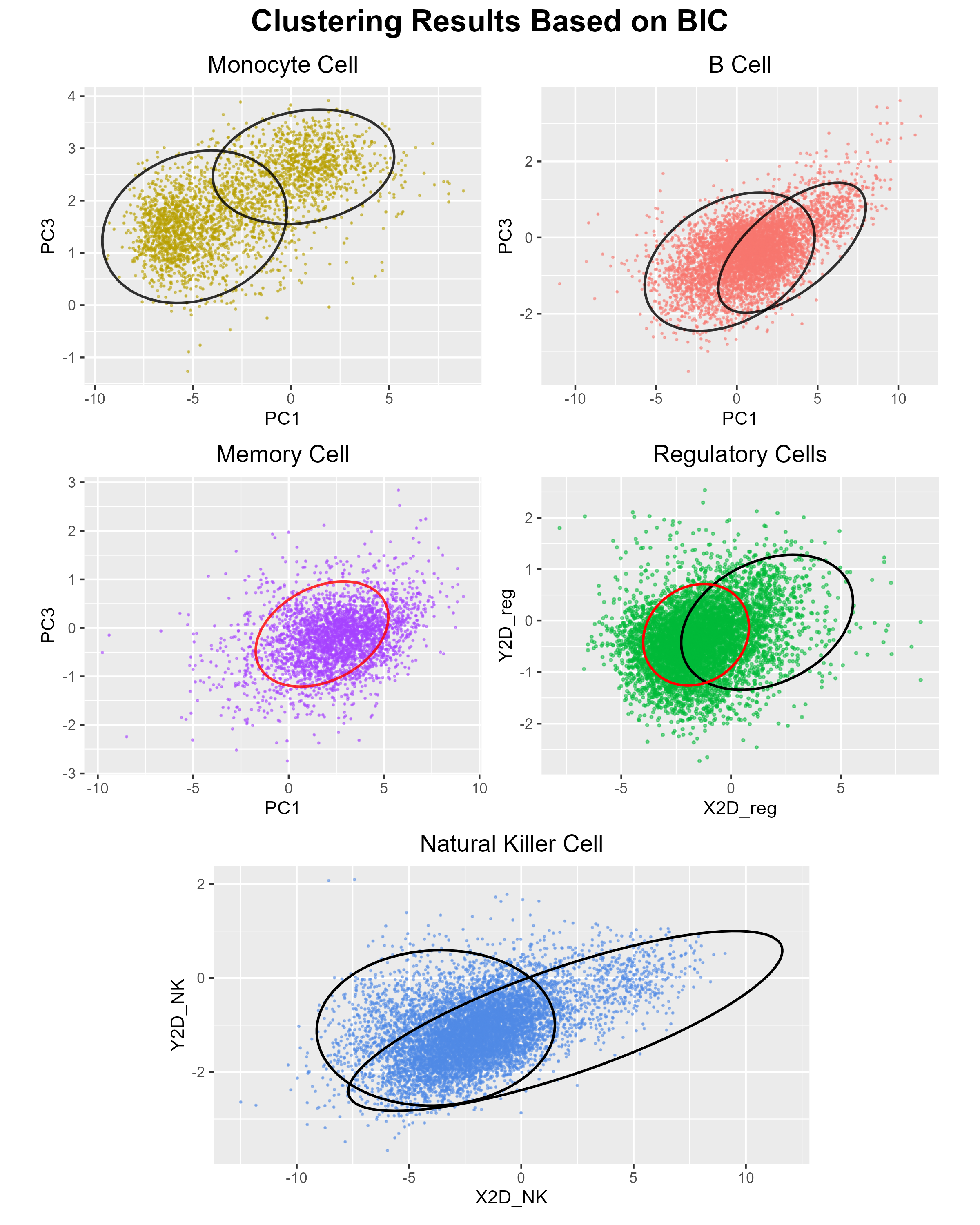}
    \end{minipage}
    
  \centering
  \begin{minipage}{0.49\textwidth}
    \centering
    \includegraphics[width=\linewidth]{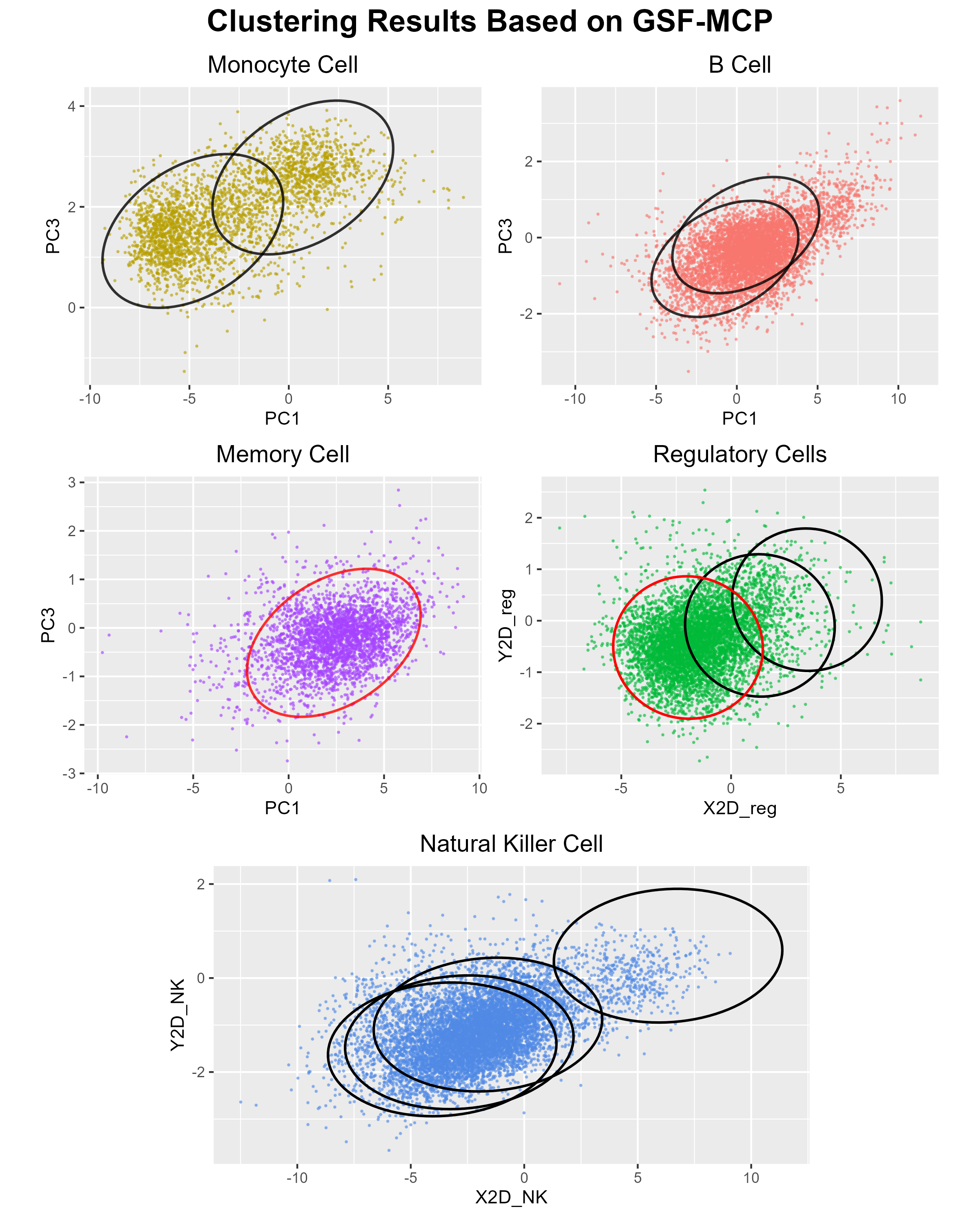}
    \end{minipage}
    \hfill
    \begin{minipage}{0.49\textwidth}
    \centering
    \includegraphics[width=\linewidth]{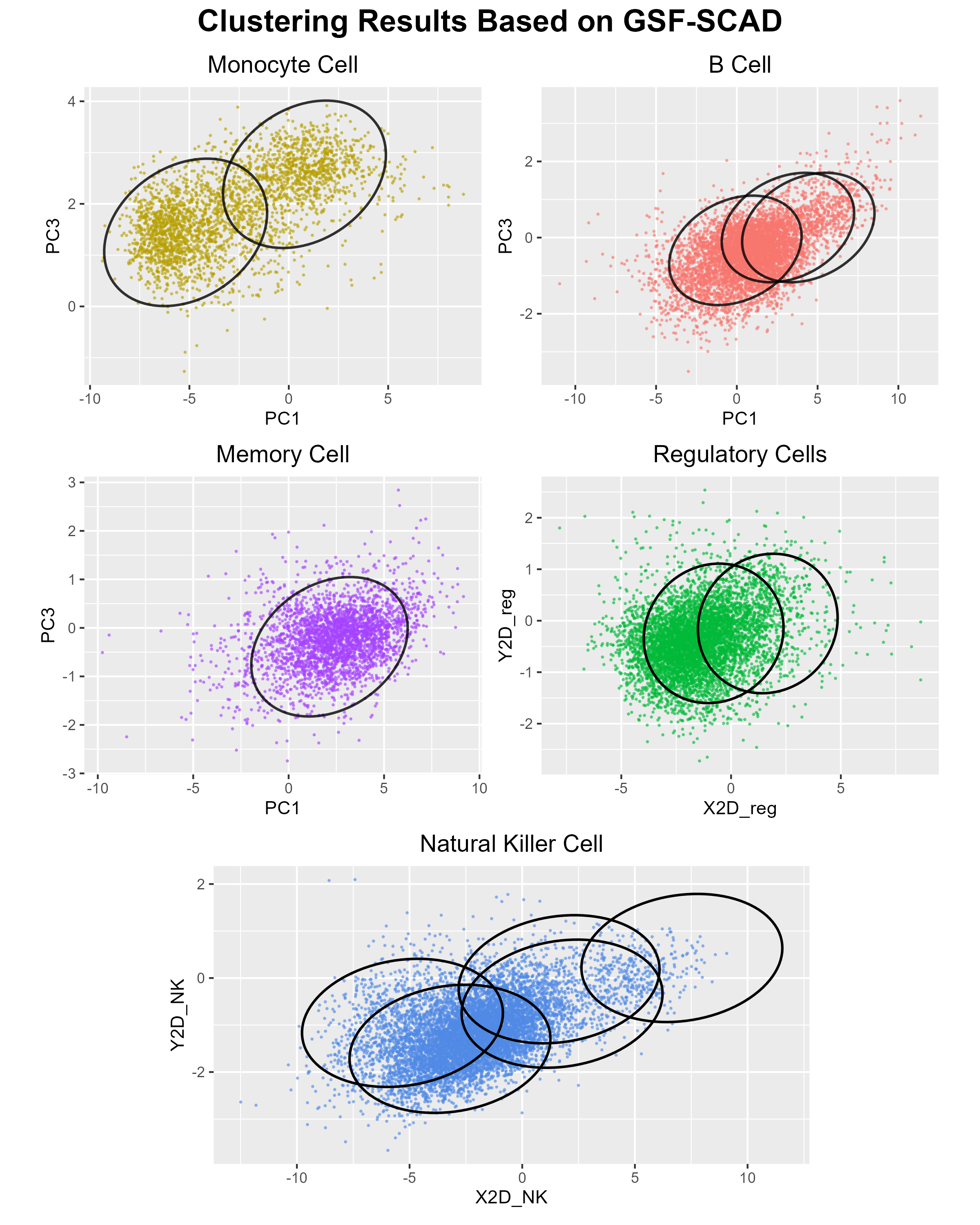}
    \end{minipage}
  \caption{Scatter plots of different cell types with 95\% confidence ellipses based on ELBO, BIC and GSF methods respectively.}
  \label{fig: fig8}
\end{figure}

\begin{figure}[]
    \centering
    \includegraphics[width=0.75\textwidth]{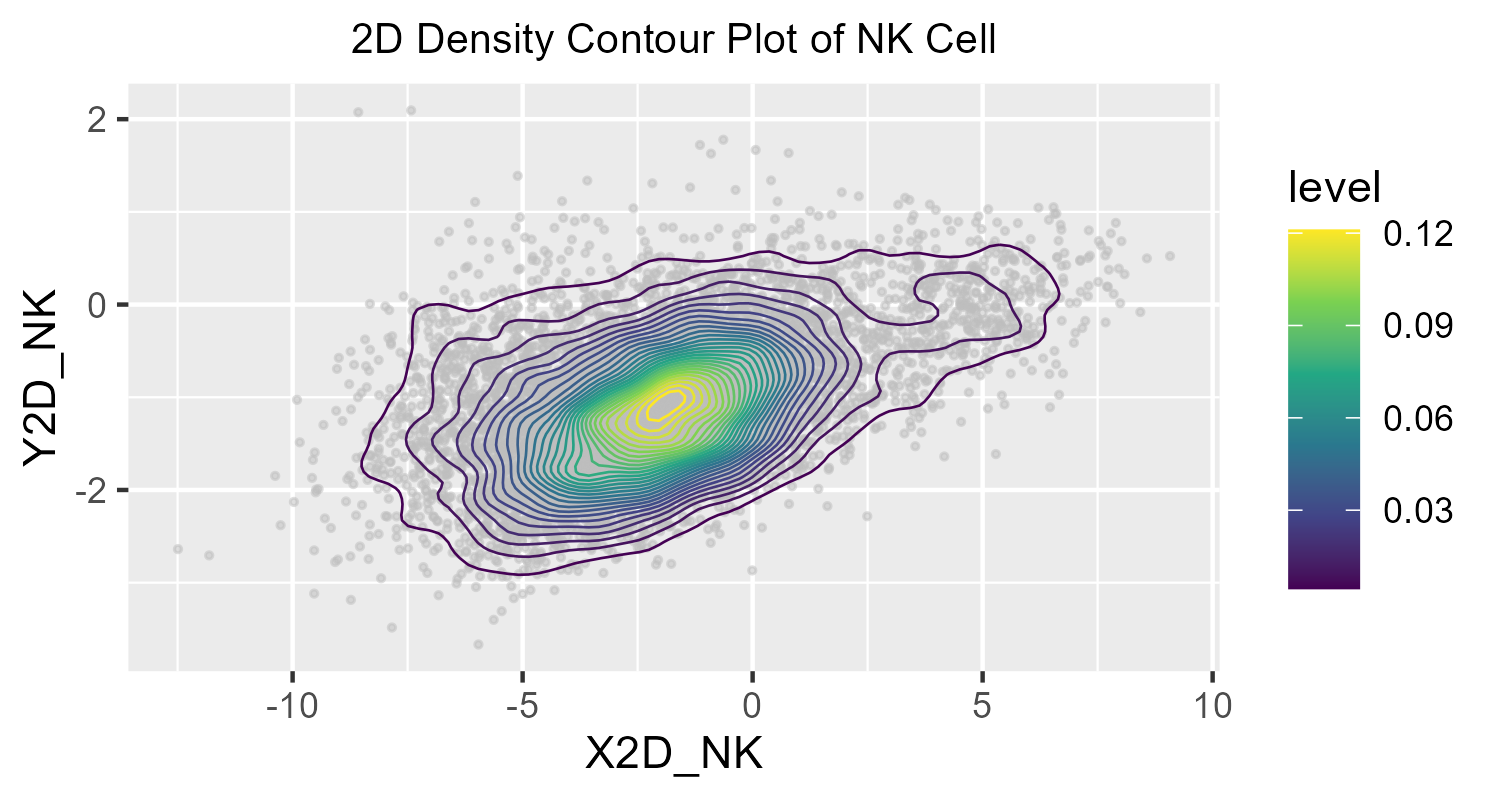}
    \caption{Density contour plot of NK cells on the given plane}
    \label{fig: fig9}
\end{figure}

{\color{black} Figure~\ref{fig: fig8} presents the clustering results obtained by ELBO, BIC, and GSF, each visualized with 95\% confidence ellipses. Monocytes, B cells, and Memory T cells are plotted in the PC1-PC3 plane; NK cells are visualized in the plane spanned by the centers of the three clusters identified by ELBO; and Regulatory T cells are plotted in the plane defined by the two ELBO-identified centers and the point $(0, 0, 5)$. These projection choices are made to enhance visualization of the clustering structures and highlight the spatial relationships among the different cell types.

It can be observed that, for each cell type, the ELBO-based clustering closely aligns with the corresponding scatter plots. The clustering results of BIC and ELBO are largely consistent for Monocytes and B cells. However, for NK cells, BIC identifies only two clusters, whereas ELBO identifies three, providing a more refined representation of the underlying structure. 
The density contour plot of NK cells (Figure~\ref{fig: fig9}) reveals a bulging contour in the lower-left vicinity of the highest-density region, indicating a local mode. Additionally, a more distant local mode is present in the upper-right region. These two local modes correspond to the two clusters identified by BIC. However, the central peak exhibits a density that clearly exceeds the combined density of these two local modes, suggesting it represents a distinct and intrinsic cluster rather than an artifact of overlapping components. This supports the conclusion that ELBO provides a more accurate and nuanced clustering solution.
Furthermore, BIC fails to distinguish between Memory T cells and Regulatory T cells. As shown in Figure~\ref{fig: fig8}, the two red ellipses represent confidence regions for the same cluster under BIC, indicating a misclassification. In contrast, ELBO successfully separates Memory and Regulatory T cells, yielding a more faithful representation of the underlying biological structure.

For the GSF methods, the results presented correspond to one of the 30 repetitions in which MCP selected 11 components and SCAD selected 13—their most frequently selected values. Overall, MCP yielded solutions ranging from 8 to 13 components, while SCAD varied between 7 and 14.
Fitting homoscedastic Gaussian mixtures to data generated from heteroscedastic mixtures poses significant challenges. These models can recover clusters with similar variances but distinct centers---for example, both methods successfully identified two Monocyte subpopulations, and SCAD separated Memory and Regulatory T cells. However, when a cluster exhibits substantially larger variance, additional components are needed to approximate its distribution, as observed for NK cells. Conversely, clusters with smaller variances tend to be merged with adjacent clusters to compensate for variance mismatches, as in MCP's merging of Memory and Regulatory T cells.
Moreover, the clustering results are highly sensitive to the initial placement of cluster centers.}

\paragraph{{\color{black} Different data cleaning strategies.}}
\begin{figure}[t]
  \centering
  \begin{minipage}{0.45\textwidth}
    \centering
    \includegraphics[width=\linewidth]{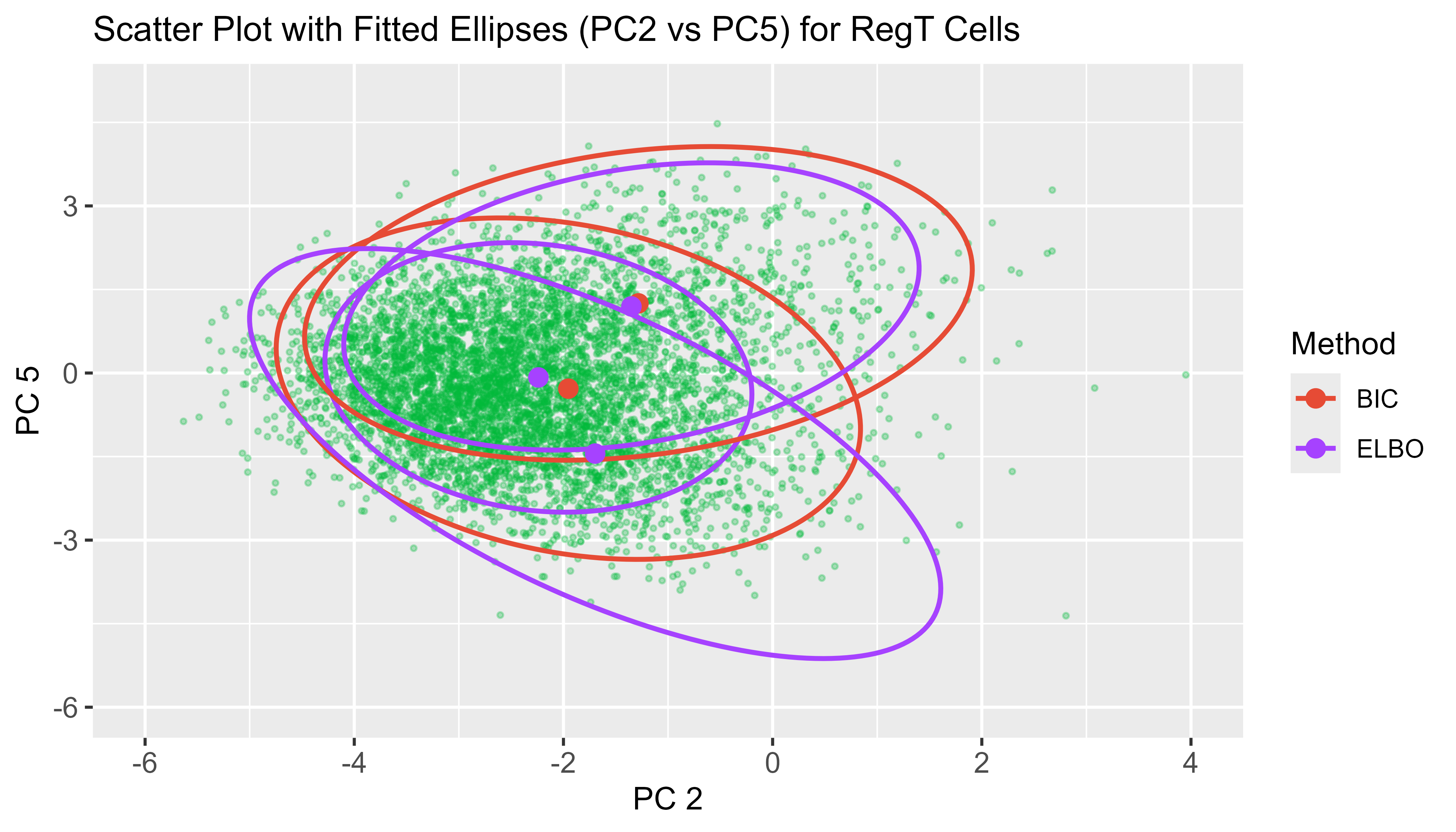}
  \end{minipage}
  \hfill
  \begin{minipage}{0.45\textwidth}
    \centering
    \includegraphics[width=\linewidth]{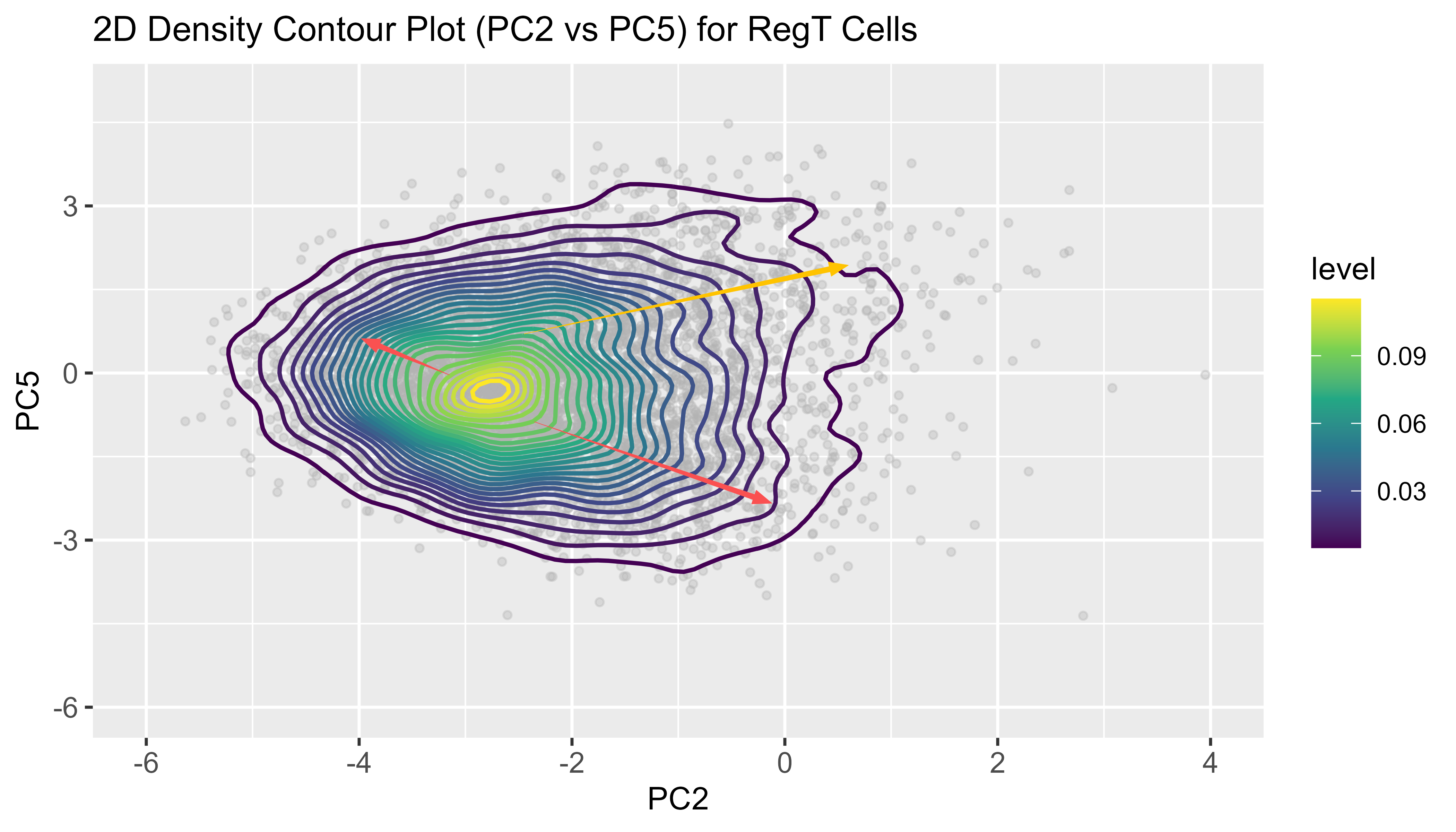}
  \end{minipage}
  \caption{Regulatory T cell under the alternative preprocessing (500 genes, 5 PCs; PC2--PC5 plane). 
  \textbf{Left:} scatter plot of cells with fitted Gaussian ellipses. 
  \textbf{Right:} 2D kernel density contour.}
  \label{fig:regT_compare_cleaning}
\end{figure}
The results of clustering and ordering methods on single-cell data can depend on how the data are preprocessed. Different choices in cell filtering, gene selection, and dimension reduction may lead to different covariance structures and therefore different model selection outcomes. When applying the method to real data, users should be aware of this sensitivity.

In the main text, we kept five selected cell types, retained the 50 most highly expressed genes, and then kept the first three principal components for downstream analysis.
Here we considered an alternative preprocessing strategy while still focusing on the same five cell types as in the main analysis. First, we removed cells with extremely small or extremely large numbers of expressed genes. Cells with very few expressed genes are often of low quality, while cells with unusually many expressed genes may correspond to doublets. We also filtered out cells with high mitochondrial gene expression, since a high proportion of mitochondrial genes is commonly used as an indicator of stressed or damaged cells in single-cell experiments. After this filtering step, 21,178 cells were retained.
We then kept the 500 most highly expressed genes, similar with the preprocessing choices in literature \citep{chen2023selective, gao2024selective, do2024dendrogram}, and performed PCA while retaining the first five principal components.

We compared the BIC and ELBO (with $\phi_0 = 1$) under this alternative preprocessing. We set $K_{\max} = 20$. For each $K$, both methods were run with 20 different initializations, and the maximum value across the 20 runs was taken as the BIC or ELBO for that $K$.
Since more information was retained (from 50 to 500 genes and from 3 to 5 principal components), the algorithms may fit a larger number of clusters. In this experiment, BIC selected 14 clusters, while ELBO selected 15 clusters. The main difference arises in the regulatory T cell region. BIC identified two clusters in this region, whereas ELBO identified three.

For visualization, we fixed the two-dimensional plane spanned by PC2 and PC5 as shown in Figure~\ref{fig:regT_compare_cleaning}. We present both scatter plots with fitted ellipses and density contour plots. In the scatter plot, the purple ellipses correspond to the ELBO fit. Two of them are close to the two clusters identified by BIC, while an additional elongated ellipse appears in the lower-left region. This additional ellipse covers the scatter points in that area more completely.

From the density contour plot, the outward extension of the contours in the upper-right direction (indicated by the yellow arrow) is well explained by the ellipses fitted by both BIC and ELBO in that region. However, in the direction indicated by the red arrow, the contours also extend outward, suggesting a covariance structure that differs from the central cluster. This pattern is consistent with the presence of a third subcluster in the regulatory T cell region. Under this preprocessing, the ELBO fit captures this structure more explicitly.

{\color{black}
\subsection{Runtime comparison across algorithms}\label{runtime}

We use the same single-cell dataset to compare the computational time of GSF, BIC, ELBO, and nested sampling (for the true evidence). Specifically, GSF is implemented using the \texttt{GroupSortFuse} package in R; BIC is computed via the EM algorithm to obtain the MLE; and ELBO is optimized using the CAVI algorithm. Except for GSF, all methods are implemented in native R code without additional computational acceleration.
For sample sizes of $320, 800, 2000, 5000$, and $12,500$, we randomly subsample the original dataset $10$ times and apply each method. We assume $K_{\max} = 10$ and record the computation time for each method. Detailed results are provided in Table~\ref{tab:computation_time}.}

\begin{table}[t]
    \centering
    \begin{tabular}{lccccc}
        \hline
        Sample size  & 320  & 800  & 2000  & 5000  & 12500 \\
        \hline
        GSF—SCAD & 11.9 & 24.9 &  66.4 & 171.9 & 421.4 \\
        GSF—MCP & 11.6 & 26.2 &  70.3 & 170.7 & 421.7 \\
        GSF—ALasso & 14.0 & 40.2 & 95.9 & 244.7 & 617.1 \\
        ELBO—CAVI &  11.3  &  42.1  &   111.1  &   496.2  & 996.2 \\
        Nested Sampling & $>10^5$ & — & — & — & —\\
        \hline
    \end{tabular}
    \caption{Average computation time (in seconds) for different methods across varying sample sizes.}
    \label{tab:computation_time}
\end{table}

{\color{black} Among the three GSF methods, SCAD and MCP exhibit comparable runtimes, while ALasso is slower, requiring approximately 1.4 times the computation time of the other two. ELBO (with CAVI) and BIC (with EM) generally have comparable runtimes. For larger sample sizes, ELBO tends to be slower. For example, at $n = 12,500$, ELBO takes $996.2$s, whereas BIC takes $770.4$s on average. This additional cost arises because each iteration of CAVI involves matrix inversion in location-scale Gaussian mixtures, a step not required for simpler distributions such as location-only Gaussians or multinomials.

Since ELBO serves as an approximation to the model evidence, we further compare it with nested sampling. Posterior sampling is computationally intensive, particularly in high-dimensional settings. In this case, with $K_{\max} = 10$ and $m = 9$, each sampling iteration involves $mK_{\max} + K_{\max} - 1 = 98$ variables, substantially increasing the computational burden. As a result, nested sampling fails to converge even after $30$ hours at $n = 320$.  
In contrast, ELBO efficiently approximates the evidence without the need for repeated sampling in high-dimensional spaces. By leveraging optimization rather than direct sampling, ELBO provides a scalable and computationally feasible alternative, making it well-suited for large datasets and complex models.

Since CAVI can be viewed as a variant of the EM algorithm, its efficiency can be further improved through algorithmic optimizations. In addition, implementing the algorithm in a more efficient programming language would enhance computational performance. Given that existing BIC implementations (e.g., \texttt{mclust}) are highly optimized, it is reasonable to expect that, with similar refinements, ELBO-based methods could achieve comparable efficiency.}

\section{Technical highlights}\label{sec:tech}
In this section, we highlight some key technical steps and results in our proofs that could be interesting in their own right.
The main goal of introducing Assumption A4 is to obtain the following Lemma~\ref{lemma 8}, which implies that when the total variation distance between a $K$ component mixture model and the true model with $K^*$ components is very close to zero, it is always possible to merge its $K$ components into $K^*$ groups $I_1,\ldots,I_{K^*} \subset [K]$, so that the aggregated mixing weights $\big\{\sum_{j\in I_k}w_j:\, k\in[K^*]\big\}$ are close to the true mixing weights $w^*=\{w^*_1,\ldots,w^*_{K^*}\}$.

\begin{lemma}\label{lemma 8}
   Under Assumptions A1 and A4, there exists a positive constant $c$ {\color{black} depending on $G^*$ through $K^*$ and $\delta \coloneqq  \min_{j\neq k}|\eta^*_j - \eta^*_k|$} so that for any $\theta$,
    $$d_{\rm TV}(p(\cdot\,|\theta),p^*(\cdot))\ge c\inf_{I_1,...I_{K^*}} \sum_{k=1}^{K^*}\Big|\sum_{j\in I_k}w_j - w_k^*\Big|,$$
    where the infimum is taken over all index sets $I_1, I_2, \ldots, I_{K^*}$ that are disjoint subsets of $[K]$. 
\end{lemma}
{\color{black} We emphasize that the strict positivity of $c$ depends on the separation $\delta$ being bounded away from zero. Although $c$ depends on $G^*$ through $K^*$ and $\delta$, it is difficult to make this dependence explicit with our current proof technique, which follows the proof-by-contradiction strategy of \citet{ho2016convergence} and only guarantees the existence of such a positive $c$.}

Note that in this lemma, $\{I_1, I_2, \ldots, I_{K^*}\}$ is not necessarily a partition of $[K]$; in other words, the union of these subsets is not necessarily $[K]$. Lemma~\ref{lemma 8} establishes a relationship between the total variation distance of two FMMs and their component weights. Since model selection focuses solely on determining the number of components, rather than estimating the parameters $\eta_k$, this inequality plays a central role and is sufficient for proving our main results on model selection consistency. Compared to standard theoretical analyses of estimation consistency, a key advantage of leveraging this lemma is that it avoids the need for strong identifiability assumptions \citep{rousseau2011asymptotic,ho2016strong,manole2021estimating}. Instead, it requires only the weaker identifiability conditions specified in Assumption A4. The proof of Lemma~\ref{lemma 8} is provided in Appendix~\ref{app: proofs}.

We remark that a key step in improving the ELBO upper bound analysis from \cite{watanabe2007stochastic} is the use of a consistency result for variational Bayes, adapted from Theorem 3.1 in \cite{pati2018statistical}, which extends to singular models. In particular, this result, combined with Lemma~\ref{lemma 8}, establishes the consistency of the estimated mixing weights $\bm w$ (after appropriately merging redundant components) under the mean-field variational approximation.

\begin{lemma}[Consistency of variational Bayes]\label{consistency Hellinger}
    Under Assumption {\color{black} A1--A3}, there exists a constant $C_4$ such that it holds with probability at least $1-n^{-1/2}$,
    \begin{align*}
        \int_\Theta h^2\big(p(x\,|\,\theta),\,p^*(x)\big) \, \widehat{q}_{\theta}(\theta)\, d\theta\le C_4\,\bigg(\,\frac{\log p^*(X^n) - \m{L}(\widehat q_{Z^n})}{n} \bigg) + \frac{\log n}{n}.
    \end{align*}
\end{lemma}
This lemma is a direct consequence of the result from \cite{pati2018statistical} since their testing condition is automatically satisfied for the Hellinger distance with a compact parameter space. By using the inequality $d_{\rm TV}\big(p(x\,|\,\theta), p^*(x)\big) \le h\big(p(x\,|\,\theta), p^*(x)\big)$, we can then obtain that, under Assumption A4, at least $K^*$ of the mixing weights are bounded away from $0$ with high probability under the variational posterior distribution $\widehat q_w$. This property leads to an improved upper bound of $\m{L}(\widehat q_{Z^n})$ (which matches the lower bound) compared to the one obtained by \cite{watanabe2007stochastic} (reviewed at the end of Section~\ref{sec:review_VB_FMM}).

{\color{black} Additionally, since we will use the Laplace approximation multiple times in the proof, primarily for computing the KL divergence between the variational posterior and the prior distribution of $\bm \eta$, we present it as a lemma to enhance readability. 

\begin{lemma}[Laplace's approximation in $\mb R^d$]\label{laplace}
Consider a Laplace-type integral 
$$
J(n) = \int_\Omega f(\eta)e^{-nh(\eta)}\,d\eta,
$$
where $n$ is a large positive parameter and $\Omega\subset\mb R^m$ is compact. Suppose $f$ is twice continuously differentiable, $h(\eta)$ has a global minimizer $\widehat \eta$, and $h$ is four times continuously differentiable with a positive definite Hessian $H = \nabla^2h(\widehat \eta)$. 
Furthermore, suppose there exist constants $r_1 = r_1(f,h)>0$ and $r_2 = r_2(f,h)>0$ such that
\begin{align}
    |f(\eta)|\exp(-n[h(\eta) - h(\widehat\eta)])\le r_2\exp(-r_1\sqrt{mn}\|\eta-\widehat\eta\|_H),\quad \forall \|\eta-\widehat\eta\|_H\ge \sqrt{m/n}, \label{growth}
\end{align}
with $\|x\|_H = \sqrt{x^THx}$.
Then it holds that
\begin{align*}
    \frac{e^{nh(\widehat\eta)}\sqrt{\det H}}{(2\pi/n)^{m/2}}\int_\Omega f(\eta)e^{-n h(\eta)} \, d\eta = f(\widehat\eta) + \frac{A_0}{n} +\m O(n^{-2}),
\end{align*}
where $A_0$ depends on $f$ and its derivatives up to the second order, and on $h$ and its derivatives up to the fourth order, all evaluated at $\widehat{\eta}$.
\end{lemma}
Condition \eqref{growth} ensures that the decay of $|f(\eta)|\exp(-nh(\eta))$ is sufficiently fast as $\eta$ moves away from $\widehat{\eta}$, allowing the integral to be approximated by its contribution in a neighborhood of $\widehat{\eta}$.
An explicit expression of $A_0$ is provided in Theorem 5.10 by \cite{katsevich2024laplace}.
According to the lemma, specifically with $h(\widehat \eta) = 0$, we have 
\begin{align*}
    \log J(n) + \frac{m}{2}\log n = \frac{m}{2}\log(2\pi) - \frac{1}{2}\log(\det H) + \log\big(f(\widehat \eta) + A_0/n + \m O(n^{-2})\big).
\end{align*}
When applied to our setting in the proof of Theorem~\ref{main theorem}, $h$ corresponds to the log-likelihood ratio function for each component, and $f$ will be either the prior or the variational posterior on $\eta_k$. Accordingly, $n$ is replaced by $n_k$, the number of observations assigned to the $k$-th component.
Under this setup, we can derive a global upper bound for $A_0$ over $\Omega$. Recall that the mixing components follow canonical forms of exponential families, $g(x;\eta) = \exp\{\eta^T T(x) - T_0(x) - A(\eta)\}$, and by Assumption A2, $A(\eta)$ is four times continuously differentiable. 
Additionally, Assumption P ensures that $\pi_\eta(\eta)$ is twice continuously differentiable. Together with the regularity conditions on $g(x;\eta)$, we can obtain that $f$ is twice continuously differentiable, as required.
In addition, $\widehat\eta_k$ is the global minimizer, so it satisfies the first order condition, i.e., $\nabla h(\widehat\eta_k) = 0$. Owing to the canonical structure of exponential families, $\log g(x;\eta) = \eta^T T(x) - T_0(x) - A(\eta)$, the second and higher-order derivatives of the log-likelihood depend solely on $A(\eta)$ and not on the data $x$, so $A_0$ is independent of $x$. Therefore, by the compactness of $\Omega$, when $f(\eta_k) = \pi_\eta(\eta_k)$, $A_0$ can be globally bounded by some constant that depends only on $\pi_\eta(\eta)$, $A(\eta)$ and $\Omega$. According to the formula in Theorem 5.10 \citep{katsevich2024laplace}, when $f(\eta_k) = \pi_\eta(\eta_k)$ and $n_kh(\eta_k)$ is the log likelihood ratio, $A_0$ is bounded by
\begin{align}
    \max_{\eta \in \Omega} \bigg\{\frac{m}{2} \big\| I(\eta)^{-1} \nabla^2 \pi_\eta(\eta) \big\|_{\infty} + \frac{m^2}{2}\big\|I(\eta)^{-2}\nabla\pi_\eta(\eta) & \nabla^3 A(\eta)\big\|_\infty + \frac{m^2 B_2}{8}\big\|I(\eta)^{-2}\nabla^4 A(\eta)\big\|_\infty \nonumber\\
    & + \frac{5m^3B_2}{12} \big\| I(\eta)^{-3/2} \nabla^3 A(\eta) \big\|_{\infty}^2\bigg\},\label{bound of A0}
\end{align}
where $I(\eta) = \nabla^2A(\eta)$ is the fisher information matrix. The notation $\|\cdot\|_{\infty}$ denotes the infinity norm, which, for a tensor (including matrices), refers to the largest absolute value among all its entries. Under Assumption A1 and A2, $A(\eta)$ is four times continuously differentiable on a compact set $\Omega$, the infinity norms admit global upper bounds on $\Omega$. The case shen $f(\eta_k) = q_{\eta_k}(\eta_k)$ will be analyzed later in the proof of Theorem~\ref{main theorem}.}

\section{Proofs of Main Results}\label{app: proofs}
In this appendix, we collect all proofs of the theoretical results in the paper.

{\color{black} \subsection{Proof of Theorem~\ref{seperation}}\label{proof of Thm1}
\begin{proof}
Recall the definition of the ELBO in \eqref{elbo}: 
\begin{align*}
\mathcal{L}(q_{Z^n}) = \int q_{Z^n}(z^n) \log \frac{p(X^n, s^n|\theta) \pi(\theta)}{q_{Z^n}(z^n)} dz^n.
\end{align*}
Under the block mean-field assumption, the variational posterior factorizes as $q_{Z^n}(z^n) = q_{\theta}(\theta)q_{S^n}(s^n)$. Substituting this into the ELBO and expanding the logarithm, we have:
$$
\m{L}(q_{Z^n}) = \mb E_{q_\theta q_{S^n}}[\log p(X^n,s^n|\theta)] + \mb E_{q_\theta}[\log \pi(\theta) - \log q_\theta(\theta)] - \mb E_{q_{S^n}}[\log q_{S^n}(s^n)].
$$
For a fixed $q_{S^n}(s^n)$, then $\m{L}(q_{Z^n})$ as a functional of $q_{\theta}(\theta)$ can be further simplified into (up to a constant independent of $q_{\theta}(\theta)$),
\begin{align*}
    \m F(q_{\theta}) = \int q_{\theta}(\theta)\Big(\mb E_{q_{S^n}}[\log p(X^n,s^n|\theta)] + \log \pi(\theta) - \log q_\theta(\theta)\Big)\, d\theta.
\end{align*}
It is then easy to see that maximizing $\m F(q_{\theta})$ is equivalent to minimizing
\begin{align*}
   \int q_{\theta}(\theta) \log\frac{q_{\theta}(\theta)}{\pi(\theta)\exp\{\mb E_{q_{S^n}}[\log p(X^n,s^n|\theta)]\}}\, d\theta = D_{\rm KL} (q_{\theta}(\theta)\|\widehat q_{\theta}(\theta)) - \log \widehat C_r,
\end{align*}
with $\widehat q_{\theta}(\theta)$ taking the form~\eqref{VBtheta} in the theorem and $\widehat C_r$ the normalizing constant independent of $q_{\theta}(\theta)$. Since the KL divergence is minimized at $q_{\theta}(\theta) = \widehat q_{\theta}(\theta)$, which leads to the optimal solution of $q_{\theta}(\theta)$ as claimed in the theorem.\\
\indent
Similarly, if we fix $q_{\theta}(\theta)$, then $\m{L}(q_{Z^n})$ as a functional over $q_{S^n}(s^n)$ is equivalent to 
\begin{align*}
    \m F(q_{S^n}) = \sum_{s^n}{{q}_{S^n}(s^n)}\Bigg[\mb E_{q_\theta} [\log p(X^n,s^n|\theta)] - \log q_{S^n}(s^n)\Bigg],
\end{align*}
and therefore maximizing $\m F(q_{S^n})$ is equivalent to minimizing 
\begin{align*}
    \sum_{s^n}{{q}_{S^n}(s^n)}\log \frac{{q}_{S^n}(s^n)}{\exp\{\mb E_{q_\theta} [\log p(X^n,s^n|\theta)]\}} = D_{\rm KL} ({q}_{S^n}(s^n)\|\widehat {q}_{S^n}(s^n)) - \log \widehat C_Q,
\end{align*}
with $\widehat {q}_{S^n}(s^n)$ taking the form~\eqref{VBsn} in the theorem 
and $\widehat C_Q$ the corresponding normalizing constant. Since the KL divergence is minimized at ${q}_{S^n}(s^n) = \widehat {q}_{S^n}(s^n)$, which leads to the optimal solution of ${q}_{S^n}(s^n)$, and completes the proof.
    
\end{proof}}

\subsection{Proof of Lemma~\ref{lemma 8}}
\begin{proof}
Consider the two mixing distributions $G = \sum_{k=1}^K w_k \delta_{\eta_k}$ and $G^* = \sum_{k=1}^{K^\ast} w^*_k \delta_{\eta^*_k}$ where $K\ge K^*$.
Let $\delta\coloneqq  \min_{j\neq k}|\eta^*_j - \eta^*_k|$ and $I_k' \coloneqq  \{j:|\eta_j - \eta_k^*|<\delta/2\}$ for $k\in[K^*]$, then $I_k'$'s are disjoint by definition. Recall that $W_r^r(G,G^*) = \inf_{q_{ij}} \sum_{i=1}^K\sum_{j=1}^{K^*} q_{ij}\big|\eta_i-\eta_j^*\big|^r$, where the infimum ranges over all $q_{ij}$ satisfying $\sum_{i\in[K]}q_{ij} =  w^*_j$ and $\sum_{j\in[K^*]}q_{ij} =  w_i$.

For each $k\in[K^*]$ and feasible $q_{ij}$, we will show 
\begin{align}\label{eqn:W_r_bound}
    W_r^r(G,G^*) \ge \Big|\sum_{i\in I_k'}w_i - w_k^*\Big|\,\delta^r.
\end{align}
To see this, we consider two cases. In the first case when $\sum_{i\in I_k'}w_i > w_k^*$, which corresponds to the weight estimate for the $k$th cluster being greater than the true $w_k^*$, we have
    \begin{align}
        W_r^r(G,G^*)&\ \ge\inf_{q_{ij}} \sum_{i\in I_k'}\sum_{j=1}^{K^*} q_{ij}\big|\eta_i-\eta_j^*\big|^r= \inf_{q_{ij}} \Big(\sum_{i\in I_k'}q_{ik}\big|\eta_i-\eta_k^*\big|^r + \sum_{i\in I_k'}\sum_{j\neq k}q_{ij} \big|\eta_i - \eta_j^*\big|^r\Big)\nonumber\\
        &\ \ge 0 + \inf_{q_{ij}}\sum_{i\in I_k'}\sum_{j\neq k}q_{ij}(\delta/2)^r.\label{eqn:wr}
    \end{align}
    Since $\sum_{i\in I_k'}\sum_{j\in[K^*]}q_{ij} = \sum_{i\in I_k'}w_i$ and $\sum_{i\in I_k'}q_{ik}\le \sum_{i\in[K]} q_{ik} = w_k^*$ for all feasible $q_{ij}$, we obtain $\sum_{i\in I_k'}\sum_{j\neq k }q_{ij}\ge \sum_{i\in I_k'}w_i - w_k^*$. This inequality combined with \eqref{eqn:wr} leads to the desired bound in~\eqref{eqn:W_r_bound}.
    
   In the second case when $\sum_{i\in I_k'}w_i \le w_k^*$, which corresponds to the weight estimate for the $k$th cluster being less than the true $w_k^*$, we have
    \begin{align*}
        W_r^r(G,G^*) \ge \inf_{q_{ij}}\sum_{i=1}^K q_{ik}\big|\eta_i - \eta_k^*\big|^r = \inf_{q_{ij}}\Bigg( \sum_{i\in I_k'}q_{ik}\big|\eta_i - \eta^*_k\big|^r + \sum_{i\notin I_k'}q_{ik}\big|\eta_i - \eta^*_k\big|^r\Bigg).
    \end{align*}
    Since $\sum_{i\in[K]}q_{ik} = w^*_k$ and $\sum_{i\in I_k'}q_{ik}\le\sum_{i\in I_k'}\sum_{j\in[K^*]}q_{ij} = \sum_{i\in I_k'}w_i$, we can obtain that $\sum_{i\notin I_k'}q_{ik} \ge w_k^* - \sum_{i\in I_k'}w_i$, which further implies bound~\eqref{eqn:W_r_bound} due to the preceding display and the definition of $I_k$.

   Finally, since $K^*$ is finite, we obtain from inequality~\eqref{eqn:W_r_bound} that
    \begin{align}
        W_r^r(G,G^*) &\ge (K^*)^{-1}  \sum_{k=1}^{K^*}\Big|\sum_{i\in I_k'}w_j - w_k^*\Big|\,\delta^r \ge (K^*)^{-1} \inf_{I_1,...,I_{K^*}} \sum_{k=1}^{K^*}\Big|\sum_{i\in I_k}w_j - w_k^*\Big|\,\delta^r.\label{lower bound of wasserstein}
    \end{align}
   The claimed bound then follows by combing above with the global inequality~\eqref{tv greater wr} between total variation and Wasserstein distance.
\end{proof}

{\color{black} \subsection{Proof of Lemma~\ref{consistency Hellinger}}
The proof of Lemma 9 closely follows that of Theorem 3.1 in \cite{pati2018statistical}, with the main difference being that they assume $\theta^*$ lies within the parameter space when applied to parametric models. This assumption is necessary in their setting because, beyond distance metrics such as the Hellinger distance and the total variation, they also analyze parameter-based metrics like the $\ell_2$ norm. In our case, since the testing condition holds naturally for the Hellinger distance, the proof of Theorem 3.1 carries over directly when focusing solely on this divergence in Lemma 9; and moreover, we no longer require the parametric model $\{p(\cdot\,|\,\theta)\}$ to contain the true data-generating distribution $p^*$. 

A more concrete justification of this claim is as follows.

\noindent
\begin{proof}
It is well known from \cite{ghosal2000convergence} that for every pair of distributions $P_0$ and $P_1$ from model $\m P$, there exists a sequence of tests $\varphi_n(P_0,P_1)$ such that
\begin{align*}
    &\ P_0^n\varphi_n(P_0,P_1)\le \exp\Big(-\frac{1}{2}nh^2(P_0, P_1)\Big)\\
    &\ P_1^n(1-\varphi_n(P_0,P_1))\le \exp\Big(-\frac{1}{2}nh^2(P_0, P_1)\Big).
\end{align*}
Moreover, since we have assumed in Assumption A1 that the parameter space $\Omega$ for family $\m G = \{g(\cdot;\eta)|\eta\in\Omega\}$ is compact, let $\m P$ here be the set of all possible FMMs with at most $K_{\max}$ components with each component from $\m G$, the covering number $N(\m P, \varepsilon_n/2, h)$ will be finite for given $n$ and $\varepsilon_n>0$. Then we are able to construct a new sequence $\tilde{\varphi}_n$ such that it holds for testing
\begin{align*}
    H_0 : p^*(\cdot) \text{ versus. } H_1: h(p^*(\cdot),\, p(\cdot|\theta))>\varepsilon_n
\end{align*}
for any $\theta\in\Theta$ with $h(p^*(\cdot),\, p(x|\theta))>\varepsilon_n$ with type-I and II error rates satisfying
\begin{align*}
    &\ \mb E_{p^*(X^n)}[\tilde{\varphi}_n] \le e^{-\frac{1}{2}n\varepsilon_n^2}\\
    &\ \mb E_{p(X^n|\theta)}[1 - \tilde{\varphi}_n] \le e^{-\frac{1}{2}nh^2(p^*(\cdot),\, p(\cdot|\theta))}.
\end{align*}
Define $\xi(\theta)\coloneqq \exp\big\{\log p(x^n|\theta) - \log p^*(x^n) + nh^2(p^*(\cdot), p(\cdot|\theta))\big\}$, the remainder of the proof follows the proof of Theorem 3.1 by \cite{pati2018statistical}, where we substitute $d(\theta, \theta^*)$ in their proof with $h(p^*(\cdot),\, p(\cdot|\theta))$ and choose $\varepsilon_n = \sqrt{\log n/n}$.
\end{proof}
}

\subsection{Proof of Theorem~\ref{main theorem}}\label{proof Thm2}

\noindent {\bf More discussions about constants in the theorem.}
The influence of $c_0$ and $r$ in Assumption A4 on the ELBO value is of the order $\sqrt{\log n / n}$, as reflected in the last big-$\m O$ term of the expansion. However, it is difficult to provide an explicit expression for the constant $c_0$ in terms of model characteristics. This is because Assumption A4 follows \cite{ho2016convergence}, whose proof technique, based on contradiction, establishes existence but does not yield an explicit construction.
Before presenting the explicit forms of the constants $C_1$ and $C_2$, we introduce the following notation for clarity: $C_{\phi_0} \coloneqq  \frac{K-1}{2} \log 2\pi + \log\frac{\Gamma(K\phi_0)}{\Gamma(\phi_0)^K}$,
$
C_{L_1} \coloneqq  \frac{m}{2} \log b - \frac{m}{2} \log(2\pi) - \log B_1
$,
$
C_{L_2} \coloneqq  \frac{m}{2} \log a - \frac{m}{2} \log(2\pi) - \log B_2 - \frac{A_2}{B_1}
$,
$\Delta K \coloneqq  K - K^*$, and $\Delta\phi \coloneqq  \phi_0 - \frac{m+1}{2}$. Here, the constant $A_2$, along with $A_3$ and $A_4$ introduced below, arise from the Laplace approximation error and are independent of both $\phi_0$ and $K$ (as can be inferred from the proof provided below). 

When $\phi_0< (m+1)/2$, $C_1$ and $C_2$ take the form:
\begin{align*}
    C_1 &\ = C_{\phi_0}+\Delta K(\phi_0-1/2)\log \phi_0 + C_{11},\mbox{ and }\; C_2 = C_{\phi_0} +K+\frac{K}{12\phi_0}- \Delta\phi\cdot K^*\log K + C_{21};
\end{align*}
while when $\phi_0\ge (m+1)/2$, $C_1$ and $C_2$ take the following form, with $w^*_{k_{\max}}\coloneqq \max_{k\in[K^*]} w_k^*$:
\begin{align*}
    C_1 &\ = C_{\phi_0} - KC_{L_1} - \Delta\phi(\Delta K+1)\log(\Delta K+1 ) - \frac{(\Delta K+1)(B_1 + A_4 + \sqrt{m}B_3A_3)}{B_1 w^*_{k_{\max}}}+C_{12},\\
    C_2 &\ = C_{\phi_0} +K+\frac{K}{12\phi_0} - KC_{L_2} - \Delta\phi K\log K,
\end{align*}
where $C_{11}, C_{12}$ and $C_{21}$ are also constants independent of $K$ and $\phi_0$.

The term $C_{\phi_0}$, which arises from the KL divergence between the variational approximation to the posterior of the mixing weights and its Dirichlet prior, is convex in $K$ for fixed $\phi_0$. It increases with $\phi_0$ when $K$ is fixed, but its dependence on $K$ is more nuanced. Notably, $C_{\phi_0} = 0$ when $K = 1$. For small values of $\phi_0$ (e.g., 0.01 or 0.1), $C_{\phi_0}$ decreases with increasing $K$, and this decay becomes more pronounced as $\phi_0$ approaches zero, reflecting the stronger sparsity imposed by the prior. In contrast, when $\phi_0 > 1$, $C_{\phi_0}$ increases with $K$, with the growth rate accelerating as $\phi_0$ becomes larger.

When $\phi_0$ is extremely small, $C_{\phi_0}$ becomes dominant and diverges, preventing the ELBO from accurately reflecting the intrinsic model complexity under limited data. As $\phi_0$ approaches $(m+1)/2$ from below, the penalty term $\lambda \log n$ aligns with that of the BIC and becomes independent of $\phi_0$ beyond this threshold. However, when $\phi_0 > (m+1)/2$, the forms of $C_1$ and $C_2$ become less stable, leading to greater ELBO instability. This instability arises because the mixing weights tend to spread across all $K$ components rather than concentrating on the true $K^*$ components (see Corollary~\ref{empty out rate}). Taken together, these observations suggest that extreme values of $\phi_0$, whether too small or too large, are not recommended.


\medskip
\noindent {\bf Proof of the theorem.}
The proof of Theorem~\ref{main theorem} is divided into three steps. First, we decompose the ELBO into two terms and derive analytically tractable approximations for each. Second, we constructively establish the lower bound by considering two specific choices of of $q_{Z^n}$, since for any $q_{Z^n}$, $\m L(q_{Z^n})$ provides a lower bound to the ELBO value. inally, we derive the upper bound using Lemma~\ref{consistency Hellinger}, together with the stated assumptions.

\medskip
\noindent \begin{proof}
\textbf{Step 1:} For the complete-data $\{X^n,S^n\}$, we denote the posterior probability of $X_i$ coming from the $k$-th component as $\widehat p_{ik} = \widehat q_{S^n}(S_i = k)$, and let $\widehat n_k = \sum_{i=1}^n \widehat p_{ik}$. Then according to the optimal form of $\widehat q_{\theta}$ given by formula~\eqref{VBtheta}, the variational posterior distribution $\widehat{q}_{\theta}(\theta) = \widehat{q}_{\bm w}(\bm w) \otimes \widehat{q}_{\bm \eta}(\bm \eta)$ can be written as
    \begin{align}
        \widehat{q}_{\bm w}(\bm w) = \frac{\Gamma(n + K\phi_0)}{\prod^K_{k=1}\Gamma(\widehat n_k+\phi_0)}\prod^K_{k=1} w_k^{(\widehat n_k+\phi_0) - 1},\label{posterior w}
    \end{align}
    and
    \begin{align}
        \widehat{q}_{\bm \eta}(\bm \eta)\propto \prod^K_{k=1}\pi_\eta(\eta_k)\exp\left[\sum_{i=1}^n \widehat p_{ik}\left(\eta_k^T T(x_i) - T_0(x_i) - A(\eta_k)\right)\right].\label{posterior eta}
    \end{align}
   From equations~\eqref{posterior w} and \eqref{posterior eta}, we observe that $\widehat{q}_{\bm w}(\bm w)$ and $\widehat{q}_{\bm\eta}(\bm\eta)$ are parameterized by (in other words, only depend on) $\{\widehat p_{ik}:\,i\in[n],k\in[K]\}$. In turn, given the optimal form of $\widehat{q}_{S^n}$ in formula~\eqref{VBsn}, the variational distribution $\widehat{q}_{S^n}(s^n)$ is fully parameterized by the set of probabilities $\{ \widehat{p}_{ik} \}$.
    Therefore, we only need to consider those $q_\theta(\theta) = q_{\bm w}(\bm w)\otimes q_{\bm\eta}(\bm\eta)$ taking the form of
    \begin{align}
        {q}_{\bm w}(\bm w) = \frac{\Gamma(n + K\phi_0)}{\prod^K_{k=1}\Gamma( n_k+\phi_0)}\prod^K_{k=1} w_k^{( n_k+\phi_0) - 1},\label{q(w)}
    \end{align}
    and
    \begin{align}
        {q}_{\bm \eta}(\bm \eta)&\ \propto \prod^K_{k=1}\pi_\eta(\eta_k)\exp\left[\sum_{i=1}^n p_{ik}\left(\eta_k^T T(X_i) - T_0(X_i) - A(\eta_k)\right)\right]\nonumber\\
        &\ \propto \prod^K_{k=1}\pi_\eta(\eta_k)\exp\left[\eta_k^T\sum_{i=1}^n p_{ik} T(X_i) - n_kA(\eta_k)\right],\label{q(eta)}
    \end{align}
    as candidates of $\widehat q_\theta$, where $n_k = \sum_{i=1}^n p_{ik}$. And given a $q_\theta(\theta)$, the optimal $q_{S^n}(s^n)$ takes the following form with a normalization constant $C_Q$ (also parameterized by $\{p_{ik}\}$),
    \begin{align}
        {q}_{S^n}(s^n) = \frac{1}{C_Q}\exp\left\{\int{{q}_\theta(\theta)}\log p(X^n, s^n\,|\,\theta)d \theta\right\}.\label{condition optimal}
    \end{align}
    In the rest of the proof, we only examine those $q_{\bm w}(\bm w)$, $q_{\bm\eta}(\bm\eta)$ and $q_{S^n}(s^n)$ having the forms as in \eqref{q(w)}, \eqref{q(eta)} and \eqref{condition optimal}. Based on these characterizations, we know that maximizing $\m{L}(q_{Z^n})$ over $q_{Z^n}(z^n) = q_\theta(\theta)\otimes q_{S^n}(S^n)$ is equivalent to maximizing $\m{L}(q_{Z^n})$ over all possible combinations of $\{p_{ik}:\,i\in[n],k\in[K]\}$ subject to $n_k = \sum_{i=1}^np_{ik}\ge 0$ for all $k$ and $\sum_{k=1}^K n_k = n$, ensuring that $q(\bm{w})$ is a valid distribution. As for $q(\bm{\eta})$, its support $\Omega^K$ is a compact set, so its normalizing constant will always be well-defined under the given restrictions on $p_{ik}$.
    In addition, as seen in equation \eqref{q(eta)}, $T_0(x_i)$ has no impact on $q_{\bm \eta}(\bm \eta)$, so for fixed $n_k$, the variational posterior mean of the $k$-th component $\overline{\eta}_k = \mb E_{q_{\eta_k}}[\eta_k]$ will be a function of $\big[\sum_{i=1}^n p_{ik}T(x_i)\big]$. As long as $T(x_i)$ is not identical for all $i$, then for any interior point $\eta_0$ of $\Omega$ and $n_k>0$, the following system of equations in terms of $p_{ik}$ has at least one solution:
    \begin{align*}
        \sum_{i=1}^np_{ik} = n_k,\, \mbox{ and } \, \, \overline{\eta}_k\left(\sum_{i=1}^n p_{ik}T(x_i)\right) = \eta_0.
    \end{align*}
    
    We now characterize the ELBO function. According to the definition of $\m{L}(q_{Z^n})$ in \eqref{elbo}, we can decompose the ELBO into two parts:
    \begin{align*}
        \m{L}(q_{Z^n}) &\ = \int q_{Z^n}(z^n)\log\frac{p(X^n,s^n\,|\,\theta)\pi(\theta)}{q_{Z^n}(z^n)}d z^n\\
        &\ = \int q_{\theta}(\theta)\log\frac{\pi(\theta)}{q_{\theta}(\theta)}d\theta + \sum_{s^n} q_{S^n}(s^n)\int q_{\theta}(\theta) \log\frac{p(X^n,s^n\,|\,\theta)}{q_{S^n}(s^n)} d\theta, 
    \end{align*}
    where the first part is the negative KL divergence between $q_\theta(\theta)$ and $\pi(\theta)$, and the second term is $\log C_Q$, with $C_Q$ being the normalization constant in \eqref{condition optimal}. Then we can reformulate $\m{L}(\widehat q_{Z^n})$ as
    \begin{align}
        \m{L}(\widehat q_{Z^n}) = \max_{\{p_{ik}\}} \Big\{-D_{\rm KL}( q_\theta(\theta)\|\pi(\theta)) + 
    \log C_Q\Big\}.\label{reform of L}
    \end{align}
    We analyze the KL divergence first. Since the priors on $\bm w$ and $\bm \eta$ are independent, we have
    \begin{align*}
        D_{\rm KL}({q}_{\theta}(\theta)\|\pi(\theta)) = D_{\rm KL}({q}_{\bm w}(\bm w)\|\pi_{\bm w}(\bm w)) + D_{\rm KL}({q}_{\bm \eta}(\bm \eta)\|\pi_{\bm \eta}(\bm \eta)).
    \end{align*}
    For $D_{\rm KL}({q}_{\bm w}(\bm w)\|\pi_{\bm w}(\bm w))$, we can use the closed forms of the prior and the variational posterior distributions of $\bm w$ to explicitly do the calculation,
    \begin{align}
        \int {q}_{\bm w}(\bm w) \log w_k dw = \Psi( n_k + \phi_0) - \Psi(n + K\phi_0)\label{intlogw},
    \end{align}
    where $\Psi(x) = \Gamma'(x)/\Gamma(x)$ is the so-called di-gamma function. This further leads to
    \begin{flalign}
        D_{\rm KL}({q}_{\bm w}(\bm w)\|\pi_{\bm w}(\bm w)) = \sum^K_{k=1} (n_k\Psi(n_k + &\ \phi_0) -  \log\Gamma(n_k + \phi_0)) - n\Psi(n+K\phi_0)\nonumber\\  &\ + \log\Gamma(n + K\phi_0) + \log\frac{\Gamma(\phi_0)^K}{\Gamma(K\phi_0)},\label{DKLw}
    \end{flalign}
    To further simplify this expressions, we resort to the following two inequalities: for any $x>0$ \citep{alzer1997some}, we have
    \begin{align}
        \frac{1}{2x} < \log x - \Psi(x) < \frac{1}{x},\label{di-gamma}
    \end{align}
    and
    \begin{align*}
        0\le \log\Gamma(x) - \left((x-\frac{1}{2})\log x - x +\frac{1}{2} \log 2\pi \right) \le \frac{1}{12x}.
    \end{align*}
    Applying the two sets inequalities to \eqref{DKLw}, we obtain a lower bound as
    {\color{black}
    \begin{align}
        D_{\rm KL}({q}_{\bm w}(\bm w)\|\pi_{\bm w}(\bm w)) &\ \ge (K\phi_0 - \frac{1}{2})\log (n + K\phi_0)  - (\phi_0 -\frac{1}{2})\sum_{k=1}^K \log (n_k + \phi_0) - \sum_{k=1}^K \frac{n_k}{n_k+\phi_0}\nonumber \\
        &\ \quad \quad \quad -\sum_{k=1}^K \frac{1}{12(n_k+\phi_0)} +\frac{n}{2(n+K\phi_0)} - \frac{K-1}{2}\log 2\pi + \log\frac{\Gamma(\phi_0)^K}{\Gamma(K\phi_0)} \nonumber\\
        &\ \ge (K\phi_0 - \frac{1}{2})\log (n + K\phi_0)  - (\phi_0 -\frac{1}{2})\sum_{k=1}^K \log (n_k + \phi_0) - K \nonumber\\
        &\ \quad \quad \quad - \frac{K}{12\phi_0} - \frac{K-1}{2}\log 2\pi  + \log\frac{\Gamma(\phi_0)^K}{\Gamma(K\phi_0)},\label{lower bound Kw}
    \end{align}
    where the second step is due to $\frac{n_k}{n_k+\phi_0}< 1$ and $\frac{n}{n+K\phi_0}>0$; and also an upper bound,
    \begin{align}
        D_{\rm KL}({q}_{\bm w}(\bm w)\|\pi_{\bm w}(\bm w)) &\ \le (K\phi_0 - \frac{1}{2})\log (n + K\phi_0) - (\phi_0 -\frac{1}{2})\sum_{k=1}^K \log (n_k + \phi_0) - \sum_{k=1}^K \frac{n_k}{2(n_k+\phi_0)}\nonumber\\
        &\ \quad \quad \quad +\frac{n}{n+K\phi_0} + \frac{1}{12(n+K\phi_0)} - \frac{K-1}{2}\log 2\pi + \log\frac{\Gamma(\phi_0)^K}{\Gamma(K\phi_0)}\nonumber\\
        &\ \le (K\phi_0 - \frac{1}{2})\log (n + K\phi_0) - (\phi_0 -\frac{1}{2})\sum_{k=1}^K \log (n_k + \phi_0) + 1 \nonumber\\
        &\ \quad \quad \quad -\frac{K-1}{2}\log 2\pi + \log\frac{\Gamma(\phi_0)^K}{\Gamma(K\phi_0)} + \m O(n^{-1}),\label{upper bound Kw}
    \end{align}
    where the second step is due to $\frac{n_k}{n_k+\phi_0}>0$, $\frac{n}{n+K\phi_0}< 1$ and $\frac{1}{12(n+K\phi_0)}$ is of order $\m O(n^{-1})$.}
    
    As for $D_{\rm KL}({q}_{\bm \eta}(\bm \eta)\|\pi_{\bm \eta}(\bm \eta))$, since both the prior and the variational posterior distributions of $\eta_k$ are factorized under our setup, we have \begin{align*}
        D_{\rm KL}({q}_{\bm\eta}(\bm\eta)\|\pi_{\bm \eta}(\bm \eta)) = \sum_{k=1}^K D_{\rm KL}({q}_{\eta_k}(\eta_k)\|\pi_\eta(\eta_k)).
    \end{align*}
    For each fixed $k\in[K]$, we denote the variational posterior mode (i.e., maximizer of its density function) associated with the $k$-th component as
    \begin{align}
        \widehat\eta_k = \mathop{\arg\max}\limits_{\eta_k}  \ \underbrace{\sum_{i=1}^n p_{ik}\left(\eta_k^T T(x_i) - T_0(x_i) - A(\eta_k)\right)}_{\ell_{n_k}(\eta_k)},\label{eqn:etahat}
    \end{align}
    Note that the critical point $\widehat\eta_k$ satisfies the first order condition $\nabla_{\eta_k}\ell_{n_k}(\widehat\eta_k) = 0$. To apply the Laplace approximation, we need to compute the second-order derivative of $-\frac{1}{n_k}\ell_{n_k}(\eta_k)$:
    \begin{align*}
        \nabla^2_{\eta_k}\Big(-\frac{1}{n_k}\ell_{n_k}(\widehat\eta_k)\Big) = \nabla^2_{\eta_k} A(\widehat\eta_k) = I(\widehat\eta_k).
    \end{align*} 
    Under this notation, with $q_{\bm \eta}(\bm \eta)$ given in \eqref{q(eta)}, we can decompose $D_{\rm KL}({q}_{\eta_k}(\eta_k)\|\pi_\eta(\eta_k))$ as
    \begin{align}
        D_{\rm KL}(q_{\eta_k}(\eta_k)\|\pi_\eta(\eta_k)) &\ = \int q_{\eta_k}(\eta_k)\log\frac{q_{\eta_k}(\eta_k)}{\pi_\eta(\eta_k)}\nonumber\\
        &\ = \int q_{\eta_k}(\eta_k) \log\frac{\pi_\eta(\eta_k)\exp \{\ell_{n_k}(\eta_k) - \ell_{n_k}(\widehat{\eta}_k)\}}{\pi_\eta(\eta_k)\int \pi_\eta(\eta'_k)\exp\{\ell_{n_k}(\eta'_k) - \ell_{n_k}(\widehat{\eta}_k)\}d\eta'_k}\nonumber\\
        &\ =\int q_{\eta_k}(\eta_k)\left(\ell_{n_k}(\eta_k) - \ell_{n_k}(\widehat{\eta}_k)\right)d\eta_k \nonumber\\ &\ \qquad - \log\int \pi_\eta(\eta_k)\exp\{\ell_{n_k}(\eta_k) - \ell_{n_k}(\widehat{\eta}_k)\}d\eta_k.\label{DKL etak}
    \end{align}
    Let $f(\eta_k) = \pi_\eta(\eta_k)$ and $h(\eta_k) = \frac{1}{n_k}(\ell_{n_k}(\widehat\eta_k) - \ell_{n_k}(\eta_k))$ in {\color{black} Lemma~\ref{laplace}}, then they satisfies condition \eqref{growth} since $\pi_\eta(\eta_k)\le B_2$ and $\frac{1}{n_k}(\ell_{n_k}(\widehat\eta_k) - \ell_{n_k}(\eta_k)) = \frac{1}{2}\nabla^2 A(\eta_k)|\eta_k -\widehat\eta_k|^2$.
    Then with $n_k$ large enough, the second integral in \eqref{DKL etak} satisfies
    \begin{align}
        \left|\frac{\sqrt{\det I(\widehat\eta_k)}\int \pi_\eta(\eta_k)\exp\{\ell_{n_k}(\eta_k) - \ell_{n_k}(\widehat\eta_k)\}d\eta_k}{(2\pi/n_k)^{\frac{m}{2}}}-\pi(\widehat\eta_k)\right|\le\frac{A_1}{n_k},\label{denominator}
    \end{align}
    where $A_1$ is given by the expression in \eqref{bound of A0}.
    Then with the bound of $\pi_\eta(\eta)$ in Assumption P we know that the absolute value of the fraction on the left hand side is in $[B_1 - A_1/n_k, B_2 + A_1/n_k]$. Therefore, for large $n_k$, we have
    \begin{align}
        &\ \log\int \pi_\eta(\eta_k)\exp\{\ell_{n_k}(\eta_k) - \ell_{n_k}(\widehat{\eta}_k)\}d\eta_k + \frac{m}{2}\log n_k\nonumber\\
        &\ \quad \quad\quad \quad \le - \frac{1}{2}\log\big(\det I(\widehat\eta_k)\big) + \frac{m}{2} \log(2\pi) + \log\big(\pi(\widehat \eta_k) + A_1/n_k\big)\nonumber\\
        &\ \quad \quad\quad \quad \le - \frac{m}{2}\log a + \frac{m}{2} \log(2\pi) + \log\big(B_2 + A_1/n_k\big),\label{up d/2 log nk}
    \end{align}
    and
    \begin{align}
        &\ \log\int \pi_\eta(\eta_k)\exp\{\ell_{n_k}(\eta_k) - \ell_{n_k}(\widehat{\eta}_k)\}d\eta_k + \frac{m}{2}\log n_k\nonumber\\
        &\ \quad \quad\quad \quad \ge - \frac{1}{2}\log\big(\det I(\widehat\eta_k)\big) + \frac{m}{2} \log(2\pi) + \log\big(\pi(\widehat \eta_k) - A_1/n_k\big)\nonumber\\
        &\ \quad \quad\quad \quad \ge - \frac{m}{2}\log b + \frac{m}{2} \log(2\pi) + \log\big(B_1 - A_1/n_k\big),\label{low d/2 log nk}
    \end{align}
    where the last step in each inequality is from the Assumptions P and A2.
    For the first integral in equation~\eqref{DKL etak}, by plugging in the explicit form of $q(\eta)$ from \eqref{q(eta)}, we obtain
    \begin{align*}
        \int q_{\eta_k}(\eta_k)\left(\ell_{n_k}(\eta_k) - \ell_{n_k}(\widehat{\eta}_k)\right)d\eta_k 
        = &\ \frac{\int \pi_\eta(\eta_k)\left(\ell_{n_k}(\eta_k) - \ell_{n_k}(\widehat{\eta}_k)\right)\exp\{\ell_{n_k}(\eta_k) - \ell_{n_k}(\widehat{\eta}_k)\}d\eta_k}{\int \pi_\eta(\eta_k)\exp\{\ell_{n_k}(\eta_k) - \ell_{n_k}(\widehat{\eta}_k)\}d\eta_k},
    \end{align*}
    whose denominator is already analyzed. Using Laplace's approximation again to the numerator and this time let $f(\eta_k) = \pi_\eta(\eta_k)\cdot\frac{1}{n_k}(\ell_{n_k}(\eta_k) - \ell_{n_k}(\widehat{\eta}_k))$ to satisfies condition \eqref{growth}, we obtain
    \begin{align*}
        \left|\frac{\sqrt{\det I(\widehat\eta_k)}\int \pi_\eta(\eta_k) \cdot\frac{1}{n_k}\left(\ell_{n_k}(\eta_k) - \ell_{n_k}(\widehat{\eta}_k)\right)\exp\{\ell_{n_k}(\eta_k) - \ell_{n_k}(\widehat{\eta}_k)\}d\eta_k}{(2\pi/n_k)^{\frac{m}{2}}}-0\right|\le\frac{A_2}{n_k}.
    \end{align*}
    To see the dependence of $A_2$, we need to evaluate $f(\eta_k) = \pi_\eta(\eta_k)\cdot\frac{1}{n_k}(\ell_{n_k}(\eta_k) - \ell_{n_k}(\widehat{\eta}_k))$ and its derivatives up to the second order at $\widehat\eta_k$. Obviously $f(\widehat \eta_k) = 0$, and
    \begin{align*}
        \nabla f(\eta_k) &\ = \frac{1}{n_k}\big[\nabla \pi_\eta(\eta_k)\cdot(\ell_{n_k}(\eta_k) - \ell_{n_k}(\widehat{\eta}_k)) + \pi_\eta(\eta_k)\cdot\nabla\ell_{n_k}(\eta_k)\big].\\
        \nabla^2 f(\eta_k) &\ = \frac{1}{n_k}\big[\nabla^2 \pi_\eta(\eta_k)\cdot (\ell_{n_k}(\eta_k) - \ell_{n_k}(\widehat{\eta}_k)) + 2 \nabla\pi_\eta(\eta_k)\cdot \nabla\ell_{n_k}(\eta_k) +\pi_\eta(\eta_k)\cdot \nabla^2\ell_{n_k}(\eta_k)\big].
    \end{align*}
    Since $\ell_{n_k}(\widehat\eta_k) - \ell_{n_k}(\widehat{\eta}_k) = 0$ and $\nabla\ell_{n_k}(\widehat\eta_k) = \bm 0$, thus $\nabla f(\widehat\eta_k) = \bm 0$ and $\nabla^2 f(\widehat\eta_k) = -\pi(\widehat\eta_k)\nabla^2A(\widehat \eta_k)$.
    Therefore $A_2$ is also determined by $\pi_\eta(\eta)$, the $A(\eta)$ and $\Omega$ and its upper bound can be established based on Theorem 5.10 in \cite{katsevich2024laplace} as $mB_2/2$.
    Noticing that $\ell_{n_k}(\eta_k) \le \ell_{n_k} (\widehat{\eta}_k)$, with the bound of the denominator in~\eqref{denominator} and the assumption $\pi(\widehat\eta_k)\ge B_1$ we have
    \begin{align}
        -\frac{A_2}{B_1 - A_1/n_k}\le\int q_{\eta_k}(\eta_k)\left(\ell_{n_k}(\eta_k) - \ell_{n_k}(\widehat{\eta}_k)\right)d\eta_k\le 0.\label{first term in KL eta}
    \end{align}
    Combining \eqref{DKL etak}, \eqref{up d/2 log nk} and \eqref{first term in KL eta} together, we finally obtain that for any sufficiently large $n_k$,
    \begin{align}
        D_{\rm KL}(q_{\eta_k}(\eta_k)\|\pi_\eta(\eta_k)) - \frac{m}{2}\log n_k\le \frac{m}{2}\log b - \frac{m}{2} \log(2\pi) - \log B_1 + \m O(n^{-1}_k),\label{up KL eta log nk}
    \end{align}
    and
    \begin{align}
        D_{\rm KL}(q_{\eta_k}(\eta_k)\|\pi_\eta(\eta_k)) - \frac{m}{2}\log n_k\ge \frac{m}{2}\log a - \frac{m}{2} \log(2\pi) - \log B_2 - \frac{A_2}{B_1} + \m O(n^{-1}_k).\label{low KL eta log nk}
    \end{align}
    For further use, we summarize the terms independent of $n_k$ as $C_{L_1}\coloneqq  \frac{m}{2}\log b - \frac{m}{2} \log(2\pi) - \log B_1$ and $C_{L_2} \coloneqq  \frac{m}{2}\log a - \frac{m}{2} \log(2\pi) - \log B_2 - \frac{A_2}{B_1}$.
    
    Before proceeding on to the two specific constructions, we need to analyze the relationship between $\overline{\eta}_k$ and $\widehat{\eta}_k$.
    The reason we need to consider $\overline{\eta}_k$ in addition to the maximizer $\widehat{\eta}_k$ is that if we rely on Laplace approximation to calculate $\mb E_{q_{\eta_k}}[\log g(x_i; \eta_k)] = \mb E_{q_{\eta_k}}[\eta_k^T T(x_i) - T_0(x_i) - A(\eta_k)]$,
    then approximation error will depend on $x_i$, making it impossible to establish a global bound. However, by directly constructing $\overline{\eta}_k$, we obtain the expectation $\mathbb{E}_{q_{\eta_k}}[\eta_k]$, thereby avoiding the issue of the approximation error depending on the observations.
     Then let us examine the distance between $\overline{\eta}_k$ and $\widehat{\eta}_k$ in order to determine the error introduced by approximating $A(\overline{\eta}_k)$ with $A(\widehat{\eta}_k)$ and further control the difference between $\mathbb{E}_{q_{\eta_k}}[\log g(x_i; \eta_k)]$ and $\log g(x_i; \overline{\eta}_k)$:
     \begin{align}
         \overline{\eta}_k = \int_\Omega\eta_k \,q_{\eta_k}(\eta_k)\, d\eta_k = \frac{\int_\Omega \eta_k\pi_\eta(\eta_k)\exp\{\ell_{n_k}(\eta_k) - \ell_{n_k}(\widehat\eta_k)\} d\eta_k}{\int_\Omega \pi_\eta(\eta_k)\exp\{\ell_{n_k}(\eta_k) - \ell_{n_k}(\widehat\eta_k)\}d\eta_k},\label{variational mean}
     \end{align}
     whose denominator has already been approximated in equation \eqref{denominator}, so we consider only the numerator. Again, using Lemma~\ref{laplace}, 
     \begin{align*}
         \left| \frac{\sqrt{\det I(\widehat\eta_k)}\int_\Omega (\eta_k - \widehat\eta_k)\pi_\eta(\eta_k)\exp\{\ell_{n_k}(\eta_k) - \ell_{n_k}(\widehat\eta_k)\} d\eta_k}{(2\pi/n_k)^{\frac{m}{2}}}\right|\le\frac{\sqrt{m}A_3}{n_k},
     \end{align*}
     where $A_3$ can be bounded by (according to Theorem 5.10 of \cite{katsevich2024laplace}), 
     \begin{align*}
         \max_{\eta \in \Omega} \bigg\{m \big\| I(\eta)^{-1} \nabla \pi_\eta(\eta) \big\|_{\infty} + \frac{m^2B_2}{2}\big\|I(\eta)^{-2}\nabla^3 A(\eta)\big\|_\infty\bigg\},
     \end{align*}
     The $\sqrt{m}$ on the right hand side appears because $\widehat \eta_k$ is a vector in $\mb R^m$ and the approximation error of each dimension is $A_3/n_k$. Therefore,
     \begin{align}
         \|\overline{\eta}_k - \widehat\eta_k\| \le \frac{\sqrt{m}A_3/n_k}{\pi(\widehat \eta_k) - A_1/n_k} \le \frac{\sqrt{m}A_3/n_k}{B_1 - A_1/n_k}.\label{mean and mode}
     \end{align}
    
    \medskip
    \noindent \textbf{Step 2:} Next we prove a lower bound for $\m{L}(\widehat q_{Z^n})$.
    According to the characterization of ELBO in \eqref{reform of L}, every instance of $\{p_{ik}\}$ gives a lower bound to $\m{L}(\widehat q_{Z^n})$. For each construction, we analyze the $D_{\rm KL}( q_\theta(\theta)\|\pi(\theta))$ and $\log C_Q$ respectively.
    Let us consider the following two special constructions:\\
    
    \noindent \underline{Construction (I)}: This construction applies to the case when $\phi_0>(d+1)/2$.  For $1\le k\le K^*-1$, we can always choose $\{p_{ik}:\,i\in[n],k\in[K^*-1]\}$ such that
    \begin{align*}
        n_k = \sum_{i=1}^n p_{ik} = w_k^*(n + K\phi_0)&\, -\phi_0,\text{ i.e. } \frac{n_k+\phi_0}{n+K\phi_0} = w_k^*,\;
       \mbox{and } \overline\eta_k = \eta_k^*,
    \end{align*}
    and for $K^*\le k\le K$, we choose $\{p_{ik}:\,i\in[n],K^*\le k\le K\}$ such that
    \begin{align*}
        n_k = w^*_{K^*}\frac{n+K\phi_0}{K-K^*+1}-\phi_0,\text{ i.e. } &\, \frac{(n_k+\phi_0)(K-K^*+1)}{n+K\phi_0} = w_{K^*}^* \; \mbox{and }\overline{\eta}_k = \eta_{K^*}^*.
    \end{align*}
    Without loss of generality, we assume the $K^*$-th component has the largest weight, i.e., $w_{K^*}^* = \max_{k\in[K^*]} w^*_k$. This construction corresponds to the scenario when the first $K^*-1$ components from the model being identical to the first $K^*-1$ components from the true model, while the weight of the $K^*$-th component---the most heavily weighted component---of the true model is uniformly distributed among the remaining $K-K^*+1$ components in the employed model.
    In this case, all $n_k$'s are proportional to $n$, so we have from \eqref{up KL eta log nk} that
    \begin{align*}
        D_{\rm KL}(q_{\bm\eta}(\bm\eta)\|\pi_{\bm \eta}(\bm \eta)) \le \frac{m}{2} \sum_{k=1}^K \log n_k + K C_{L_1},
    \end{align*}
    where $C_{L_1}$ is the summarized error of Laplace approximation defined above. Combining the inequality above with the approximation~\eqref{upper bound Kw} to $D_{\rm KL}(q_{\bm w}(\bm w)\|\pi_{\bm w}(\bm w))$, we have the following upper bound for the KL divergence between $q_\theta(\theta)$ and $\pi(\theta)$,
    {\color{black}
    \begin{align}
        &\ D_{\rm KL} (q_\theta(\theta)\|\pi(\theta)) \le (K\phi_0 - \frac{1}{2})\log (n + K\phi_0) + (\frac{1}{2}-\phi_0) \sum_{k=1}^K \log (n_k + \phi_0) + \frac{m}{2}\sum_{k=1}^K \log n_k\nonumber\\
        &\ \quad\quad\quad\quad\quad\quad\quad\quad\quad\quad\quad\quad\quad\quad +\bigg[1 -\frac{K-1}{2}\log 2\pi + \log\frac{\Gamma(\phi_0)^K}{\Gamma(K\phi_0)} + KC_{L_1}\bigg] + \m O(n^{-1})\nonumber\\
        \le &\ (K\phi_0 - \frac{1}{2})\log n + \Big(\frac{m + 1}{2} - \phi_0\Big) \sum_{k=1}^K \log n_k + C + \m O(n^{-1})\nonumber\\
        = &\ (K\phi_0 - \frac{1}{2})\log n + \Big(\frac{m + 1}{2} - \phi_0\Big)\bigg[\sum_{k = 1}^{K^*-1} \log (w_k^*n) + (K-K^*+1)\log\frac{w^*_{K^*}n}{K-K^* +1}\bigg]  + C + \m O(n^{-1})\nonumber \\
        = &\ \frac{mK+K-1}{2}\log n + \Big(\frac{m + 1}{2} - \phi_0\Big)\bigg[\sum_{k = 1}^{K^*-1} \log w_k^* + (K-K^*+1)\log\frac{w^*_{K^*}}{K-K^* +1}\bigg] + C + \m O(n^{-1}).\label{KL phi large}
    \end{align}
    where $C$ denotes the terms inside the brackets in the first step.}
    As for the lower bound of $\log C_Q$, we can first rewrite $C_Q$ as
    \begin{flalign}
        C_Q &\ = \prod_{i=1}^n\sum_{s_i}\exp\int{q}_{\theta}(\theta)\log p(x_i, s_i|\theta) d\theta \nonumber\\
        &\ = \prod_{i=1}^n\sum_{k=1}^K\exp\left(\int {q}_{w_k}(w_k)\log w_k dw_k + \int{q}_{\eta_k}(\eta_k)\log g(x_i;\eta_k) d\eta_k\right).\label{CQ}
    \end{flalign}
    Using equations~\eqref{intlogw} and \eqref{di-gamma} again, we obtain
    \begin{align*}
        \int{q}_{w_k}(w_k) \log w_k dw_k = \Psi(n_k + \phi_0) - \Psi(n + K\phi_0)\ge \log \frac{n_k+\phi_0}{n+K\phi_0}-\frac{1}{n_k+\phi_0}.
    \end{align*}
    For the second part, since we have $\overline\eta_k = \eta_k^*$ for $1\le k\le K^*-1$ and $\overline\eta_k = \eta^*_{K^*}$ for $K^*\le k\le K$, we can write $\log g(x_i;\eta_k)$ explicitly in the integral:
    \begin{align*}
        \int{q}_{\eta_k}(\eta_k)\big(\eta_k^T T(x_i) - T_0(x_i) - A(\eta)\big)\, d\eta_k = \eta_k^{*T}T(x_i) - T_0(x_i) - \int{q}_{\eta_k}(\eta_k)A(\eta_k)\, d\eta_k.
    \end{align*}
    Concretely, we can expand $q_{\eta_k}(\eta_k)$ similar as above,
    \begin{align*}
        \int{q}_{\eta_k}(\eta_k)\big(A(\eta_k) - A(\widehat\eta_k)\big) \, d\eta_k &\ = \frac{\int \pi_\eta(\eta_k)(\eta_k)\big(A(\eta_k) - A(\widehat\eta_k)\big)\exp\{\ell_{n_k}(\eta_k) - \ell_{n_k}(\widehat \eta_k)\} \, d\eta_k}{\int \pi_\eta(\eta_k)\exp\{\ell_{n_k}(\eta_k) - \ell_{n_k}(\widehat \eta_k)\} \, d\eta_k}.
    \end{align*}
    We've seen the denominator in equation~\eqref{denominator} and apply Lemma~\ref{laplace} again to the numerator it shows
    \begin{align*}
        \left| \frac{\sqrt{\det I(\widehat\eta_k)}\int_\Omega \pi_\eta(\eta_k)\big(A(\eta_k) - A(\widehat\eta_k)\big)\exp\{\ell_{n_k}(\eta_k) - \ell_{n_k}(\widehat\eta_k)\} d\eta_k}{(2\pi/n_k)^{\frac{m}{2}}}\right|\le\frac{A_4}{n_k},
    \end{align*}
    where $A_4$ can be bounded by (according to Theorem 5.10 of \cite{katsevich2024laplace}), 
     \begin{align*}
         \max_{\eta \in \Omega} \bigg\{\frac{m}{2} \big\| I(\eta)^{-1} \big(I(\eta) + 2\nabla \pi_\eta(\eta)\nabla A(\eta)^T\big) \big\|_{\infty} + \frac{m^2B_2}{2}\big\|I(\eta)^{-2}\nabla A(\eta)\nabla^3 A(\eta)\big\|_\infty\bigg\}.
     \end{align*}
    Then $\int{q}_{\eta_k}(\eta_k)\big(A(\eta_k) - A(\widehat\eta_k)\big) \, d\eta_k\ge -\frac{A_4/n_k}{B_1 - A_2/n_k}$.
    Additionally, we've shown at the end of Step 1 that the distance between $\overline\eta_k$ and $\widehat \eta_k$ is no greater than $\frac{\sqrt{m}A_3/n_k}{B_1 - A_2/n_k}$. Therefore, by the mean value theorem and the bound $B_3$ on $\|\nabla_\eta A(\eta)\|$ in Assumption A2, we obtain that
    \begin{align*}
        \int{q}_{\eta_k}(\eta_k)\big(\eta_k^T T(x_i) - T_0(x_i) - A(\eta)\big)\, d\eta_k \ge \eta_k^{*T}T(x_i) - T_0(x_i) - A(\eta_k^*) - \frac{(A_4 + \sqrt{m}B_3A_3)/n_k }{B_1 - A_2/n_k}.
    \end{align*}
    Similarly, for $K\ge K^*$ we have
    \begin{align*}
        \int{q}_{\eta_k}(\eta_k)\big(\eta_k^T T(x_i) - T_0(x_i) - A(\eta)\big)\, d\eta_k \ge \eta_{K^*}^{*T}T(x_i) - T_0(x_i) - A(\eta_{K^*}^*) - \frac{(A_4 + \sqrt{m}B_3A_3)/n_k }{B_1 - A_2/n_k}.
    \end{align*}
    Substituting the lower bounds when $k<K^*$ of the two integrals in Equation~\eqref{CQ} yields,
    \begin{align}
        &\ \sum_{k=1}^{K^*-1}\exp \left(\int{q}_{w_k}(w_k) \log w_k dw_k + \int{q}_{\eta_k}(\eta_k)\log g(X_i;\eta_k) d\eta_k\right)\nonumber\\
        \ge &\ \sum_{k=1}^{K^*-1}\exp\left(\log w_k^* -\frac{1}{n_k+\phi_0} +\log g(X_i;\eta_k^*)- \frac{(A_4 + \sqrt{m}B_3A_3)/n_k }{B_1 - A_2/n_k}\right)\nonumber\\
        = &\ \sum_{k=1}^{K^*-1} w_k^* g(X_i;\eta_k^*)\exp\left(-\frac{1}{n_k+\phi_0}- \frac{(A_4 + \sqrt{m}B_3A_3)/n_k }{B_1 - A_2/n_k}\right).\label{lower CQ}
    \end{align}
    The same argument can applies to $k\ge K^*$, where by using the definition of $n_k$ for $K\ge K^*$,
    \begin{align*}
        &\ \sum_{k=K^*}^K \exp\left(\int{q}_{w_k}(w_k) \log w_k dw_k + \int{q}_{\eta_k}(\eta_k)\log g(X_i;\eta_k) d\eta_k\right)\\
        \ge &\ \sum_{k=K^*}^K \exp\left(\log \frac{w_{K^*}^*}{K-K^*+1} -\frac{1}{n_k+\phi_0} +\log g(X_i;\eta_{K^*}^*)- \frac{(A_4 + \sqrt{m}B_3A_3)/n_k }{B_1 - A_2/n_k}\right)\\
        \ge &\ w_{K^*}^* g(X_i;\eta_{K^*}^*) \exp\left(-\frac{1}{n_{K^*}+\phi_0}- \frac{(A_4 + \sqrt{m}B_3A_3)/n_{K^*} }{B_1 - A_2/n_{K^*}}\right).
    \end{align*}
    According to form of $C_Q$ in \eqref{CQ} and the fact $e^{-x}\ge 1-x$, $\log C_Q$ can be bounded from below by
    \begin{align}
        \log C_Q &\ \ge \sum_{i=1}^n \log \left[\sum_{k=1}^{K^*} w_k^* g(X_i;\eta_k^*) \exp \left(-\frac{1}{n_{k_{\min}}+\phi_0} - \frac{(A_4 + \sqrt{m}B_3A_3)/n_{k_{\min}}}{B_1 - A_2/n_{k_{\min}}} \right)\right]\nonumber\\
        &\ \ge \log p^*(X^n) - \frac{\Delta K+1}{w^*_{K^*}} - \frac{(\Delta K+1)(A_4 + \sqrt{m}B_3A_3)}{B_1 w^*_{K^*}}  - \frac{1}{w^*_{k_{\min}}} - \frac{A_4 + \sqrt{m}B_3A_3}{B_1 w^*_{k_{\min}}} - \m O(n^{-1}), \label{CQ phi large}
    \end{align}
    where $\Delta K = K-K^*+1$, $n_{k_{\min}} \coloneqq  \min_{1\le k\le K} n_k$ and $w^*_{k_{\min}} \coloneqq  \min_{1\le k\le K^*} w_k^*$. The second step follows from the definition of $n_k$ under this construction, $n_{k_{\min}} + \phi_0\ge \min\{ w^*_{K^*}(n + K\phi_0)/(K-K^*+1), w^*_{k_{\min}}(n + K\phi_0)\}$.
    Finally, given the formulation of $\m{L}( q_{Z^n})$ as in \eqref{reform of L}, we can add \eqref{KL phi large} and \eqref{CQ phi large} together to get a lower bound when $\phi_0>(m+1)/2$ as
    \begin{align}
        \m{L}(\widehat q_{Z^n}) - \log p^*(X^n)\ge -\frac{mK+K-1}{2}\log n + C - \m O(n^{-1}),\label{lower bound phi large}
    \end{align}
    where $C$ corresponds to $C_1$ in Theorem~\ref{main theorem} when $\phi_0\ge(m+1)/2$.

     \noindent \underline{Construction (II)}: This construction applies to the case when  $\phi_0\le(m+1)/2$. For each $k$ such that $1\le k\le K^*$, we can always choose  $\{p_{ik}:\,i\in[n],k\in[K^*]\}$ such that
    \begin{align*}
        n_k = \sum_{i=1}^n p_{ik} = w_k^*(n + K\phi_0)& -\phi_0, \ \ \text{ i.e. } \frac{n_k +\phi_0}{n+K\phi_0} = w_k^*,\:\mbox{and }\overline{\eta}_k = \eta_k^*,
    \end{align*}
    and for $K^*< k\le K$, we set $p_{ik} = 0$ for all $i$. Then for each $k>K^*$, $q_{\eta_k}(\eta_k)$ is the same as prior density $\pi_\eta(\eta_k)$, so we have $D_{\rm KL}(q_{\eta_k}(\eta_k)\|\pi_\eta(\eta_k)) = 0$ for all $k>K^*$. For each $k\le K^*$, we have $n_k\asymp n$, so \eqref{up KL eta log nk} leads to
    \begin{align*}
        D_{\rm KL}(q_{\bm \eta}(\bm\eta)\|\pi_{\bm \eta}(\bm \eta))  \le \sum_{k=1}^{K^*} \frac{m}{2}\log n_k + K^*C_{L_1} + \m O(n^{-1}).
    \end{align*}
   Combining the preceding display with \eqref{upper bound Kw}, and noting that $n_k = 0$ for $K > K^*$, we obtain:
   \begin{align}
        &\ D_{\rm KL} (q_\theta(\theta)\|\pi(\theta)) \le (K\phi_0 - \frac{1}{2})\log (n + K\phi_0) + (\frac{1}{2}-\phi_0) \sum_{k=1}^{K^*} \log (n_k + \phi_0) + \frac{m}{2}\sum_{k=1}^{K^*} \log n_k \nonumber\\
        &\ \quad\quad\quad\quad\quad\quad +\bigg[1 -\frac{K-1}{2}\log 2\pi + \log\frac{\Gamma(\phi_0)^{K}}{\Gamma(K\phi_0)} + (\frac{1}{2} - \phi_0)(K-K^*)\log \phi_0 + K^*C_{L_1}\bigg] + \m O(n^{-1})\nonumber\\
        \le &\ (K\phi_0 - \frac{1}{2})\log n + \Big(\frac{m + 1}{2} - \phi_0\Big) \sum_{k=1}^{K^*} \log n_k + C + \m O(n^{-1})\nonumber\\
        = &\ (K\phi_0 - \frac{1}{2})\log n + \Big(\frac{m + 1}{2} - \phi_0\Big) \sum_{k=1}^{K^*} \log (w^*_kn) + C + \m O(n^{-1})\nonumber\\
        = &\ \left[(K - K^*)\phi_0 + \frac{mK^* + K^* -1}{2}\right]\log n + \Big(\frac{m + 1}{2} - \phi_0\Big) \sum_{k=1}^{K^*} \log w^*_k + C + \m O(n^{-1}),
        \label{KL phi small}
    \end{align}
    where $C$ denotes the terms inside the brackets in the first step.
    Now since the summation over $k\in[K]$ for the $C_Q$ in \eqref{CQ} is dominated by its first $K^*$ terms, we can lower bound this summation to the sum of the first $K^*$ terms. Similarly as we derived a lower bound of $C_Q$ as in \eqref{lower CQ}, we 
    \begin{align*}
         \sum_{k= 1}^{K^*}\exp\left(\int q_{\theta_k}(\theta_k) \log \left(w_kg(x_i;\eta_k)\right)d\theta_k \right) \ge &\ \sum_{k=1}^{K^*}w_k^*g(x_i;\eta^*_k)\exp\left(-\frac{1}{n_k+\phi_0}- \frac{(A_4 + \sqrt{m}B_3A_3)/n_k }{B_1 - A_2/n_k}\right).
    \end{align*}
    Under this construction, $n_{k_{\min}} = w_{k_{\min}}^*(n+K\phi_0) - \phi_0$.
    Putting all the pieces together, we obtain
    \begin{align}
        \log C_Q &\ \ge \sum_{i=1}^n\log\left(\sum_{k=1}^{K^*} w^*_kg(x_i;\eta^*_k)\right) - n\left(\frac{1}{n_{k_{\min}}+\phi_0} + \frac{(A_4 + \sqrt{m}B_3A_3)/n_{k_{\min}} }{B_1 - A_2/n_{k_{\min}}}\right)\nonumber\\
        &\ \ge \log p^*(X^n) - \frac{1}{w_{k_{\min}}^*} - \frac{A_4 + \sqrt{m}B_3A_3}{B_1 w^*_{k_{\min}}} - \m O(n^{-1}).\label{log CQ phi small}
    \end{align}
    Adding up \eqref{KL phi small} and \eqref{log CQ phi small}, we can obtain a lower bound to $\m{L}(q_{Z^n}(z^n))$ for all $\phi_0\le(m+1)/2$ as
    \begin{align}
        \m{L}(\widehat q_{Z^n}) - \log p^*(X^n)\ge -\left(\frac{mK^*+K^*-1}{2} + (K-K^*)\phi_0\right)\log n + C - \m O(n^{-1}), \label{lower bound phi small}
    \end{align}
    where $C$ corresponds to $C_1$ in Theorem~\ref{main theorem} when $\phi_0<(m+1)/2$.
    Finally, combining \eqref{lower bound phi large} and \eqref{lower bound phi small} leads to the claimed lower bound of $\m{L}(\widehat q_{Z^n})$.

    \medskip
   \noindent \textbf{Step 3:}
    Now we derive an upper bound of the ELBO. We first tackle the $\log \widehat C_Q$ term, which is the normalization constant for the optimal distribution $\widehat q_{S^n}(s^n)$ in \eqref{VBsn}. Note that $\log w_k$ is concave with respect to $w_k$. For $\log g(x;\eta_k)$, the second order derivative with respect to $\eta_k$ is the negative fisher information matrix for the mixture component
    \begin{align*}
        \nabla^2_{\eta_k}\log g(x;\eta_k) = -\nabla^2_{\eta_k}A(\eta_k) = -I(\eta_k),
    \end{align*}
    which is negative definite according to Assumption A1. Therefore, $\log g(x;\eta)$ is concave with respect to $\eta_k$. By applying Jensen's inequality, we can then obtain
    \begin{align*}
        \int\widehat{q}_{w_k}(w_k) \log w_k\, dw_k + \int\widehat{q}_{\eta_k} (\eta_k) \log g(x;\eta_k)\, d\eta_k\le \log (\overline{w}_kg(x;\overline\eta_k)),
    \end{align*}
    which, combined with the expression of $C_Q$ in Equation \eqref{CQ}, leads to:
    \begin{align}
        \log \widehat C_Q \le \sum_{i=1}^n\log\left(\sum_{k=1}^K \overline{w}_kg(x_i;\overline{\eta}_k)\right) = \log p(x^n\,|\,\overline{\theta}).\label{upper bound log CQ}
    \end{align}
    The bounds of KL divergence $D_{\rm KL}(q_{w_k}(w_k)\|\pi_w(w_k))$ has been given in \eqref{lower bound Kw} and \eqref{upper bound Kw}. As for the $D_{\rm KL}(q_{\eta_k}(\eta_k)\|\pi_\eta(\eta_k))$ term, note that for a large $n_k$, we can apply the approximation \eqref{low KL eta log nk}. Consequently, we define a set $\widehat {\m F}\coloneqq  \{k \,| \,1 \le k \le K, \widehat{n}_k \ge N \}$, where $N$ is a sufficiently large (truncating) constant, to select those indices $k$ for which the approximation is accurate. This approximation, together with the non-negativity of the KL divergence, gives
    \begin{align*}
        D_{\rm KL}(\widehat q_{\bm\eta}(\bm\eta)\|\pi_{\bm \eta}(\bm \eta)) = \sum_{k=1}^K D_{\rm KL}(\widehat q_{\eta_k}(\eta_k)\|\pi_\eta(\eta_k))\ge \frac{m}{2} \sum_{k\in\widehat {\m F}}\big(\log \widehat n_k- \m O(\widehat n_k^{-1})\big) + |\widehat{\m F}|C_{L_2}.
    \end{align*}
    {\color{black} Using approximation \eqref{lower bound Kw}, we can bound the entire KL divergence term from below as
    \begin{align}
        &\ D_{\rm KL}(\widehat{q}_{\theta}(\theta)\|\pi(\theta))\nonumber\\ 
        &\ \ge (K\phi_0-\frac{1}{2})\log n + (\frac{1}{2}-\phi_0)\sum_{k \notin \m{F}}\log \widehat n_k + \frac{m}{2}\sum_{k \in \widehat {\m F}}\big(\log \widehat n_k- \m O(\widehat n_k^{-1})\big) + C + |\widehat{\m F}| C_{L_2}  \nonumber\\
        &\ = (K\phi_0-\frac{1}{2})\log n + \left(\frac{m+1}{2}-\phi_0\right)\sum_{k \in \widehat {\m F}}\big(\log \widehat n_k- \m O(\widehat n_k^{-1})\big) + C + |\widehat{\m F}| C_{L_2},\label{lower of KL}
    \end{align}}
    where $C$ here is short for $-K-\frac{K}{12\phi_0} - \frac{K-1}{2}\log 2\pi + \log \frac{\Gamma(\phi_0)^K}{\Gamma(K\phi_0)}$ from equation \eqref{lower bound Kw}.
    In the last part of this proof, we will show that {\color{black} with high probability}, at least $K^*$ of the $\widehat n_k$'s are proportional to $n$, that is, $\widehat n_k\ge \gamma_k n$ for some constant $\gamma_k$. Without loss of generality, we assume they are the first $K^*$ components, then $[K^*] \subset \widehat{\m{F}}$. Using this fact, we can obtain that for each $\phi_0<(m+1)/2$,
    \begin{align*}
        &\ D_{\rm KL}(\widehat{q}_{\theta}(\theta)\|\pi(\theta)) \\
        &\ \ge(K\phi_0-\frac{1}{2})\log n + \left(\frac{m+1}{2}-\phi_0\right) \Big(K^*\log n + \sum_{k=1}^{K^*}\log\gamma_k\Big) + C + K^* C_{L_2}- \m O(n^{-1})  \\
        &\ =\left(\frac{mK^*+K^*-1}{2} + \phi_0(K-K^*)\right)\log n + \left(\frac{m+1}{2}-\phi_0\right) \Big(\sum_{k=1}^{K^*}\log\gamma_k\Big)+ C + K^* C_{L_2}- \m O(n^{-1}),
    \end{align*}
    where the constant $C$ here is the same from that in the previous equation.
    When $\phi_0 > (m+1)/2$, by using the fact $\sum_{k = 1}^K \widehat n_k = n$ so that $\sum_{k=1}^K \log \widehat n_k\le K\log(n/K)$ (the equality is attained when $\widehat{n}_k = n/K$ for all $k$), and the lower bound in \eqref{lower of KL}, we obtain
    \begin{align*}
        D_{\rm KL}(\widehat{q}_{\theta}(\theta)\|\pi(\theta))
        &\ \ge (K\phi_0-\frac{1}{2})\log n + \left(\frac{m+1}{2}-\phi_0\right) K\log (n/K) + C + KC_{L_2} + \m O(n^{-1})\\
        &\ = \frac{mK+K-1}{2}\log n + \Big(\phi_0 - \frac{m + 1}{2}\Big)K\log K + C + KC_{L_2} + \m O(n^{-1}),
    \end{align*}
    where the constant $C$ here is the same from that in the previous equation.
    Combining the two lower bounds of $D_{\rm KL}(\widehat{q}_{\theta}(\theta)\|\pi(\theta))$ obtained above and the bound in \eqref{upper bound log CQ}, we can get an upper bound to $\m{L}(\widehat q_{Z^n}) - \log p^*(X^n)$ as 
    \begin{align*}
        \m{L}(\widehat q_{Z^n}) - \log p(X^n\,|\,\overline\theta) \le -\frac{mK + K-1}{2} \log n + C + \m O(n^{-1}),
    \end{align*}
    where $C$ corresponds to $C_2$ in Theorem~\ref{main theorem} when $\phi_0\ge (m+1)/2$.

    It remains to show that at least $K^*$ of the $\widehat n_k$'s are of the same magnitude of $n$. Noticing that the Hellinger distance
    \begin{align*}
        h^2\big(p(x\,|\,\theta),\,p^*(x)\big) &\ = \int\bigg[\Big(\sum_{k=1}^K w_kg(x;\eta_k)\Big)^{1/2} - \sqrt{p^*(x)}\bigg]^2\, d x\\
        &\ = \int \bigg\{\sum_{k=1}^K w_kg(x;\eta_k) + p^*(x) - 2\Big(\sum_{k=1}^K w_kg(x;\eta_k)\Big)^{1/2}\sqrt{p^*(x)}\bigg\} dx
    \end{align*}
    is a convex function with respect to every $w_k$, then using the global inequality between total variation and $W_r^r$ in \eqref{tv greater wr}, Lemma~\ref{lemma 8}, Lemma~\ref{consistency Hellinger}, Jensen's inequality and the inequality $d_{\rm TV}\big(p(x\,|\,\theta),\, p^*(x)\big) \le h\big(p(x\,|\,\theta),\, p^*(x)\big)$, we can get
    \begin{align*}
        \bigg(\inf_{\{I_1,...I_{K^*}\}}\sum^{K^*}_{k=1}\Big|\sum_{j\in I_k} \overline w_j - w_k^*\Big|\bigg)^2&\ \lesssim \int_{\Omega^K} h^2\big(p(x\,|\,\overline w,\eta), p^*(x)\big) \widehat q(\eta)\,d\eta\\
        &\ \le \int_\Theta h^2\big(p(x\,|\, w,\eta), p^*(x)\big) \widehat q(\theta)\,d\theta.\\
        &\ \le C_4\,\bigg(\,\frac{\log p^*(X^n) - \m{L}(\widehat q_{Z^n})}{n} \bigg) + \frac{\log n}{n},
    \end{align*}
    where the convexity is used in the second inequality. Recall that, in Step 2, we proved a lower bound to $\m{L}(\widehat q_{Z^n}) - \log p^*(X^n)$, which is $-\lambda\log n + C_1$ with $C_1$ independent of $n$. Thus we obtain
    \begin{align*}
        \left(\inf_{\{I_1,...I_{K^*}\}}\sum_{k = 1}^{K^*}\Big|\sum_{j\in I_k}\overline w_j - w^*_k\Big|\right)^2 \le C_5 \cdot\frac{\lambda\log n + \log n}{n}
    \end{align*}
    {\color{black} for sufficiently large $n$ and some constant $C_5$ depending on $K$ and $G^*$ through $c_0$ in Assumption A4.}
    Since the cardinality of $I_k$ is at most $K$, we have that for every $k\le K^*$, there is at least one $\overline w_j,j\in I_k$ such that $\overline w_j\ge (w^*_k - \sqrt{C_5(\lambda +1)\log n/n})/K$. Therefore, it holds with high probability,
    \begin{align*}
        \widehat n_j = \overline w_j(n+K\phi_0) - \phi_0\ge\left(w^*_k - \sqrt{C_5(\lambda +1)\log n/n}\right)(n + K\phi_0)/K -\phi_0,
    \end{align*}
    then we have $\widehat n_k\ge \gamma_k n$ with $\gamma_j = w_k^*/K - \m O(\sqrt{\log n/n})$
    which completes the proof. Moreover, assume they are the first $K^*$ components, $\sum_{k=1}^{K^*}\log \gamma_k = \sum_{k=1}^{K^*}\log w_k^* - K^*\log K -\m O(\sqrt{\log n/n})$. It can be seen that the constants $c_0$ and $r$ in Assumption A4 influence the weights only at the $\sqrt{\log n / n}$ order, which also validates our discussion on constants in Theorem~\ref{main theorem}.
\end{proof}

\begin{remark}[Extension to general prior]\label{rmk: extended prior}
    {\color{black} As discussed in Section~3.1, although Theorem 2 is stated for the symmetric Dirichlet prior $\pi_{\bm w} = \text{Dir}(\phi_0)$, the results naturally extend to any prior $\pi'_{\bm w}$ satisfying the boundedness condition $L \le \pi'_{\bm w}(\bm w) / \pi_{\bm w}(\bm w) \le R$ for some constants $R > L > 0$.
For any given variational distribution $q \in \Gamma_{\rm MF}$, the difference between the ELBO objectives under the two priors is given by:
\begin{equation*}
    \mathcal{L}_{\pi'}(q) - \mathcal{L}_{\pi}(q) = \mathbb{E}_q [\log \pi'_{\bm w}(\bm w)] - \mathbb{E}_q [\log \pi_{\bm w}(\bm w)] = \mathbb{E}_q \left[ \log \frac{\pi'_{\bm w}(\bm w)}{\pi_{\bm w}(\bm w)} \right].
\end{equation*}
The boundedness condition implies that this difference lies in $[\log L, \log R]$. Let $\widehat q_{Z^n,\pi}$ and $\widehat q_{Z^n,\pi'}$ denote the optimal approximations under $\pi$ and $\pi'$, respectively. We then have the following inequalities:
\begin{align*}
    \m L_{\pi'}(\widehat q_{Z^n,\pi'}) & \ge \m L_{\pi'}(\widehat q_{Z^n,\pi}) \ge \m L_{\pi}(\widehat q_{Z^n,\pi}) + \log L, \\
    \m L_{\pi'}(\widehat q_{Z^n,\pi'}) & \le \m L_{\pi}(\widehat q_{Z^n,\pi'}) + \log R \le \m L_{\pi}(\widehat q_{Z^n,\pi}) + \log R.
\end{align*}
Since $\log L$ and $\log R$ depend only on the prior $\pi'$, they are negligible compared to the dominant penalty term $-\lambda \log n$. Consequently, the two-sided bound in Theorem 2 continues to hold for $\pi'_{\bm w}$, with the constants $C_1$ and $C_2$ shifted by $\log L$ and $\log R$, respectively.}
\end{remark}

\subsection{Proof of Corollary~\ref{coro:ELBO}}\label{app:coro:ELBO}

{\color{black} In this subsection, we prove Corollary~\ref{coro:ELBO}, with a focus on handling the log-likelihood ratio test (LRT) statistic (due to its relevance to the gap $\Delta_n$, as noted in the discussion prior to this corollary) and providing relevant background on singular learning theory, primarily based on \cite{watanabe2009algebraic}. 

It is well known that for regular models, the test statistic of the LRT statistic converges to a chi-squared distribution as the sample size increases. However, this result does not extend to singular models. In general, the asymptotic behavior of the LRT statistic for mixture models is closely related to a limiting Gaussian process to the log-likelihood ratio process. In addition, for singular models, the compactness of the parameter space is critical and often assumed to prevent the maximum likelihood estimator and the LRT statistic from divergence. Below, we briefly outline how \cite{watanabe2009algebraic} addresses the asymptotics of the LRT statistic in the context of singular models.

First, in a singular model, there are generally multiple parameters in the parameter space $\Theta$ that correspond to the true model. We define the set of all such parameters as $\Theta_0$, which may consist of a union of manifolds with different dimensions. Now, consider a compact parameter space $\Theta$ and define the KL divergence as $\m{K}(\theta) = D_{\rm KL}\big(p^*(\cdot)\,|\,p(\cdot\,|\,\theta)\big)$ for a parametric family $\{p(\cdot\,|\,\theta):\,\theta\in\Theta\}$. The set of true parameters is then given by $\Theta_0 \coloneqq  \{\theta \in \Theta \,|\, \m{K}(\theta) = 0\}$.
By the well-known Hironaka's theorem (Theorem 2.3 in \cite{watanabe2009algebraic}), the function $\m{K}(\theta)$ can be transformed (via an real analytic map) into its normal crossing form, which is uniquely determined by the analytic properties of $\m{K}(\theta)$ around any of its zero points $\theta^* \in \Theta_0$. Specifically, as established in Theorem 6.1 of \cite{watanabe2009algebraic}, there exist an $m$-dimensional manifold $\m{M}$ and a real-analytic resolution map $g: \m{M} \to \Theta$ such that, in every local coordinate $U$ of $\m{M}$,
\begin{align*}
    \m K(g(u)) = u^{2k} = u_1^{2k_1}u_2^{2k_2}\cdots u_m^{2k_m},
\end{align*}
where $k_1,k_2,\dots,k_d$ are nonnegative integers. Moreover, by denoting $r(x,\theta) = \log p^*(x)- \log p(x|\theta) $, there exists a real analytic function $a(x,u)$ such that
\begin{align*}
    r(x,g(u)) = u^k a(x,u) \quad\text{and}\quad
    \int a(x,u) p^*(x)\, dx = u^k \quad(u\in U).
\end{align*}
Using these notations, the empirical KL divergence can be written as $\m K_n(\theta) = n^{-1}\sum_{i=1}^n r(X_i,\theta)$. Now let us define the following stochastic process on $\m M$ (even for $\m K(g(u)) =0$),
\begin{align*}
    \zeta_n(u) = \frac{1}{\sqrt n}\sum_{i=1}^n \big(u^k - a(X_i,u)\big).
\end{align*}
Then it is easy to obtain a representation of the (rescaled) empirical KL divergence as
\begin{align*}
    n\m K_n(g(u)) = nu^{2k} - \sqrt{n} u^k\zeta_n(u).
\end{align*}
From its definition, we see that $\zeta_n(u)$ satisfies 
\begin{align*}
    \mb E[\zeta_n(u)] = 0,\quad \mb E[\zeta_n(u)\zeta_n(v)] = \mb E_X[a(X,u)a(X,v)] - u^kv^k,\,\forall u,v\in\m M.
\end{align*}
It is then shown by that $\zeta_n(u)$ converges in law to a limiting (tight) Gaussian process $\zeta(u)$ on $\m M$ with zero mean function and covariance function $\mb E[\zeta(u)\zeta(v)] = \mb E_X[a(X,u)a(X,v)] - u^kv^k$.

It's well known that the MLE is invariant under any one-to-one reparameterization of the model, a property known as the invariance of the MLE. Consequently, the likelihood ratio statistic ${\rm LRT}$, as defined after Theorem~\ref{main theorem}, corresponds to the maximum of $\m{K}(g(u))$ over all $u$ across the local coordinate charts ${\m{M}_\alpha}$ that cover the manifold $\m{M}$.
Leveraging this connection, Main Theorem 6.4 in \cite{watanabe2009algebraic} establishes that, under certain mild fundamental regularity conditions on $p(x|\theta)$ and $p^*(x)$, the following convergence in probability holds:
\begin{align*}
    n\m K_n(\widehat\theta) \overset{p}{\rightarrow}-\frac{1}{4}\mb \max_\alpha \max_{u\in\m M_{\alpha 0}}\Big(\max\big\{0,\zeta(u)\big\}\Big)^2\coloneqq W_0, \ \ \mbox{as }n\to\infty,
\end{align*}
where $\max_\alpha$ corresponds to the maximization over all local coordinates that covers the manifold $\m M =\bigcup_{\alpha} \m M_{\alpha}$ and
\begin{align*}
    \m M_{\alpha 0} = \{u\in\m M_\alpha|\m K(g(u)) = 0\}
\end{align*}
denotes the submanifold of $\m M_\alpha$ corresponding to the set $\Theta_0$; moreover, $n\m K_n(\widehat\theta)$ is asymptotically uniformly integrable. This result implies that the LRT statistic in our model, when defined on a compact parameter space, will converge in probability to a non-degenerate random variable $W_0$ for any given $K \geq K^*$. Since we assume $K \leq K_{\max}$, and the maximum of a finite number of random variables remains a random variable, Corollary~\ref{coro:ELBO} follows.

To verify the conditions under which the above conclusions hold, as discussed in Section 7.3 of \cite{watanabe2009algebraic}, for mixture models where $x$ takes values within a bounded range, such as beta and multinomial distributions, these general conclusions hold. For the location family of Gaussians, the same results can be established with additional effort, because although the log density ratio $r(x, \theta)$ does not satisfy the integrability condition, $r(x, \theta) - 1$ does with respect to a certain analytically transformed density function;(see Section 7.8 of \cite{watanabe2009algebraic}). For the location-scale family and more general exponential families, \cite{azais2009likelihood} considers the case in which the variance-covariance matrix is treated as a structural parameter (i.e., homoscedastic across all components). They demonstrate that the likelihood ratio test (LRT) converges to the supremum of a centered Gaussian process with a compact index set and finite covariance functions. The general conditions required for their results include the identifiability of the mixing measure $G(\theta)$, the compactness of the parameter space, and a H\"older continuity condition on the normalized directional derivatives with respect to $\theta$ around the true model, whose modulus of continuity is controlled by a data-dependent, square-integrable function $C(x)$,
which can be verified for homoscedastic location-scale Gaussian mixtures (see Section 4.3.1 of \citep{azais2009likelihood}).}

\subsection{Proof of Corollary~\ref{consistency in model selection}}
\begin{proof}
    According to the expansion of ELBO derived in Theorem~2 and the convergence of $\Delta_n$ to a random variable for all $K = K^*, \ldots, K_{\max}$ as $n \to \infty$, we can conclude that for sufficiently large $n$, the probability that the ELBO $\widehat{\mathcal L}_K$ strictly decreases with $K$ over $K^* \le K \le K_{\max}$ approaches one as $n \to \infty$. Particularly, for $K = K^*$, we have
    \begin{align*}
        \log p^*(X^n) - \frac{mK^* + K^* - 1}{2}\log n + C_1\le \m{L}(\widehat q_{Z^n})\le \log p(X^n\,|\,\overline{\theta}) - \frac{mK^* + K^* - 1}{2}\log n+C_2.
    \end{align*}
    On the other hand, when $K<K^*$ and $G(\theta) = \sum_{j=1}^Kw_j\delta_{\eta_j}$, we will show below that the Wasserstein distance will be bounded from below by a constant; then we can apply Lemma~\ref{consistency Hellinger} to show that the ELBO value $\widehat {\m L}_K$ is also strictly less than $\widehat {\m L}_{K^*}$. Concretely, recall that in the proof of Lemma~\ref{lemma 8}, we defined $\delta = \min_{j\neq k}|\eta^*_j - \eta^*_k|>0$ and $I_k = \{j:|\eta_j - \eta_k^*|<\delta/2\}$ for $k\in[K]$. Since $K<K^*$ and $I_k$'s are disjoint, there is at least one $I_k$ that is empty. Without loss of generality, we may assume $I_1 = \emptyset$. This then leads to (according to the proof of Lemma~\ref{lemma 8})
    \begin{align*}
        W_r^r(G,G^*)\ge\inf_{q_{i1}}\sum_{i=1}^K q_{i1}|\eta_i-\eta_1^*|^r \ge \inf_{q_{i1}} \sum_{i=1}^K q_{i1}(\delta/2)^r = w_1^*(\delta/2)^r.
    \end{align*}
    Applying Lemma~\ref{consistency Hellinger}, Assumption A4 and the fact $h\big(p(x\,|\,\theta),\,p^*(x)\big)\ge d_{\rm TV}(p(x\,|\,\theta),\,p^*(x)\big)$ leads to, with probability at least $1-n^{-1/2}$,
    \begin{align*}
        C_4\left(\frac{\log p^*(X^n) - \m{L}(\widehat q_{Z^n})}{n}\right) + \frac{\log n}{n} \ge \int_\Theta d_{\rm TV}^2\big(p(x\,|\,\theta),\,p^*(x)\big) \, \widehat{q}_{\theta}(\theta) \ge c_0'{w_1^*}^2(\delta/2)^{2r},
    \end{align*}
    with $C_4$ from Lemma~\ref{consistency Hellinger} and $c_0'$ from the global relationship of $d_{\rm TV}$ and $W_r^r$ in equation~\eqref{tv greater wr}.
    Therefore, we have with high probability
    \begin{align*}
        \m{L}(\widehat q_{Z^n})\le \log p^*(X^n) + \log n - C\,n{w_1^*}^2(\delta/2)^{2r}
    \end{align*}
    for some positive constant $C$ depending on $C_4$ and $c_0'$. On the other hand, the ELBO value $\widehat {\m L}_{K^*}$ is at least $\log p^*(X^n) - \m O_P(\log n)$, which  implies that the probability of $\widehat {\m L}_K$ at $K<K^*$ is strictly smaller than that at $K=K^*$ will approach $1$ as $n$ increases.
\end{proof}

\subsection{Proof of Corollary~\ref{empty out rate}}
\begin{proof}
    With the assumption $\Delta_n\le C_3\log\log n$ holds with probability at least $1-\kappa_n$, we have with probability at least $1-n^{-1/2} - \kappa_n$,
    \begin{align}
        -\lambda\log n - C_3\log\log n - C_1\le \m{L}(\widehat q_{Z^n}) - \log p(X^n\,|\,\overline\theta)\le-\lambda\log n + C_2.\label{eqn:two_bod}
    \end{align}
    Let us start with the case of $\phi_0<(m+1)/2$.
    In the proof of Theorem \ref{main theorem}, we showed that at least $K^*$ of the $\widehat n_k$'s are proportional to $n$ when $\phi_0<(m+1)/2$. We can then without loss of generality assume that it is the first $K^*$ of $\widehat n_k$'s. 
    We will prove the claimed bound by contradiction. Suppose there exists some $k'>K^*$ such that $\overline w_{k'}\ge [(\log n)^{\rho_1} + \phi_0]/n$, then $\widehat n_{k'} = \overline w_{k'}(n + K\phi_0) - \phi_0 \ge (\log n)^{\rho_1}$, implying $k'\in\widehat{\m{F}}$ (defined in the proof of Theorem~\ref{main theorem}). Now, using the lower bound of the KL divergence in \eqref{lower of KL} and incorporating all the terms independent of $n$ into $C$, we obtain
    \begin{align*}
        D_{\rm KL}(\widehat{q}_{\theta}(\theta)\|\pi(\theta)) &\ \ge (K\phi_0-\frac{1}{2})\log n + \left(\frac{m+1}{2}-\phi_0\right)\sum_{k \notin F}\log n_k -C - \m O (n^{-1})\\
         &\ \ge \left[(K-K^*)\phi_0 + \frac{mK^*+K-1}{2} \right]\log n + (C_3+1)\log\log n- C' - \m O (n^{-1}).
    \end{align*}
    Therefore, with the decomposition of $\m L(\widehat q_{Z^n})$ in \eqref{reform of L} and the upper bound of $\log \widehat C_Q$ in \eqref{upper bound log CQ},
    \begin{align*}
        \m{L}(\widehat q_{Z^n}) - \log p(X^n\,|\,\overline\theta) &\ \le -  D_{\rm KL}(\widehat{q}_{\theta}(\theta)\|\pi(\theta)) + C + \m O(n^{-1/2})\\
        &\ \le - \lambda_1 \log n - (C_3+1)\log\log n - C  +\m O(n^{-1/2}),
    \end{align*}
    where $\lambda_1 = (K-K^*)\phi_0 + (mK^*+K^*-1)/2$. Note that this upper bound is smaller than the corresponding lower bound as in inequality~\eqref{eqn:two_bod} for all sufficiently large $n$, which is a contradiction. Therefore, we must have $\overline w_k \le [(\log n)^{\rho_1} + \phi_0]/n$ for all $k>K^*$, which leads to
    \begin{align*}
        \inf_{\sigma\in\m{S}_k} \sum_{k = K^*+1}^K \overline w_{\sigma(k)} \le \sum_{k = K^*+1}^K \overline w_k <\frac{(K-K^*)[(\log n)^{\rho_1} + \phi_0]}{n}.
    \end{align*}

    \medskip
    For the other case where $\phi_0>(m+1)/2$, suppose there exists some $k''$ such that $\overline w_k\le 1/(\log n)^{\rho_2} + \phi_0/n$, then we have $\widehat n_{k''}\le n/(\log n)^{\rho_2}$. By further using inequality \eqref{lower of KL}, we get
    \begin{align*}
        D_{\rm KL}(\widehat{q}_{\theta}(\theta)\|\pi(\theta)) &\ \ge (K\phi_0-\frac{1}{2})\log n + \left(\frac{m+1}{2}-\phi_0\right)(K\log n - \rho_2\log\log n) -C - \m O(n^{-1/2}),
    \end{align*}
    where $C$ varies from the previous one. It implies that, with $\lambda = (mK+K-1)/2$, we have
    \begin{align*}
        \m{L}(\widehat q_{Z^n}) - \log p(X^n\,|\,\overline\theta)\le - \lambda \log n - (C_3 + 1)\log\log n + C + \m O(n^{-1/2}).
    \end{align*}
    This upper bound is again smaller than its corresponding lower bound from~\eqref{eqn:two_bod}, which cause a contradiction. Therefore, for all $k\in\{1,...,K\}$, we must have $\overline{w}_{k} \ge 1/(\log n)^{\rho_2} + \phi_0/n$ and this completes the proof.
\end{proof}

\subsection{Proof of Theorem~\ref{para root n}} 

\smallskip
\begin{proof}
The proof will be divided into two steps. Since in Corollary~\ref{empty out rate}, we see that when $\phi_0 < (m+1)/2$, the decay rate of the weights of the redundant components is sufficiently fast, at the order of $n^{-1}$. Therefore, to some extent, if we retain only the $K^*$ components that do not empty out, we can roughly treat it as an exactly-fitted problem. We first establish a relationship between the total variation distance of two finite mixture distributions and a discrepancy metric between their respective parameters following similar steps as in the proof of Theorem 3.1 by \cite{ho2016convergence}. In the second step, we utilize the fast decay rate of the weights and the upper bound of the total variation distance to control the newly defined metric.

    \noindent \textbf{Step 1:} Concretely, we consider a general finite mixture $p(x\,|\,\theta) = \sum_{k=1}^{K^*} w_kg(x;\eta_k)$, with $\theta=(w_1,\ldots,w_{K^*},\eta_1,\ldots,\eta_{K^*})$, and $\bm {w} = (w_1,\ldots,w_{K^*})$, $\bm {\eta} = (\eta_1,\ldots,\eta_{K^*})$ where the mixing weights satisfy $w_k\ge 0$ and $\sum_{k=1}^{K^*}w_k \le 1$, and every $\eta_k$ is from the compact set $\Omega\in\mb R^m$ with diameter $L<\infty$. Note that $p(x|\theta)$ has $K^*$ components but is not necessarily a density function since the total mixing weight can be smaller than 1. Also recall that $p^*(x) = \sum_{k=1}^{K^*} w_k^* g(x;\eta_k^*)$ is the truth. We consider the following discrepancy metric between the parameters of $p(\cdot\,|\,\theta)$ and $p^*(\cdot)$,
    $$D(\theta) \coloneqq  \inf_{\sigma\in\m{S}_{K^*}} \sum_{k=1}^{K^*}\big(|w_{\sigma(k)} - w_k^*| + w_{\sigma(k)}|\eta_{\sigma(k)} - \eta_k^*|\big).$$
    We will show below that there exist positive constants $\varepsilon$ and $c$ depending on $G^*$, such that for any $\theta$ satisfying $D(\theta) < \varepsilon$, it holds 
    \begin{align}
        \frac{1}{2}\int \big|p(x|\theta) - p^*(x)\big|dx > cD(\theta).\label{geq w1}
    \end{align}
    Similar as the proof provided by \cite{ho2016convergence}, we will show this inequality by contradiction. If this is not the case, then there exists some sequence $\{\theta^s:\,s\ge 1\}$ such that $D(\theta^s)\rightarrow 0$ as $s\to\infty$ and 
    \begin{align}
        \lim_{m\rightarrow\infty} \frac{1}{ D(\theta^s)} \,\int |p(x|\theta^s) - p^*(x)|dx = 0\label{limit 0}.
    \end{align}
    Since $D(\theta^s)\rightarrow 0$ as $s\to\infty$ and the number of possible permutations of $[K^*]$ is finite, there is a sub-sequence $\{\theta^{s_\ell}\}$ and a permutation $\sigma'\in \m{S}_{K^*}$ such that $w^{s_\ell}_{\sigma'(k)}\rightarrow w^*_k$ and $\eta^{s_\ell}_{\sigma'(k)} \rightarrow \eta^*_k$ for all $k = 1,..., K^*$ as $\ell\to\infty$. Without loss of generality, we may assume that $\sigma'$ is the identity permutation and substitute the sequence with its sub-sequence, so that we can write $D(\theta^s) = \sum_{k=1}^{K^*}\big(|w_k - w_k^*| + w_k|\eta_k - \eta_k^*|\big)$ for sufficiently large $s$. Now we can apply a first order Taylor expansion as follows:
    \begin{align*}
        p(x|\theta^s) - p^*(x) &\ = \sum_{k=1}^{K^*} w_k^s \big(g(x;\eta^s_k) - g(x;\eta_k^*)\big) + \sum_{k=1}^{K^*}(w_k^s - w_k^*)g(x;\eta_k^*)\\
        &\ = \sum_{k=1}^{K^*} w_k^s \Big[\big(\eta^s_k - \eta_k^*\big)\nabla g(x;\eta_k^*) + \m O\big(\eta^s_k - \eta_k^*\big)^2 \Big] + \sum_{k=1}^{K^*}(w_k^s - w_k^*)g(x;\eta_k^*)\\
        &\ = \sum_{k=1}^{K^*} w_k^s \big(\eta^s_k - \eta_k^*\big)\nabla g(x;\eta_k^*) + \sum_{k=1}^{K^*}(w_k^s - w_k^*)g(x;\eta_k^*) + R_s,
    \end{align*}
   where the remainder term satisfies $R_s/D(\theta^s)\rightarrow 0$ as $s\rightarrow\infty$ due to the convergence of $\eta_k^s$ to $\eta_k^*$. Then equation~\eqref{limit 0} will imply
    \begin{align*}
        \lim_{m\rightarrow\infty} \frac{p(x|\theta^s) - p^*(x)}{D(\theta^s)} = \lim_{m\rightarrow\infty} \bigg\{\sum_{k=1}^{K^*} \frac{w_k^s \big(\eta^s_k - \eta_k^*\big)}{D(\theta^s)} \,\nabla g(x;\eta_k^*) + \sum_{k=1}^{K^*}\frac{w_k^s - w_k^*}{D(\theta^s)}\,g(x;\eta_k^*)\bigg\} = 0 
    \end{align*}
    for almost every $x$. 
    Note that the whole term inside the second limit is a linear combination of finite many $g(x;\eta_k^*)$ and $\nabla g(x;\eta_k^*)$ terms.
    Also, since $\Omega$ is compact under Assumption A1, each linear combination coefficient has a convergent sub-sequence. Without loss of generality, we may substitute the sequence with its sub-sequence, and denote the respective limits as
    \begin{align*}
        \alpha_k = \lim_{s\to\infty}\frac{w_k^s \big(\eta^s_k - \eta_k^*\big)}{D(\theta^s)} \ \ \mbox{and} \ \ \beta_k = \lim_{s\to\infty}\frac{w_k^s - w_k^*}{D(\theta^s)}.
    \end{align*}
    We claim that at least one of these limiting coefficients $\{\alpha_k\}_{k=1}^{K^*}$ and $\{\beta_k\}_{k=1}^{K^*}$ does not vanish, otherwise, we will have
    \begin{align*}
       1= \lim_{m\rightarrow\infty} \frac{D(\theta^s)}{D(\theta^s)} = \lim_{m\rightarrow\infty} \frac{\sum_{k=1}^{K^*}w_k^s|\eta^s_k - \eta_k^*| + \sum_{k=1}^{K^*}|w_k^s - w_k^*|}{D(\theta^s)} = \sum_{k=1}^{K^*} \big(|\alpha_k| + |\beta_k|\big) =  0,
    \end{align*}
    which is a contradiction. Therefore, we have two lists of coefficients $\{\alpha_k\}_{k=1}^{K^*}$ and $\{\beta_k\}_{k=1}^{K^*}$ not both identically zero and 
    \begin{align*}
         \sum_{k = 1}^{K^*} \Big[\alpha_k g(x;\eta_k^*) + \beta_k^T \nabla g(x;\eta_k^*)\Big] = \lim_{m\rightarrow\infty} \frac{p(x|\theta^s) - p^*(x)}{D(\theta^s)} = 0
    \end{align*}
    for almost every $x$. However, the preceding display and the weak identifiability condition together imply that $\alpha_k = 0$, $\beta_k = \vec{0}$ for all $k = 1,..., K^*$, which contradicts with the fact that at least one of the coefficients is not zero. Therefore, \eqref{geq w1} holds. 
    With the first half of Assumption A4, it's easy to see that $\frac{1}{2}\int |p(x|\theta) - p^*(x)|dx = 0$ if and only if $D(\theta) = 0$. Because both $D(\theta)$ and $\frac{1}{2}\int |p(x|\theta) - p^*(x)|dx$ are continuous in $\theta$, we can further obtain (following the same argument from which we derived the global relationship \eqref{tv greater wr}) from equation~\eqref{geq w1} that for each possible $\theta$, there exists a positive number $C_D$ depending on $G^*$ and $\Omega$ such that
    \begin{align}\label{eqn:TV_D}
            \frac{1}{2}\int |p(x|\theta) - p^*(x)|dx \ge C_D\cdot D(\theta).
    \end{align}

    \medskip
     \noindent \textbf{Step 2:} In this step, we use the relationship obtained in \eqref{eqn:TV_D} and the consistency from Lemma~\ref{consistency Hellinger} to prove that every maximizer $\widehat{\eta}_j$ of a non-emptying component falls within a neighborhood of radius $\m O(\log n / \sqrt{n})$ around some true parameter $\eta^*_k$. Moreover, since we have already established the distance relationship between $\overline{\eta}_j$ and $\widehat{\eta}_j$ in equation~\eqref{mean and mode}, it follows that $\overline{\eta}_k$ also lies within a neighborhood of the same order.
     
     As given in Corollary~\ref{empty out rate}, when $\phi_0<(m+1)/2$, there are $K^*$ of the mixing weights $\overline w_j$'s being bounded away from 0, and the rest $K-K^*$ of $\overline w_j$'s will empty out at the rate $(\log n)^{\rho_1}/n$, where $\rho_1$ depends on $\phi_0$. Without loss of generality, we again assume that the first $K^*$ of the weights $\overline w_j$'s do not vanish and denote $\bm w_{K^*} = (w_j)_{j = 1}^{K^*}$ and $\bm \eta_{K^*} = (\eta_j)_{j = 1}^{K^*}$. Then for any fixed $\phi_0<(m+1)/2$ we have, with $C_D$ defined in equation~\eqref{eqn:TV_D},
    \begin{align*}
        d_{\rm TV}(p(\cdot\,|\,\overline{\bm w},\bm\eta),p^*(\cdot)) &\ = \frac{1}{2}\int \Big|\sum_{j=1}^K \overline w_jg(x;\eta_j) - \sum_{k=1}^{K^*} w_k^* g(x;\eta_k^*) \Big| dx\\
        &\ \ge \frac{1}{2}\int \Big|\sum_{j=1}^{K^*} \overline w_jg(x;\eta_j) - \sum_{k=1}^{K^*} w_k^* g(x;\eta_k^*) \Big| dx - \frac{1}{2}\sum_{j = K^* + 1}^K \overline w_j\\
        &\ \ge \frac{1}{2}\int \Big|\sum_{j=1}^{K^*} \overline w_jg(x;\eta_j) - \sum_{k=1}^{K^*} w_k^* g(x;\eta_k^*) \Big| dx - \frac{(K-K^*)[(\log n)^{\rho_1} + \phi_0]}{n}\\
        &\ \ge C_D\cdot D(\overline {\bm w}_{K^*},\bm\eta_{K^*}) - \frac{(K-K^*)[(\log n)^{\rho_1} + \phi_0]}{n}.
    \end{align*}
    In Assumption A1 we assumed $\sup_{\eta,\eta'\in\Omega}|\eta-\eta'|\le L$, which leads to the bound $D(\overline {\bm w}_{K^*},\bm\eta_{K^*})\le \sum_{k=1}^{K^*}(w_k + w_K^* + w_kL)\le 2 + L$. Now since $\big((\log n)^{\rho_1}/n\big)^2>0$, squaring both sides gives us
    \begin{align}
        d^2_{TV}(p(\cdot\,|\,\overline{\bm w},\bm\eta),p^*(\cdot))\ge  C_D^2\cdot D^2(\overline {\bm w}_{K^*},\bm\eta_{K^*}) - \frac{2(L + 2)(K-K^*)[(\log n)^{\rho_1} + \phi_0]}{n}.\label{TV_squred}
    \end{align} 
    With this result, we are able to show that with high probability, for any $k \leq K^*$, there will be some $\overline{\eta}_j$, $j \leq K^*$, concentrated within a ball of radius $M(\log n)^{\rho_1'/2}/\sqrt{n}$ around $\eta_k^*$. The general idea is as follows: First, for the components whose $\overline{w}_j$ are bounded away from 0, since the number of assigned samples $\widehat{n}_j$ is sufficiently large, the posterior distribution $\widehat{q}_{\eta_j}(\eta_j)$ concentrates most of its mass within a region of radius $\Phi_kn^{-1/2}$ around $\widehat{\eta}_j$ with some constant $\Phi_k$. Then, if there exists some $\eta_k^*$ such that no $\widehat{\eta}_j$ falls within the ball, the integral of the metric $D(\overline{\bm{w}}, \bm{\eta})$ with respect to the variational posterior distribution $\widehat{q}_{\bm{\eta}}(\bm{\eta})$ will have a lower bound of order $(\log n)^{\rho_1'/2} / \sqrt{n}$. Using the relationship established in \eqref{eqn:TV_D} and the connection between total variation and Hellinger distance, the integral of $h^2(\overline{\bm{w}},\bm{\eta})$ with respect to $\widehat{q}_{\bm{\eta}}(\bm{\eta})$ will correspondingly be bounded from below with order $(\log n)^{\rho_1'}/n$. Then, together with the convexity of the Hellinger distance with respect to each $w_k$, this lower bound contradicts the upper bound in Lemma 9, which implies that there must exist some $\widehat{\eta}_j$ falling in the small neighborhood of $\eta_k^*$.
    
    From Theorem~\ref{seperation}, we have that the expression of the optimal variational posterior as
    \begin{align*}
        \widehat q_{\eta_k}(\eta_k)\propto \pi_\eta(\eta_k)\exp\left[\sum_{i=1}^n \widehat p_{ik}\left(\eta_k^T T(x_i) - T_0(x_i) - A(\eta_k)\right)\right],
    \end{align*}
    where $\sum_{i=1}^n \widehat p_{ik} = \widehat n_k$, and $\widehat \eta_k$, defined in \eqref{eqn:etahat}, denotes the maximizer of the weighted log-likelihood function. In \eqref{denominator} we have examined the normalization constant of $\widehat q_{\eta_k}(\eta_k)$ using Lemma~\ref{laplace}:
    \begin{align*}
        \int \pi_\eta(\eta_k)\exp\left[\sum_{i=1}^n \widehat p_{ik}\left(\eta_k^T T(x_i) - T_0(x_i) - A(\eta_k)\right)\right] d\eta_k \le \left[B_2 + \frac{A_2}{\widehat n_k}\right]\cdot\frac{(2\pi/\widehat n_k)^{m/2}}{a^{m/2}},
    \end{align*}
    where $a$ is the lower bound for the eigenvalues of $I(\widehat\eta_k)$ as stated in Assumption A2. We also need to control the numerator to obtain the lower bound for the integral over a neighborhood of $\widehat \eta_k$:
    \begin{align*}
        \pi_\eta(\eta_k)\exp\left[\sum_{i=1}^n \widehat p_{ik}\left(\eta_k^T T(x_i) - T_0(x_i) - A(\eta_k)\right)- \ell_{n_k}(\widehat{\eta}_k)\right]\ge B_1\exp\Big[-\frac{b\widehat n_k}{2}(\eta_k-\widehat\eta_k)^2\Big],
    \end{align*}
    which implies that for any positive number $C$,
    \begin{align*}
        \int_{\{|\eta_k-\widehat\eta_k|_\infty \le C(b\widehat n_k)^{-1/2}\}}\widehat q_{\eta_k}(\eta_k)d\eta_k \ge \frac{B_1}{B_2 + A_2/\widehat n_k}\Bigg(\frac{a\big(2\Phi(C) - 1\big)}{b}\Bigg)^{m/2},
    \end{align*}
    where $\Phi$ denotes the CDF of the standard normal distribution and in particular, by choosing $C = \Phi^{-1}(3/4)$, we have $2\Phi(C) - 1 = \frac{1}{2}$. Recall from the proof of Theorem~\ref{main theorem}, for each non-emptying out component, we have $\widehat n_k\ge \gamma_kn$ for some positive constant $\gamma_k$ that only depends $w_k^*$ when $n$ is sufficiently large, then
    the integral above can be bounded from below by some positive constant we denote as $c_\pi$. Therefore, by selecting $\Phi_k = \Phi^{-1}(3/4)\cdot(b\gamma_k)^{-1/2}$, we have the following inequality,
    \begin{align*}
        \int_{\{|\eta_k-\widehat\eta_k|_\infty \le \Phi_k/\sqrt{n}\}}\widehat q_{\eta_k}(\eta_k)d\eta_k \ge c_\pi.
    \end{align*}
    We can now prove that for any $\eta^*_k$, there is some $\widehat\eta_j$, $1\le j\le K^*$ such that $|\eta^*_k - \widehat\eta_j|\lesssim(\log n)^{\rho_1'/2}/\sqrt{n}$ by contradiction.
    Suppose $\min_{1\le j\le K^*}|\widehat \eta_j - \eta^*_{k'}|\ge \max_{1\le j\le K^*}\Phi_j(\log n)^{\rho_1'/2} /\sqrt{n}$ holds for some $k'\le K^*$, then each $\bm\eta$ in the ball $\m{B} \coloneqq  \bigcap_{j=1}^{K^*}\{\bm\eta:|\eta_j - \widehat\eta_j|\le \Phi_j/ \sqrt{n}\}$ cannot be close to $\eta_k^*$ by the triangle inequality, i.e.,
    \begin{align*}
        \min_{1\le j\le K^*}|\eta_j - \eta^*_k|\ge\min_{1\le j\le K^*} \big(|\widehat\eta_j - \eta^*_{k'}| - |\eta_j - \widehat \eta^*_k|\big)\ge\min_{1\le j\le K^*} \frac{\Phi_j\big((\log n)^{\rho_1'/2} - 1\big)}{\sqrt{n}}
    \end{align*}
    holds for any $\bm\eta \in \m{B}$. Moreover, on $\m B$, it always holds that, by the definition of $D(\theta)$,
    \begin{align*}
        D(\overline{\bm{w}}, \bm{\eta}) \geq \inf_{j \in [K^*]} \overline{w}_j\cdot\frac{\Phi_j\big((\log n)^{\rho_1'/2} - 1\big)}{\sqrt{n}}.
    \end{align*}
    Then the lower bound of $d^2_{\rm TV}(p(x|\overline{\bm w},\bm\eta),p^*(x))$ in equation~\eqref{TV_squred} further implies
    \begin{align*}
        &\ \int_{\Omega^K} h^2\big(p(x\,|\,\overline{\bm w},\bm\eta), p^*(x)\big) \widehat q(\bm\eta)\,d\bm\eta  \ge \int_{\m{B}} d_{\rm TV}^2\big(p(x\,|\,\overline{\bm w},\bm\eta), p^*(x)\big) \widehat q(\bm\eta)\,d\bm\eta \\
        \ge &\ \int_{\m{B}} D^2(\overline{\bm w}_{K^*},\bm\eta_{K^*}) \widehat q(\bm\eta)\,d\bm\eta - \frac{2(L + 2)(K-K^*)[(\log n)^{\rho_1} + \phi_0]}{n}\\
        \ge &\ \frac{\inf_{j\in[K^*]} (\overline w_j \Phi_j)^2 \big((\log n)^{\rho_1'/2} - 1\big)^2n^{-1}\widehat Q(\m{B})}{n} - \frac{2(L + 2)(K-K^*)[(\log n)^{\rho_1} + \phi_0]}{n}\\
        \ge &\ \frac{\inf_{j\in[K^*]} (\overline w_j \Phi_j)^2\big((\log n)^{\rho_1'/2} - 1\big)^2\big(c_\pi(a/2b)^{m/2}\big)^{K^*}}{n} - \frac{2(L + 2)(K-K^*)[(\log n)^{\rho_1} + \phi_0]}{n},
    \end{align*}
    which is a contradiction with Lemma~\ref{consistency Hellinger} since $\rho_1'> 1\vee\rho_1$. This further implies that $\min_{1\le j\le K^*}|\widehat \eta_j - \eta^*_{k}|< \max_{1\le j\le K^*}\Phi_j(\log n)^{\rho_1'/2} /\sqrt{n}$ for any $k\le K^*$.
    Lastly, we showed in eqaution~\eqref{mean and mode} the distance between $\widehat \eta_j$ and $\overline\eta_j$ if of order $\sqrt{m}\cdot \widehat n_k^{-1}$ for each $j\in[K^*]$. Since $\widehat n_k>\gamma_k n$ with high probability, we obtain the desired result by selecting $M = \max_{1\le j\le K^*}\Phi_j + 1$:
    \begin{align*}
        \min_{1\le j\le K^*}|\overline\eta_j - \eta_k^*|&\ \le \min_{1\le j\le K^*}\big(|\overline\eta_j - \widehat\eta_j| + |\widehat\eta_j - \eta_k^*|\big)\\
        &\ \le \max_{1\le j\le K^*}\Phi_j(\log n)^{\rho_1'/2} /\sqrt{n} + \frac{\sqrt{m}A_3/\widehat n_k}{B_1 - A_1/\widehat n_k}\le M(\log n)^{\rho_1'/2} /\sqrt{n}.
    \end{align*}.
\end{proof}
\end{document}